\documentclass[twocolumn]{aastex631}

\usepackage{amsmath}
\usepackage{ragged2e}
\usepackage{booktabs, makecell, tabularx}

\setcellgapes{3pt}
\newcolumntype{L}{>{\RaggedRight}X}
\newcolumntype{P}[1]{>{\raggedright\arraybackslash}p{#1}}
\usepackage{xcolor}
\usepackage{xspace} 
\usepackage{hyperref}
\usepackage{soul}
\usepackage{multirow}

\newcommand{\Msol}{\ensuremath{M_{\mathrm{\sun}}}\xspace}

\newcommand{\fasthyphenmode}{fast-mode\xspace}
\newcommand{\fastmode}{fast mode\xspace}
\newcommand{\slowhyphenmode}{slow-mode\xspace}
\newcommand{\slowmode}{slow mode\xspace}
\newcommand{\MBfastminusslow}{\ensuremath{\scriptM^{\mathrm{fast}\; -\; \mathrm{slow}}}\xspace}
\newcommand{\MBslow}{\ensuremath{\scriptM^{\mathrm{slow}}}\xspace}

\newcommand{\Rc}{\ensuremath{R_{c_B}}\xspace}
\newcommand{\tauc}{\ensuremath{\tau_{c_R}}\xspace}
\newcommand{\fslow}{\ensuremath{f^{\mathrm{slow}}}\xspace}

\newcommand{\alphafast}{\ensuremath{\alpha^{\mathrm{fast}}}\xspace}
\newcommand{\alphaslow}{\ensuremath{\alpha^{\mathrm{slow}}}\xspace}
\newcommand{\Rxonefast}{\ensuremath{R_{x_1}^{\mathrm{fast}}}\xspace}
\newcommand{\Rxoneslow}{\ensuremath{R_{x_1}^{\mathrm{slow}}}\xspace}
\newcommand{\Rcfast}{\ensuremath{R_{c_B}^{\mathrm{fast}}}\xspace}
\newcommand{\Rcslow}{\ensuremath{R_{c_B}^{\mathrm{slow}}}\xspace}
\newcommand{\xonestarfast}{\ensuremath{x_1^{*, \;\mathrm{fast}}}\xspace}
\newcommand{\xonestarslow}{\ensuremath{x_1^{*, \;\mathrm{slow}}}\xspace}
\newcommand{\cstarfast}{\ensuremath{c_B^{*, \;\mathrm{fast}}}\xspace}
\newcommand{\cstarslow}{\ensuremath{c_B^{*, \;\mathrm{slow}}}\xspace}
\newcommand{\sigmaunexplfast}{\ensuremath{\sigma^{\mathrm{unexpl.,\; fast}}}\xspace}

\newcommand{\cstar}{\ensuremath{c^*_B}\xspace}

\newcommand{\lighthypencurveshape}{light-curve shape\xspace}
\newcommand{\lighthypencurvehypencshape}{light-curve-shape\xspace}

\newcommand{\evidencestrength}{mild\xspace}

\newcommand{\xonetrue}{\ensuremath{x_{1}^{\mathrm{true}}}\xspace}
\newcommand{\cBtrue}{\ensuremath{c^{\mathrm{true}}_B}\xspace}
\newcommand{\cRtrue}{\ensuremath{c^{\mathrm{true}}_R}\xspace}
\newcommand{\mBobs}{\ensuremath{m_{B}^{\mathrm{obs}}}\xspace}
\newcommand{\xoneobs}{\ensuremath{x_{1}^{\mathrm{obs}}}\xspace}
\newcommand{\cobs}{\ensuremath{c^{\mathrm{obs}}}\xspace}

\newcommand{\mBmodel}{\ensuremath{m_{B}^{\mathrm{model}}}\xspace}
\newcommand{\xonemodel}{\ensuremath{x_{1}^{\mathrm{model}}}\xspace}
\newcommand{\cmodel}{\ensuremath{c^{\mathrm{model}}}\xspace}

\newcommand{\Pslow}{\ensuremath{P^{\mathrm{slow}}}\xspace}
\newcommand{\Pfast}{\ensuremath{P^{\mathrm{fast}}}\xspace}

\newcommand{\Phigh}{\ensuremath{P^{\mathrm{high}}}\xspace}
\newcommand{\Phigheff}{\ensuremath{P^{\mathrm{high}}_{\mathrm{eff}}}\xspace}

\newcommand{\betaRL}{\ensuremath{\beta_{R}^{\mathrm{low}}}\xspace}
\newcommand{\betaRH}{\ensuremath{\beta_{R}^{\mathrm{high}}}\xspace}
\newcommand{\betaR}{\ensuremath{\beta_{R}}\xspace}

\newcommand{\Om}{\ensuremath{\Omega_m}\xspace}
\newcommand{\OL}{\ensuremath{\Omega_{\Lambda}}\xspace}
\newcommand{\LCDM}{$\Lambda$CDM\xspace}
\newcommand{\Omw}{$\Omega_m$-$w$\xspace}
\newcommand{\scriptM}{\ensuremath{\mathcal{M}_B}\xspace}

\newcommand{\nTotUNITY}{2085}

\newcommand{\flatLCDMOmSNe}{\ensuremath{0.334^{+0.025}_{-0.024}}\xspace}
\newcommand{\wOSNBAOCMB}{\ensuremath{-0.760^{+0.084}_{-0.082}}\xspace}
\newcommand{\waSNBAOCMB}{\ensuremath{-0.79^{+0.28}_{-0.30}}\xspace}
\newcommand{\DeltaChiSqBAOCMB}{\ensuremath{6.8}\xspace}
\newcommand{\DeltaChiSqSNeBAOCMB}{\ensuremath{9.6}\xspace}
\newcommand{\DeltaChiSqSNeBAOCMBHOT}{\ensuremath{7.4}\xspace}
\newcommand{\SigmaTwoDBAOCMB}{\ensuremath{2.1}\xspace}
\newcommand{\SigmaTwoDSNeBAOCMB}{\ensuremath{2.6}\xspace}
\newcommand{\SigmaTwoDSNeBAOCMBHOT}{\ensuremath{2.2}\xspace}

\newcommand{\deltaOUNITYOneSeven}{\ensuremath{0.037 \pm 0.014}\xspace}
\newcommand{\deltaOUNITYOneEight}{\ensuremath{-0.002^{+0.013}_{-0.012}}\xspace}

\begin{document}

\title{Banana Split: Improved Cosmological Constraints with Two Light-Curve-Shape and Color Populations Using Union3.1+UNITY1.8}

\newcommand{\uhawaii}{\affiliation{Department of Physics and Astronomy, University of Hawai`i at M{\=a}noa, Honolulu, Hawai`i 96822}}
\newcommand{\stsci}{\affiliation{Space Telescope Science Institute, 3700 San Martin Drive Baltimore, MD 21218, USA}}
\newcommand{\lbnl}{\affiliation{Physics Division, E.O. Lawrence Berkeley National Laboratory, 1 Cyclotron Rd., Berkeley, CA, 94720, USA}}
\newcommand{\ANUrsaa}{\affiliation{Research School of Astronomy and Astrophysics, The Australian National University, Canberra, ACT 2601, Australia}}
\newcommand{\ANUcga}{\affiliation{Centre for Gravitational Astrophysics, College of Science, The Australian National University, ACT 2601, Australia}}
\newcommand{\lancaster}{\affiliation{Physics Department, Lancaster University, Lancaster LA1 4YB, United Kingdom}}
\newcommand{\ucberkeley}{\affiliation{Department of Physics, University of California Berkeley, Berkeley, CA 94720, USA}}

\newcommand{\pilarone}{\affiliation{Instituto de F\'{\i}sica Fundamental, Consejo Superior de 
Investigaciones Cient\'{\i}ficas, E-28006, Madrid, Spain}}
\newcommand{\pilartwo}{\affiliation{Institute of Cosmos Sciences (UB--IEEC),  c/. Mart\'{\i} i Franqu\'es 1, E-08028, Barcelona, Spain}}

\newcommand{\fsu}{\affiliation{Department of Physics, Florida State University, 77 Chieftan Way, Tallahassee, FL 32306, USA}}

\newcommand{\usf}{\affiliation{Department of Physics and Astronomy, University of San Francisco, San Francisco, CA 94117-108, USA}}

\newcommand{\LPNHE}{\affiliation{LPNHE, CNRS/IN2P3 \& Sorbonne Universit\'e, 4 place Jussieu, 75005 Paris, France}}

\author[0000-0001-5402-4647]{David Rubin}
\uhawaii
\lbnl

\author[0000-0001-9664-0560]{Taylor Hoyt}
\ucberkeley
\lbnl

\author{Greg Aldering}
\lbnl

\author{Saul Perlmutter}
\lbnl
\ucberkeley

\begin{abstract}
SNe~Ia have been used to provide key constraints on the equation-of-state parameter of dark energy. They are generally standardized under the assumption that they belong to a single population, with luminosities standardized in a continuous (roughly linear) fashion using the observed light-curve timescale. 
We update the Union3+UNITY1.5 SN cosmology analysis in light of increasing evidence for at least two core populations of SNe~Ia and apply this ``UNITY1.8'' model to the updated ``Union3.1'' compilation \citep{Hoyt2026Mass}. In addition to finding evidence for two different \lighthypencurvehypencshape ($x_1$) distributions, we also find that the color distributions are different, that the \lighthypencurvehypencshape/magnitude standardization relations are different, and that these populations have different distributions across host-galaxy stellar mass and redshift. Importantly, we find that the residual host-mass luminosity step found in prior SN~Ia cosmology analyses is now consistent with zero for unreddened SNe. We report a significantly tightened constraint on the split in the red-color standardization between SNe in low- and high-mass galaxies. We find that the estimated uncertainties shrink on cosmological parameters when fitting the same SNe assuming two modes versus one mode. We confirm similar trends in simulated data when running both versions of UNITY on the same (two-mode) simulations. For a flat~\LCDM cosmology, we find $\Omega_m = $\flatLCDMOmSNe from SNe alone; for a flat $w_0$-$w_a$ cosmology, we find $w_0=\wOSNBAOCMB$ and $w_a=\waSNBAOCMB$ when including SNe, BAO, and CMB. In the 2D $w_0$-$w_a$ plane, UNITY1.8 decreases the tension with flat \LCDM by $0.1 \sigma$ to $\sim 3 \sigma$ compared to single-mode UNITY.
\end{abstract}


\section{Introduction} \label{sec:intro}

For the first time since the discovery of the accelerated expansion of the universe \citep{Riess1998, Perlmutter1999}, the redshift dependence of the dark energy equation of state parameter ($w$) seems to be in $\sim$~3--4$\sigma$ tension with that of a cosmological constant (i.e., $w=-1$ for all time). This was shown at $\sim 2\sigma$ significance in \citet{Rubin_2025} who combined Union3 SNe with Planck CMB and 6dF+SDSS+BOSS+eBOSS BAO \citep{PlanckCollaboration2020, Beutler2011, Ross2015, Alam2017, duMasdesBourboux2020, Neveux2020, Bautista2021, Raichoor2021, Hou2021, deMattia2021}. The statistical significance increased to $4.2 \sigma$ after combining the Dark Energy Survey (DES) SN Year~5 results \citep{DES5YR} with the Dark Energy Spectroscopic Instrument (DESI) collaboration Baryon Acoustic Oscillation (BAO) measurements \citep{Adame2025DR1, DESICollaboration2025DR2}. That statistical significance decreased with a recalibration (and bug fix) of the DES Year~5 results, but remains at $3.2 \sigma$ for DES+CMB+DESI \citep{Popovic2025Dovekie, Popovic2025DESCosmo}.

Since the publication of the Union3 SN\,Ia cosmology analysis, a few developments and improvements motivated us to update its underlying cosmological framework (Unified Nonlinear Inference for Type Ia cosmologY, UNITY, \citealt{Rubin2015, Rubin_2025}) and the supporting data. 
First, as previously mentioned, the DESI BAO results supersede the earlier SDSS+BOSS+eBOSS \citep{Alam2017, Alam2021} measurements used in Union3.\footnote{Union3 also used BAO from 6dF \citep{Beutler2011} which is mostly independent of DESI and thus should not be replaced.} 
Second, a growing body of evidence suggests that standardizing SNe\,Ia using their light-curve widths is more complex than the simple linear standardization historically used and also used in Union3 \citep{Burns2014, Wojtak2023, Garnavich2023, Ginolin2025a, Wojtak2025}.
Third, the correlation of host-galaxy properties with SN~Ia characteristics has emerged as a significant, underreported systematic discrepancy between independent analyses of the same SNe, thereby also impacting the combined interpretation of BAO and CMB with SNe \citep{Efstathiou_2025, Hoyt_2025arXiv250311769H, Vincenzi_2025}. In this work, we use an update to the Union3 host-galaxy stellar masses from \citet{Hoyt2026Mass}, referred to as ``Union3.1'', as inputs to our model.

This paper is organized as follows. Section~\ref{sec:UNITYchanges} details the updates implemented in UNITY since \citet{Rubin_2025} (``UNITY1.7''). Section~\ref{sec:TwoModeUNITY} outlines the evidence for two light-curve-shape modes and updates UNITY1.7 to fit for this (``UNITY1.8''). Section~\ref{sec:TwoModeUNITY} also validates our updates against both real and simulated data. Section~\ref{sec:SNresults} shows the results for SN standardization from applying UNITY1.8 to Union3.1. Section~\ref{sec:cosmology} shows our Union3.1+UNITY1.8 cosmological constraints with and without external probes. Finally Section~\ref{sec:conclusion} summarizes and concludes.

\section{UNITY1.7: Improvements since UNITY1.5} \label{sec:UNITYchanges}

\subsection{UNITY1.5 Review}

The Unified Nonlinear Inference for Type Ia cosmologY framework (UNITY) is a Bayesian hierarchical model for computing SN cosmological constraints. It simultaneously models SN standardization, \lighthypencurvehypencshape and color population distributions, outliers, unexplained dispersion, selection effects, systematic uncertainties, and cosmological parameters. The underlying approach (described and motivated in more detail in \citealt{Rubin2015, Rubin_2025}) is to describe SN cosmology as a regression problem where the results of light-curve fits can be forward modeled in an analysis that treats all parameters simultaneously. When there is ambiguity about the correct model in way that affects the cosmological results (for example, whether the host-galaxy relations should evolve with redshift), UNITY parameterizes and marginalizes out these choices.

To review, the UNITY1.5 standardization equation (Equation~5 from \citealt{Rubin_2025}) models the rest-frame $B$-band magnitude ($m_B$) for each SN as:
\begin{eqnarray}
    m^{\mathrm{model}}_B & = & M_B + \mu(z,\ \mathrm{cosmology}) \nonumber \\
    & & -\alpha\, x^{\mathrm{true}}_1 + \beta_B\, c^{\mathrm{true}}_B \nonumber \\
    & & + [\betaRL \, (1 - \Phigheff) +  \betaRH \, \Phigheff] c^{\mathrm{true}}_R  \nonumber \\
    &  & - \delta(z=0) \, \Phigheff  \;.  \label{eq:tripponefive}
\end{eqnarray}
$M_B$ is the rest-frame $B$-band absolute magnitude, $\mu$ is the distance modulus, $\alpha$ is the $x_1$ standardization coefficient, $\beta_B$ is the color standardization coefficient for Gaussian core of the color distribution, \betaRL and \betaRH are the color standardization coefficient of the red exponential tail in low-mass and high-mass hosts respectively, and $\delta(z=0)$ is the host-mass standardization coefficient at low redshift. \Phigheff is the probability of a host-galaxy behaving like a low-redshift host galaxy having high stellar mass:
\begin{eqnarray}
    \Phigheff   \equiv & \Phigh & \Biggl[   \frac{1.9}{0.9 \cdot 10^{0.95 \,z} + 1} \left[ 1 - \frac{\delta(z=\infty)}{\delta(z=0)}\right]  \nonumber \\ 
    & &  + \frac{\delta(z=\infty)}{\delta(z=0)} \Biggr] \; ; \label{eq:Phigheff}
\end{eqnarray}
the functional form of the redshift dependence is motivated in \citet{Rigault2013} and is also similar to the curve in Figure~9 of \citet{Childress2014}.  It assumes the host-stellar-mass step is due to progenitor age, and models how (mostly) high-stellar-mass galaxies converge to the same mean progenitor age as low-stellar-mass galaxies at high redshift. The same \Phigheff was also used for the \betaR step, as this is part of the host-stellar-mass step. \Phigh was computed assuming Gaussian stellar-mass uncertainties and a low/high-stellar-mass cutoff of $10^{10} \Msol$.

There are three latent per-SN parameters (\xonetrue, \cBtrue, and \cRtrue) that are explicitly marginalized over (as opposed to analytically marginalized over) for every SN in the analysis, and these parameters have population-distribution priors with their own parameters (``hyperparameters''). The $x_1$ distribution is modeled as an exponentially modified Gaussian distribution:
\begin{equation}
        \xonetrue  \sim \mathrm{ExpModNormal}(x_1^{*},\   R_{x_1}^2, 1/\tau_{x_1}) \;.
\end{equation}
The red colors $c^{\mathrm{true}}_R$ are assumed to be distributed as an exponential while the blue colors $c^{\mathrm{true}}_B$ are assumed to be Gaussian distributed:
\begin{eqnarray}
    \cRtrue & \sim & \mathrm{Exp}(1/\tauc) \\
    \cBtrue & \sim & \mathcal{N}(\cstar,\   \Rc^2) \;.
\end{eqnarray}
Each of these hyperparameters above really consists of six numbers, as each is allowed to vary independently in bins of low/high host-galaxy stellar mass and low/mid/high redshift. The motivation for the two sources of color variation is intrinsic ($c_B$) and extinction ($c_R$), although UNITY is an empirical model and this correspondence does not have to be exact. Of course, if the $c^{\mathrm{true}}_R$ values really do have a good correspondence with extinction, the distribution will likely be more sharply peaked than an exponential \citep{Hatano1998, Hallgren2025}; we validate our assumptions against the data in Section~\ref{sec:PPD}.\footnote{We also perform a toy-model test where we generate some fraction of the SN population at zero extinction and the rest with an exponential distribution. This extinction is convolved with a Gaussian intrinsic color distribution and Gaussian uncertainties to get the observed color distribution. We cut at an observed color of 0.3, as we do for the real data. When fitting with our Gaussian \cBtrue/exponential \cRtrue model, we find a bias in the recovered mean distance modulus of 0.0016 magnitudes per percent of SNe in the zero-extinction spike. This bias is generally small compared to other uncertainties and would partially cancel in redshift if similar fractions were in the spike as a function of redshift. Models that do a better job modeling intrinsic and extrinsic color at the light-curve-fitting stage are an important area of ongoing research \citep{Mandel2022, Hand2025, Kenworthy2025SALT3plus}.}$^,$\footnote{Note that some models allow for the opposite: a two-tailed extinction distribution, where the modal extinction is greater than zero \citep{Wojtak2023}. One practical difficulty with this is that a two-tailed extinction distribution can be similar to a Gaussian-convolved exponential, which would make it difficult to assign the $c_B$ (Gaussian distributed) and $c_R$ (extinction-like, here assumed exponentially distributed) colors; one can see this in the \cite{Wojtak2023} model in the degeneracy between the mean intrinsic color and the degree that the extinction distribution is two-tailed. Another factor is that, for some of the \citet{Wojtak2023} models, there is no unexplained dispersion other than the dispersion in extinction $R_B$, which may favor a two-tailed distribution so that all SNe have some extinction (and thus some unexplained dispersion). Note that there is no unexplained dispersion in color for any of the \citet{Wojtak2023} models. In any case, \citet{Wojtak2023} finds $\lesssim$ 3-to-1 posterior density in favor of a two-tailed distribution (depending on SN sample) so we assume exponentially distributed $c_R$ for now.}

The model for $x_1$ is simply 
\begin{equation}
    x_1^{\mathrm{model}} = \xonetrue
\end{equation} for each SN and the model for $c$ is 
\begin{equation}
    c^{\mathrm{model}} = \cBtrue + \cRtrue
\end{equation}
for each SN. The observed $\{\mBobs, \xoneobs, \cobs\}$ is compared to the model $\{\mBmodel, \xonemodel, \cmodel\}$ assuming multi-dimensional (correlated) Gaussian uncertainties (including unexplained dispersion in $m_B$, $x_1$, and $c$) modified with a model of the selection effects (\citealt{Rubin2015} Appendix~B). The inlier distribution is mixed with an outlier distribution which is also assumed to be Gaussian (but is constrained to be broader to capture any SNe not well modeled by the inlier distribution).

\subsection{UNITY Updates}

This update paper provides an opportunity to make some smaller, technical tweaks to UNITY. The current best version, referred to as UNITY1.5, was developed for the Union3 analysis \citep{Rubin_2025} and we begin our updates from the version with the corrected parameter limit for the width of the outlier distribution in exponentially distributed latent color ($\sigma^{\mathrm{outl}}_{c_R}$, we refer to the corrected version as ``UNITY1.6''). Table~\ref{tab:UNITYVersions} outlines the versions of UNITY referenced and developed in this paper.

\begin{deluxetable*}{lp{3 cm}p{12 cm}}[htbp]
\caption{Table comparing recent UNITY versions. \label{tab:UNITYVersions}}
    \tablehead{
\colhead{UNITY Version} & \colhead{Reference} & \colhead{Summary}}
\startdata
1.5 & \citet{Rubin_2025} & Used for Union3 analysis. Models one inlier population of SNe with Gaussian $*$ exponential $x_1$ and $c$ populations.\\
1.6 & \citet{Rubin_2025} & Same as UNITY1.5 above, but with improved \cRtrue outlier parameter limit discussed in \citet{Rubin_2025} Appendix~B. \\
1.7 & \makecell[lt]{This work and \\ \citet{Hoyt2026Mass}} & Also includes fit for mass-step location  and distance-ladder capability. \\
1.8 & This work & Union3.1 model that includes two separate SN populations with different (Gaussian) \xonetrue and \cBtrue populations, different $\alpha$ values, and extra unexplained dispersion for the faster population. \\
\enddata
\end{deluxetable*}

UNITY1.7 now fits the mass of the host-mass step instead of fixing it to $10^{10}$~\Msol. This mass affects both the luminosity step $\delta(z=0)$ and the $\beta_R$ step (in other words, the difference of $\beta_R$ with host-galaxy stellar mass, $\betaRL-\betaRH$); see Equation~\ref{eq:tripponefive}. Differences in the galaxy photometry and underlying galaxy SED model can lead to systematic offsets in estimates of galaxy stellar mass as large as 0.5~dex, so even if one knew {\it a priori} that the step was at $10^{10} \Msol$ (as was assumed in UNITY1.5 and many prior analyses), the apparent location could be different, making it an important value to solve for. Including it as a fit parameter naturally propagates the corresponding uncertainties into the other parameters. See \cite{Hoyt2026Mass} for a more detailed discussion.

UNITY now supports distance-ladder constraints by replacing the cosmological integral with a distance modulus from another measurement (e.g., from Cepheids or Tip of the Red-Giant Branch). These distances will have their own distance-modulus covariance matrix from the lower rungs of the distance ladder (e.g., \citealt{Riess_2022} Figure 11), and UNITY can now include this covariance matrix. UNITY can also include unexplained dispersion from the lower rungs, in case the calibrators have distance-modulus dispersion larger than their uncertainties. Making full use of this capability on real data requires collecting light curves for many calibrator SNe which are not in Union3, so we pursue this in an upcoming $H_0$ paper (Union3.2, \citealt{TaylorUNITYHubble}).

We also made a mild UNITY update to the generation of the distance moduli. As part of its set of parameters, UNITY infers a distance-modulus $\mu(z)$ function. This function is either a parameterized cosmological model (e.g., flat \LCDM) or is constructed from a set of distance moduli. As discussed in Section~\ref{sec:SNCosmology}, the set of distance moduli is parameterized by adding the distance modulus of a flat\LCDM model with $\Omega_m=0.3$ and spline-interpolated values as a function of redshift. However, when not including a distance ladder, there is a degeneracy between adding a constant to all distance moduli and adjusting \scriptM. In UNITY1.5, this degeneracy was broken by adding a virtual spline node at $z=0$ and fixing this node to 0. In the UNITY versions presented here, we instead fix this node to $-$(the sum of all other nodes). This is more effective at breaking the degeneracy, so there is less covariance between the quoted distance moduli. This covariance does not have an impact on the cosmological constraints, so this is a cosmetic improvement.

As mentioned in the Introduction, we also update some of the external cosmological constraints. We replace the SDSS+BOSS+eBOSS Baryon Acoustic Oscillation (BAO) measurements with DESI DR2 distances \citep{DESICollaboration2025DR2} which are higher precision over the same sky. However, we keep 6dF \citep{Beutler2011}, which has little sky overlap with DESI. Planck CMB \citep{PlanckCollaboration2020} and Big Bang nucleosynthesis \citep{Cooke2016} constraints carry over from Union3.\footnote{We also consider some newer approaches to modeling the Planck data and their impact on cosmology in \autoref{sec:cosmoexternal}.}

We also update the Cepheid-based Hubble constant measurement to the latest from SH0ES ($73.17\pm 0.86$~km/s/Mpc, \citealt{Breuval2024}), and we update the TRGB measurement to the latest from CCHP ($70.39 \pm 1.80$~km/s/Mpc, \citealt{Freedman2025}). Of course, applying these measurements as priors in the analysis ignores the covariance between $H_0$ and other cosmological parameters; as mentioned above, we will explore this in a subsequent paper that correctly incorporates the lower rungs of the distance ladder into UNITY \citep{TaylorUNITYHubble}. Nevertheless, it gives an approximate sense of how the cosmological constraints are affected by external Hubble-constant measurements of different values and precisions.

\section{UNITY1.8: Standardization with Multiple Populations}{\label{sec:TwoModeUNITY}}

This Section describes the evidence for multiple populations of SNe~Ia (Section~\ref{sec:motivation}) and the updates to UNITY that we make (Section~\ref{sec:UNITYoneeight}).
All of these UNITY changes were made with the cosmology hidden (a ``blinded'' analysis, \citealt{Maccoun2015}) until after the analysis had been validated on simulated data (Section~\ref{sec:simdata}) and by examining the predictive posterior distributions (Section~\ref{sec:PPD}). We revisit alternate models as part of the pre-unblinding tests described in Section~\ref{sec:unblinding}.

\subsection{Motivation} \label{sec:motivation}

A key component in the empirical standardization of SNe\,Ia is the light curve shape, as parameterized by its rate of decline after maximum ($\beta$ in \citealt{Pskovskii1977}, $\Delta m_{15}$ in \citealt{Phillips1993, Tripp1997, Tripp1998}), light-curve width ``stretch'' \citep{Perlmutter1997a, Perlmutter1997}, or a \lighthypencurvehypencshape template parameter ($\Delta$ in \citealt{Riess1996} or $x_1$ in \citealt{Guy2007}). In the canonical single-degenerate SN\,Ia progenitor model---where a WD accretes matter from a donor star until it reaches the Chandrasekhar mass, at which point a thermonuclear runway is triggered---the rise time of the light curve is ostensibly proportional to the amount of radioactive $^{56}$Ni generated \citep{Arnett1982, Arnett1985, Pinto2000}. However, when broadening the SN~Ia progenitor landscape to include WD mergers, the light-curve width reflects a combination of the ejecta mass and the amount of $^{56}$Ni \citep{Scalzo2014}. Thus, the $\Delta m_{15}$ and $x_1$ parameters may be trying to account for two physical effects at once---much as the SALT $c$ parameter tries to account simultaneously for both intrinsic color-luminosity correlations and reddening by line-of-sight dust. Here we briefly review evidence for nonlinear and multi-mode models for the \lighthypencurvehypencshape standardization, then use that information to develop a suitable model for UNITY.

There has long been evidence that standardization is not a simple linear relation with \lighthypencurveshape. If nothing else, $x_1$ is commonly substituted for the $\Delta m_{15}$ in the original \citet{Tripp1998} linear-standardization relation, yet the relationship between these two \lighthypencurvehypencshape parameters is itself nonlinear \citep{Guy2007, Kessler2009}, so they cannot both be optimally treated with linear standardization. Indeed, \citet{Phillips1999, Wang2006, Amanullah2010, Sullivan2010, Scolnic2014, Rubin2015, Burns2018, Garnavich2023, Ginolin2025b} show evidence for nonlinear standardization relations. Even so, merely expanding to a nonlinear standardization with respect to $x_1$ assumes a single unique mapping between SN~Ia luminosity and $x_1$. This does not have to be case if, for example, there are multiple subpopulations, each with their own relation.

The evidence for discontinuous or multi-modal luminosity standardization at a given $x_1$ is  growing. The $x_1$ distribution changes across host-galaxy types \citep{Hamuy2000, Howell2001, Sullivan2006, Smith2020, Wiseman2021, Garnavich2023, Larison2024, Senzel2025}, with fast decliners more preferentially located in passive galaxies and slow decliners preferentially located in star-forming galaxies. As the mix of host-galaxy types changes with redshift, the $x_1$ distribution is thus expected to change too. Indeed, \citet{Nicolas2021} find that the $x_1$ is better described by two Gaussian modes than with a skewed Gaussian,
and that the relative strengths of the modes evolves with redshift. \citet{Rigault2013} suggested a bimodality to Hubble residuals, with the second mode showing up only in passive host environments. 
\cite{Wojtak2023} incorporated this into their standardization model, having two subpopulations---``fast'' and ``slow'' decliners---described by Gaussian distributions in $x_1$. They found similar standardization luminosities, significantly different mean colors and reddening distributions, but found consistent luminosity-width slopes ($\alpha$). They discussed a possible young vs. old interpretation of the two modes and expanded on it in \citet{Wojtak2025}. 

Apart from indirect evidence related to different delay-time distributions, there are some more direct clues to the differences in the progenitor systems. \citet{Maguire2013} finds time-varying, blueshifted Na\,\textsc{i}\,D absorption in SNe only hosted by late-type galaxies, possibly indicating that these are single-degenerate systems that have undergone mass loss. \citet{Tucker2025} finds that bimodal nebular phase emission, a sign of WD mergers, only occurs in passive galaxies. If SNe in old and young environments arise from different projector channels, there is no expectation that the population distributions, luminosity distribution, or standardization relations will be the same.

\subsection{Defining the new standardization model} \label{sec:UNITYoneeight}
We therefore construct a two-mode UNITY model which we refer to as ``UNITY1.8.'' We provide a list of parameters with their priors in Appendix~\ref{sec:priors}, but summarize the changes here. UNITY1.8 includes separate $x_1$ mean and dispersion ($x_1^*$ and $R^{x_1}$) parameters for each of two modes, separate mean and dispersion of the blue-color  (\cstar and \Rc) parameters, separate $\alpha$ values, separate absolute magnitudes ($M_B$), and separate unexplained dispersions. We investigated allowing $\beta_B$ or $\beta_R$ to be different between modes, but found almost identical $\beta_B$ values and that $\beta_R$ seems to correlate more with host-galaxy stellar mass (as in the UNITY1.5 model), so we do not include these splits as part of our nominal model. As with UNITY1.5, we split the red-color-population scale length (\tauc) in bins of both host-galaxy stellar mass and redshift, helping to lessen any biases if the extinction distribution changes with these parameters. We treat the blue-color population and $x_1$ distributions as associated with the modes, and so these do not get different values depending on redshift and host mass. However, the relative fractions of the modes are allowed to vary with both host-galaxy stellar mass and redshift.

In equation form, we update the UNITY1.5 standardization equation (Equation~5 of \citealt{Rubin_2025}, reproduced in Equation~\ref{eq:tripponefive}) to read:
\begin{eqnarray}
    & & m^{\mathrm{model,\ fast\ or\ slow}}_B \label{eq:tripp} \\
    & = & -\alpha^{\mathrm{fast\ or\ slow}}\, \left(x^{\mathrm{true}}_1 - x_1^{*,\ \mathrm{fast\ or\ slow}}\right) \nonumber \\
    & + & \beta_B\, c^{\mathrm{true}}_B \nonumber \\
    & + & [\betaRL \, (1 - \Phigheff) +  \betaRH \, \Phigheff] c^{\mathrm{true}}_R  \nonumber \\
    & - &  \delta(z=0) \, \Phigheff \nonumber \\ 
                     & & + M_B^{\mathrm{fast\ or\ slow}} + \mu\left(\{z,\ \mathrm{cosmology}\}\ \mathrm{or\ calibrator}\right) \nonumber \;,  
\end{eqnarray}
where ``fast'' and ``slow'' refer to the lower- and higher-$x_1$ population, respectively. While the naming convention is derived from the $x_1$ parameter (a strong motivation of the two-population model), UNITY1.8 also allows the two populations to have distinct $c^{\mathrm{true}}_B$ distributions, if supported by the data. These are combined in a mixture model (along with the outlier distribution) so each SN is only probabilistically assigned into each population's mode. Each $x_1$ mode is modeled as Gaussian:
\begin{equation}
        \xonetrue  \sim \mathcal{N}(x_1^{*,\ \mathrm{fast\ or\ slow}},\   (R_{x_1}^{\mathrm{fast\ or\ slow}})^2) \;.
\end{equation}
As in UNITY1.5, the red colors $c^{\mathrm{true}}_R$ are assumed to be distributed as an  exponential while the blue colors $c^{\mathrm{true}}_B$ are assumed to be Gaussian distributed:
\begin{eqnarray}
    \cRtrue & \sim & \mathrm{Exp}(1/\tauc) \\
    \cBtrue & \sim & \mathcal{N}(c_B^{*,\ \mathrm{fast\ or\ slow}},\   (\Rc^{\mathrm{fast\ or\ slow}})^2) \;.
\end{eqnarray}
In other words
\begin{widetext}
\begin{align}
p(\xonetrue,\cBtrue,\cRtrue)
&=
(1-f^{\rm slow})\,
p(\xonetrue,\cBtrue,\cRtrue \mid {\rm fast})
+
f^{\rm slow}\,
p(\xonetrue,\cBtrue,\cRtrue \mid {\rm slow}) ,
\\[4pt]
p(\xonetrue,\cBtrue,\cRtrue \mid {\rm fast})
&=
\mathcal{N}\!\left(\xonetrue \mid \xonestarfast, (\Rxonefast)^2\right)
\,
\mathcal{N}\!\left(\cBtrue \mid \cstarfast, (\Rcfast)^2\right)
\,
\mathrm{Exp}(\cRtrue \mid 1/\tauc),
\\[4pt]
p(\xonetrue,\cBtrue,\cRtrue \mid {\rm slow})
&=
\mathcal{N}\!\left(\xonetrue \mid \xonestarslow, (\Rxoneslow)^2\right)
\,
\mathcal{N}\!\left(\cBtrue \mid \cstarslow, (\Rcslow)^2\right)
\,
\mathrm{Exp}(\cRtrue \mid 1/\tauc).
\end{align}
\end{widetext}
With the assumed Gaussian outlier model separate. We define \Pslow, \Pfast, and the outlier probability by computing the median of the slow/fast/outlier log likelihood for each SN, exponentiating, and normalizing the sum of the three to 1.

Note that we continue to use Equation~\ref{eq:Phigheff} to capture any redshift evolution in the host-stellar-mass correlations (both the luminosity step and the \betaR step). We do this even though this functional form was motivated by progenitor age and the two modes we already model likely capture young and old SNe \citep{Wojtak2023}; here it simply serves as a two-parameter redshift-dependent function not very different than one linear in scale factor $a(t)$.

Table~\ref{tab:UNITY17vs18} shows an accounting of the difference in parameter count between UNITY1.7 and UNITY1.8 (13 fewer parameters in UNITY1.8). UNITY1.7 has one set of \cstar, \Rc, \tauc, $x_1^*$, $R_{x_1}$, and $\tau_{x_1}$ for each redshift (low/mid/high) and host stellar mass (low/high) bin for a total of 36 population parameters. UNITY1.8 has one set of \cstar, \Rc, $x_1^*$, and $R_{x_1}$ for each mode (slow/fast). The number of \tauc parameters is unchanged (at six) and there are also the same six bins (low/mid/high redshift and low/high mass) for the fraction of SNe in each mode. In addition, UNITY1.8 has two extra standardization coefficients as $\alpha$ and $M_B$ are now split by slow/fast mode instead of having just one value, and there is an extra unexplained-dispersion parameter (\sigmaunexplfast).

\begin{deluxetable*}{lccc}
\tablecaption{Comparison of changed parameter groups in UNITY1.7 and UNITY1.8\label{tab:UNITY17vs18}}
\tablehead{
\colhead{Parameter group} & \colhead{UNITY1.7\tablenotemark{a}} & \colhead{UNITY1.8\tablenotemark{b}} & \colhead{Count change}
}
\startdata
Latent color-location parameters, \cstar 
& 6 ($3$ $z$ bins $\times$ $2$ host-mass bins) 
& 2 (slow, fast) 
& $-4$ \\
Latent color-width parameters, \Rc
& 6 ($3$ $z$ bins $\times$ $2$ host-mass bins) 
& 2 (slow, fast) 
& $-4$ \\
Latent color-skew parameters, $\tauc$
& 6 ($3$ $z$ bins $\times$ $2$ host-mass bins) 
& 6 ($3$ $z$ bins $\times$ $2$ host-mass bins) 
& 0 \\
Latent stretch-location parameters, $x_1^*$
& 6 ($3$ $z$ bins $\times$ $2$ host-mass bins) 
& 2 (slow, fast) 
& $-4$ \\
Latent stretch-width parameters, $R_{x_1}$
& 6 ($3$ $z$ bins $\times$ $2$ host-mass bins) 
& 2 (slow, fast) 
& $-4$ \\
Latent stretch-skew parameters, $\tau_{x_1}$
& 6 ($3$ $z$ bins $\times$ $2$ host-mass bins) 
& 0 
& $-6$ \\
Mode-fraction parameters
& N/A 
& 6 ($3$ $z$ bins $\times$ $2$ host-mass bins) 
& $+6$ \\
Stretch standardization coefficients, $\alpha$
& 1 
& 2 ($\alpha_{\mathrm{slow}}$, $\alpha_{\mathrm{fast}}$) 
& $+1$ \\
Absolute-magnitude parameters, $M_B$
& 1 
& 2 ($M_{B,\mathrm{slow}}$ and $\Delta M_{B,\mathrm{fast-slow}}$) 
& $+1$ \\
Extra unexplained-dispersion parameter
& N/A 
& 1 ($\sigma_{\mathrm{unexpl,fast}}$) 
& $+1$ \\
\hline
\textbf{Subtotal of changed parameter groups}
& \textbf{38}
& \textbf{25}
& $\textbf{-13}$
\enddata
\tablecomments{This table counts only the parameter groups that differ between UNITY1.7 and UNITY1.8; parameters unchanged between the two model versions are not included. \tablenotetext{a}{UNITY1.7 was designed to flexibly capture the changing population distributions with redshift and host-galaxy stellar mass. However, it only modeled one population of SNe, with one $\alpha$ and $M_B$.} \tablenotetext{b}{UNITY1.8 models the SN population as two modes where the relative proportion of each mode varies with redshift and host-galaxy stellar mass. The exponentially distributed color ($c_R$) distribution is still allowed to depend on redshift and host-galaxy stellar mass because it may be associated with SN-extrinsic host-galaxy properties. UNITY1.8 generally describes the data better despite having fewer parameters, see Section~\ref{sec:PPD}.}}
\end{deluxetable*}

To summarize, unlike models that split samples on the {\it observed} value of \xoneobs, in this case $x_1$ influences the relative probability of being in the fast or slow decliner subpopulations (with the subpopulations based on the preceding equations). When the standardizations are different for the two subpopulations, this has the interesting feature that a unique value of $x_1$ predicts two alternative standardization corrections (in some proportion). This uncertainty may explain part of the so-called ``unexplained dispersion'' encountered when attempting to standardize SNe~Ia using light curve shape and color alone, but our simulated-data testing (Section~\ref{sec:simdata}) shows that UNITY1.7's single-population model does not increase the unexplained dispersion much over UNITY1.8's two-population model. We present a detailed discussion of our SN results in Section~\ref{sec:SNresults}.

\subsection{Simulated-Data Testing} \label{sec:simdata}

Simulated-data testing is an essential part of a complex analysis to ensure the correctness of the results. We performed our simulated-data testing and validation before unblinding the cosmology results for real data. Our simulated-data testing is based on that done for Union3+UNITY1.5 \citep{Rubin_2025}, updated to be based on the new two-mode UNITY1.8 model. Detailed results are provided in Appendix~\ref{sec:simdataappendix}. Note we do not simulate from the UNITY model itself, but instead construct an independent simulation. This provides a more realistic (and more difficult) test for UNITY to pass than if the test data were simulated from UNITY and fed right back into it.

We ran 100 realizations of simulated data (each one with four simulated datasets) based on the UNITY1.8 model with two populations in $x_1$. For each realization, we ran four UNITY variants: UNITY1.8 and 1.7 with flat~\LCDM and flat~$w_0$-$w_a$ (with $\Omega_m$ fixed so that external data need not be included). For each UNITY run, we used four chains with 2500 samples/chain, and (to be conservative) discarded the first half of each chain. This is sufficient sampling such that sampling errors on parameter estimates fell more than an order of magnitude below the typical uncertainty $(N_{\mathrm{eff}} \sim 1000)$. We restarted runs that got stuck or that did not sample well.

We summarize here key findings from the simulated-data testing, with more details provided in Appendix B.

\begin{enumerate}
    \item The testing shows little evidence of bias  ($\lesssim  0.2\sigma \pm 0.1 \sigma$) in the inference of final cosmological parameters ($H_0$, $\Omega_m$, $w_0$, $w_a$) using the nominal UNITY1.8 model, indicating our cosmological parameter inference from the real data is likely also minimally biased.

    \item However, the cosmological-parameter inference is noticeably biased in the previous, single-mode-$x_1$ model (UNITY1.7) when fitting these two-mode simulations, with this bias ranging from $0.7 \sigma$ on \Om to 0.3--$0.5 \sigma$ on the dark energy equation-of-state parameters. The simulations are designed for testing whether the outputs match the inputs and differ from the data in some important respects. For example, the {\it HST}-like sample is larger in the simulation to make for a more sensitive test and the population parameters do not vary with redshift in the way the real data do (Figure~\ref{fig:redshiftmass}). So the exact cosmological change from UNITY1.7 to UNITY1.8 should not be expected to perfectly match the real data. That being said, the change from UNITY1.7 to UNITY1.8 is fairly similar for the simulated data and the real data (Section~\ref{sec:cosmology}). The flat-\LCDM \Om value changes by $-0.009$ on the simulated data when running UNITY1.8 vs. UNITY1.7. On the real data, this change is also $-0.009$ (using the UNITY1.7 results from \citealt{Hoyt2026Mass}). For flat $w_0$-$w_a$ (with fixed $\Omega_m$), $w_0$ changes by $+0.041$ for UNITY1.8 vs. UNITY1.7 on the simulated data and a similar $+0.036$ on the real data. For the same model, $w_a$ changes by $-0.396$ on the simulated data and $-0.328$ on the real data.

    \item UNITY1.8 estimates slightly smaller cosmological-parameter uncertainties when run on the same input simulated data. Note that the input simulations are based on a two-mode model.

    \item Both UNITY1.7 and UNITY1.8 misestimate the outlier fraction present in the simulations, indicating that the outlier model needs more fidelity to match the data. This was also seen in Union3 \citep{Rubin_2025} and is generally not a cause for concern for two reasons. 1) There is very little covariance between the outlier-distribution parameters and the cosmological parameters. 2) Because the inlier distribution(s) are much narrower than the outlier distribution, the normalization of the outlier distribution does not strongly affect which SNe are recognized as outliers.
    
    \item A non-zero correlation of SN luminosity with host mass $\delta(z=0)$ is introduced when attempting to model a two-mode population with UNITY1.7, suggesting at least part of the signal for the characteristic host-mass step seen in SN luminosities is a result of incomplete modeling of SN population diversity along the \lighthypencurvehypencshape axis. After unblinding, we ran an additional set of simulations with $\delta(z=0)=0.1$~mag to verify that UNITY1.8 would be able to recover a non-zero host-mass step. Encouragingly, we find that it does; the mean of the realizations is $0.099 \pm 0.002$~mag for UNITY1.8 while it is $0.127 \pm 0.002$~mag for UNITY1.7.
    
    \item The unexplained-dispersion values are biased somewhat high for UNITY1.7 and somewhat low for UNITY1.8. This bias is detectable using the aggregated 100 realizations, but is mild in any one realization. The fact that the bias is not large for UNITY1.7 (when fitting data generated with a two-mode model) implies that not much of the UNITY1.7 unexplained dispersion is due to the ambiguity over which mode a given SN is in and this likely carries over to the real data.

    \item The absolute value of $\alpha$ is biased low by about $1.3 \sigma$ in either cosmology considered, while the difference $\alphafast - \alphaslow$ exhibits negligible bias ($<0.1\sigma$). Thus, the evidence UNITY finds for a difference in $\alpha$ between the two $x_1$ populations is very likely real and not very biased in size. Other $x_1$-related parameters are also biased, especially for the \fastmode. A similar (but smaller) bias on standardization coefficients was seen by \citet{Rubin_2025}. We also see biases in \fslow for the high-$z$ simulated dataset (which is much smaller than the others. These biases are generally small in absolute terms (e.g., population parameters biased by about $0.1$ in $x_1$) and are thus difficult to diagnose. As these biases are still mild in any one realization, and these parameters do not have much correlation with cosmological parameters, we decided to accept UNITY1.8 as an overall improvement over current models and proceed to unblinding, although more investigation is needed.
\end{enumerate}

\subsection{Checks with the Predictive Posterior Distributions}
\label{sec:PPD}

One way to look for any improvement in model fidelity between UNITY1.7 and 1.8 is to test with the predictive posterior distributions. These are datasets that are simulated from the model posterior distribution. For each MCMC sample, we simulate a dataset from the UNITY model, including simulating which SNe are discovered and selected to be in the final sample according to the modeled efficiency of each survey. As in \citet{Rubin_2025}, we approximate the light-curve-fit uncertainties by constructing an interpolation for each SN of $m_B$, $x_1$, $c$ uncertainties based on the true $m_B$ and $c$ ($x_1$ has a smaller effect on the uncertainties, which we ignore).

\newcommand{\PPDxoneSentence}{These results show that UNITY1.8 standardizes almost the entire $x_1$ range well (i.e., $x_1 > -2$), and even $x_1 \sim -2.5$ SNe are standardized much better with UNITY1.8 than 1.7.\xspace}

\newcommand{\PPDcSentence}{These results show a good match to the data for both UNITY1.7 and 1.8.\xspace}

\newcommand{\PPDchisquaresentence}{To give a sense of the improvement, the total $\chi^2$ value improves by 29 from UNITY1.7 to 1.8 across all panels (although the exact value will depend on the binning), so UNITY1.8 models the data much better than UNITY1.7. UNITY1.8 still misses some features of the data (notably the very low redshift $x_1$ distribution where the low-$x_1$ mode moves to even lower $x_1$, c.f., \citealt{Larison2024}), but we have tested that adding yet another Gaussian mode does not affect the cosmology; see Section~\ref{sec:unblinding}. \xspace}

\begin{figure*}
    \centering
    \includegraphics[width= 0.85\textwidth]{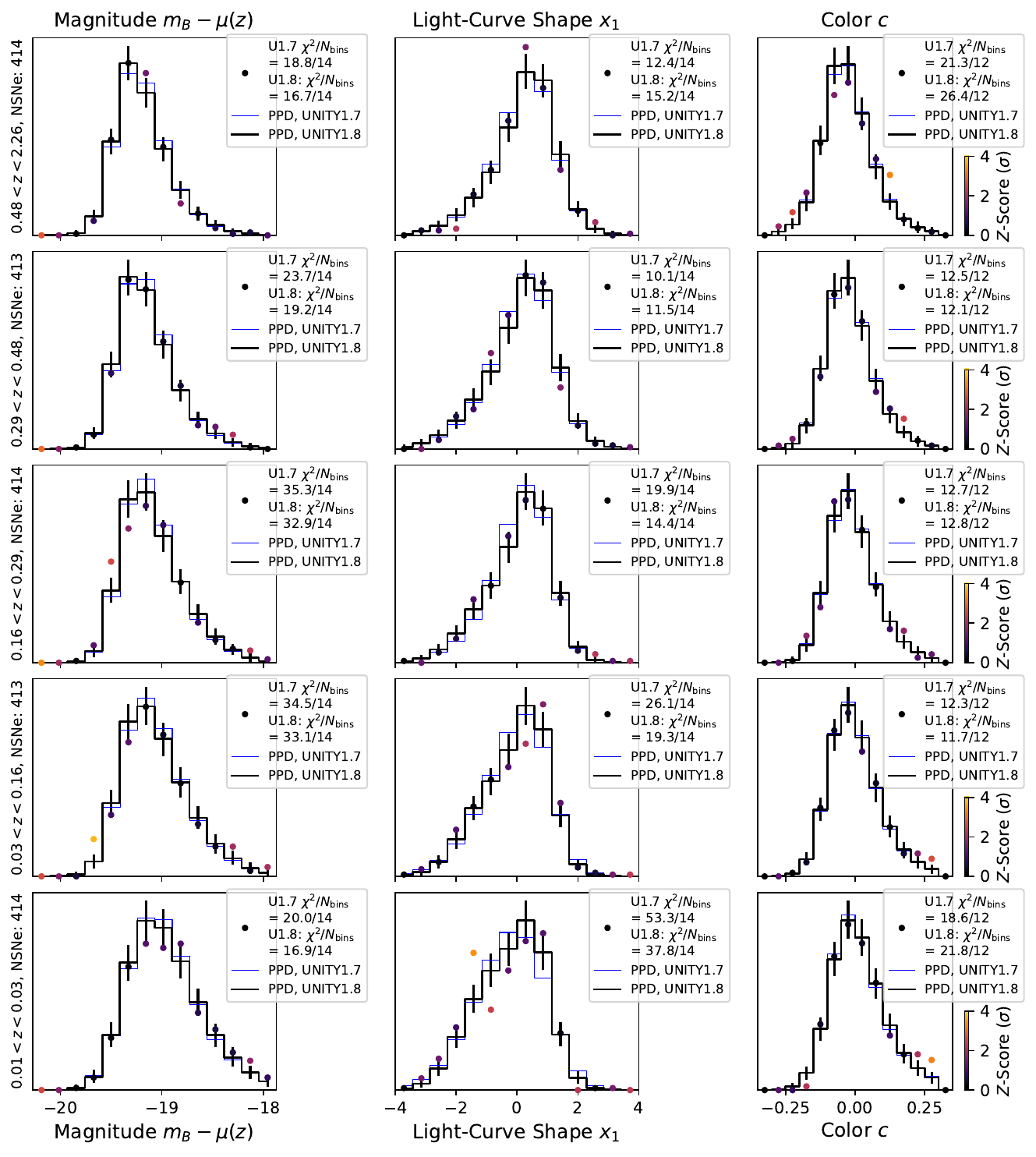}
    \caption{Histograms of the predictive posterior distribution $m_B-\mu(z)$, $x_1$, $c$ ({\bf columns}) in bins of redshift ({\bf rows}) compared to the actual data. The UNITY1.7 distributions are plotted with thin blue lines; the UNITY1.8 distributions are plotted with thick black lines with error bars. The data are plotted with points, color-coded by the $z$ score of the point and the legends give the total $\chi^2 = \sum z^2$. As seen in \citet{Rubin_2025}, many of the points making large contributions to the $\chi^2$ are from bins with a small number of events and so will not have much weight in the analysis. UNITY1.8 cuts off at high $x_1$ faster than UNITY1.7 and correctly models a flatter-top distribution at low redshift, despite having fewer parameters. \PPDchisquaresentence \label{fig:PPDdistr}}
\end{figure*}

\begin{figure*}
    \centering
    \includegraphics[width= 0.96\textwidth]{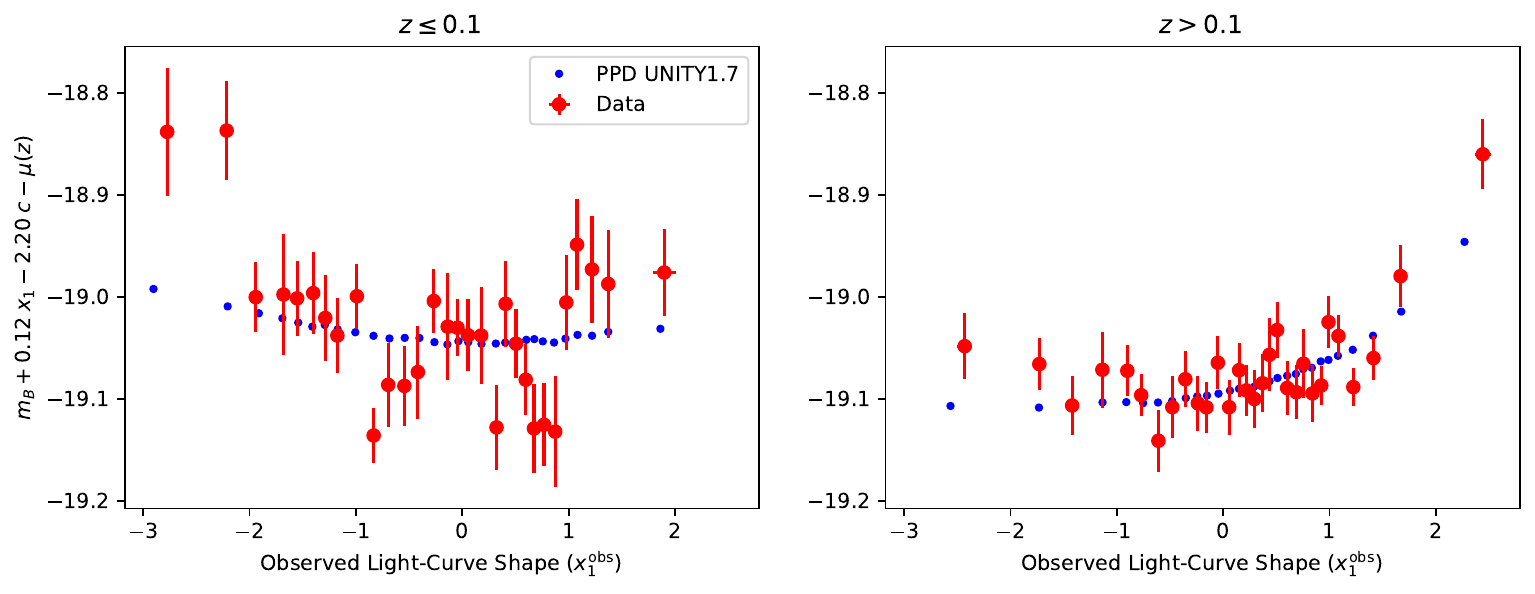}

    \includegraphics[width= 0.96\textwidth]{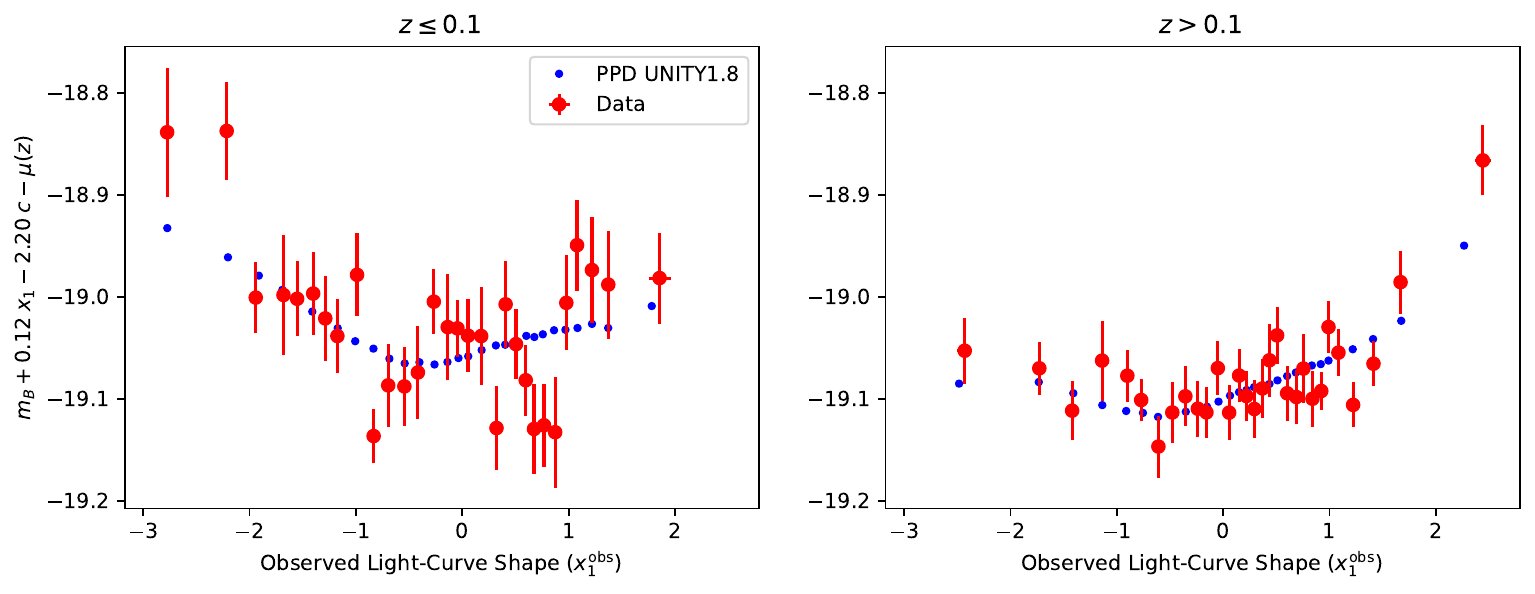}

    \caption{A comparison of the mean observed \citet{Tripp1998} Hubble residuals (red points with uncertainties) and the mean posterior-predictive Hubble residuals (blue points) in 30 $x_1$ intervals whose boundaries are chosen to contain roughly equal numbers of SNe in observed $x_1$ (\xoneobs). We choose $x_1$ and $c$ coefficients that give roughly flat relations when Hubble residual is plotted against \xoneobs and \cobs (these coefficients will be smaller than the $\alpha$'s and $\beta$'s due to regression dilution). To give a sense of the impact on cosmology, the {\bf left panels} show $z \leq 0.1$ while the {\bf right panels} show $z>0.1$. The {\bf top panels} show the single-mode UNITY1.7 results while the {\bf bottom panels} show the two-mode UNITY1.8 results. \PPDxoneSentence \label{fig:PPDHR}}
\end{figure*}

\begin{figure*}
    \centering
    \includegraphics[width= 0.96\textwidth]{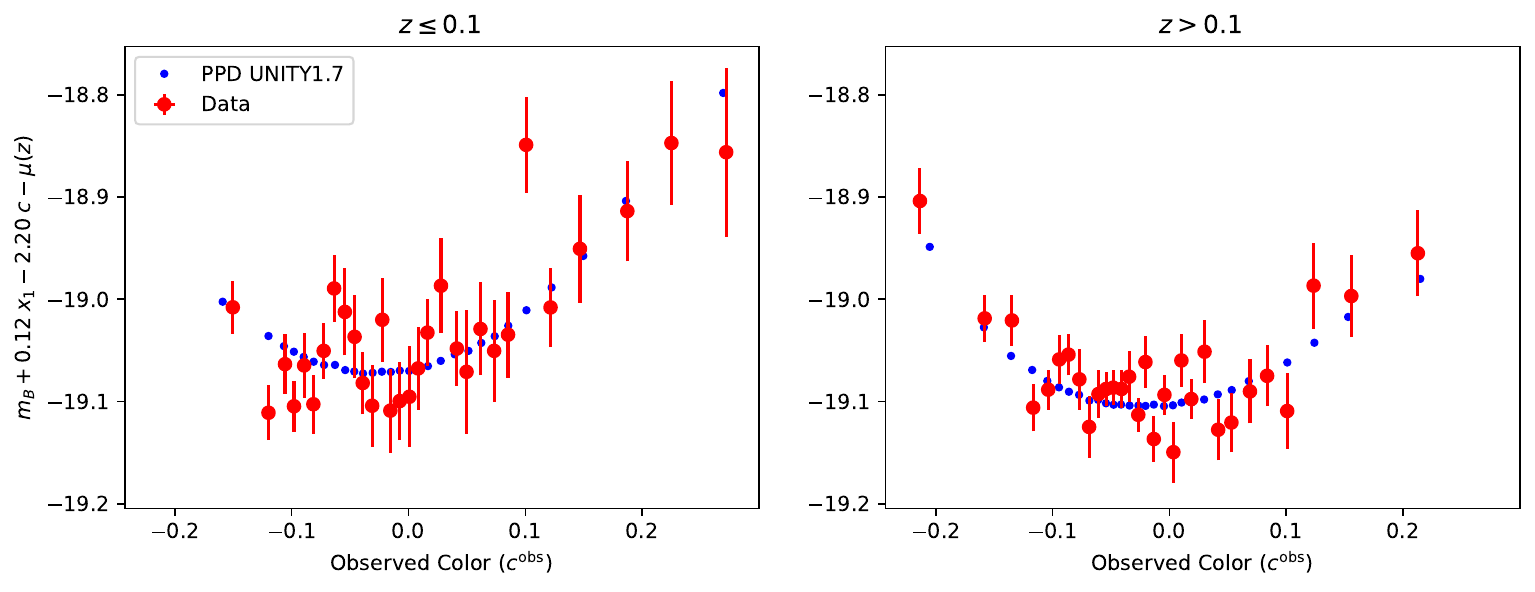}

    \includegraphics[width= 0.96\textwidth]{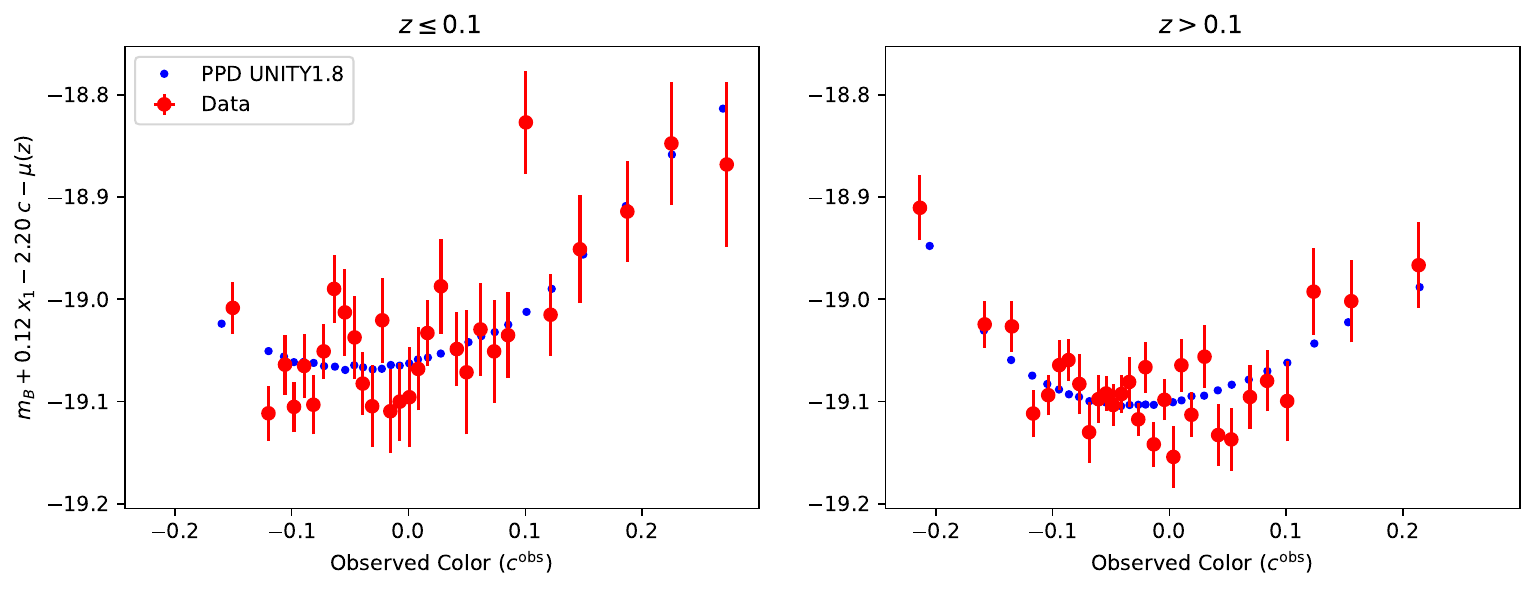}

    \caption{A comparison of the mean observed \citet{Tripp1998} Hubble residuals (red points with uncertainties) and the mean posterior-predictive Hubble residuals (blue points) in 30 color intervals whose boundaries are chosen to contain roughly equal numbers of SNe in observed color (\cobs). To give a sense of the impact on cosmology, the {\bf left panels} show $z \leq 0.1$ while the {\bf right panels} show $z>0.1$. The {\bf top panels} show the single-mode UNITY1.7 results while the {\bf bottom panels} show the two-mode UNITY1.8 results. \PPDcSentence \label{fig:PPDHRc}}
\end{figure*}

We make three figures comparing these predictive-posterior distributions with the actual data. Figure~\ref{fig:PPDdistr} compares the $m_B$ (minus distance modulus $\mu(z)$ for flat \LCDM), $x_1$, and $c$ predictive posterior and data distributions in five redshift bins containing equal numbers of SNe. It compares the single-mode UNITY1.7 with the two-mode UNITY1.8, with UNITY1.8 matching the data generally better. \PPDchisquaresentence Figure~\ref{fig:PPDHR} shows Hubble residuals plotted against observed $x_1$ in two redshift bins for both UNITY1.7 and UNITY1.8. \PPDxoneSentence Figure~\ref{fig:PPDHRc} parallels the plots for $c$. \PPDcSentence

\subsection{Cosmology Unblinding Procedure} \label{sec:unblinding}

Before unblinding, the co-authors agreed upon a set list of variants on the nominal UNITY1.8 model. These variations of the nominal model were designed primarily to gauge the sensitivity of cosmological parameters to slight adjustments in how we modeled the $x_1$ and color distributions, as well as the resultant standardization. The variants of the nominal UNITY1.8 model we considered were decided upon as (nearly) equally well motivated equivalents to the nominal model. These were:
\begin{enumerate}
    \item The addition of a third superposed Gaussian to the nominal, two-mode description of the $x_1/c$ distributions. This was motivated by the large width of the \fasthyphenmode $x_1$ distribution and the observation that the mean of the \fasthyphenmode $x_1$ distribution shifts to lower values for the oldest populations \citep{Larison2024}. We linked the $\alpha$ values of the two lower-$x_1$ modes, essentially using the new mode to help model the $x_1$ distribution of the \fasthyphenmode SNe.
    
    \item Splitting $\beta_B$ by \Pslow instead of having only one $\beta_B$ value for both modes. We found consistency between the \fasthyphenmode $\beta_B$ and the value from the \slowmode, and so did not choose to make this split in the nominal model.

    \item Splitting $\beta_R$ by \Pslow instead of \Phigheff. This is similar to the \citet{Wojtak2023} model. For this model, the host-stellar-mass step is not found to be zero. Ultimately, this motivated us to run another variant (after unblinding) which splits $\beta_R$ on both mode and host stellar mass, discussed below.
\end{enumerate}

\begin{figure*}
    \centering
    \includegraphics[width=0.7\textwidth]{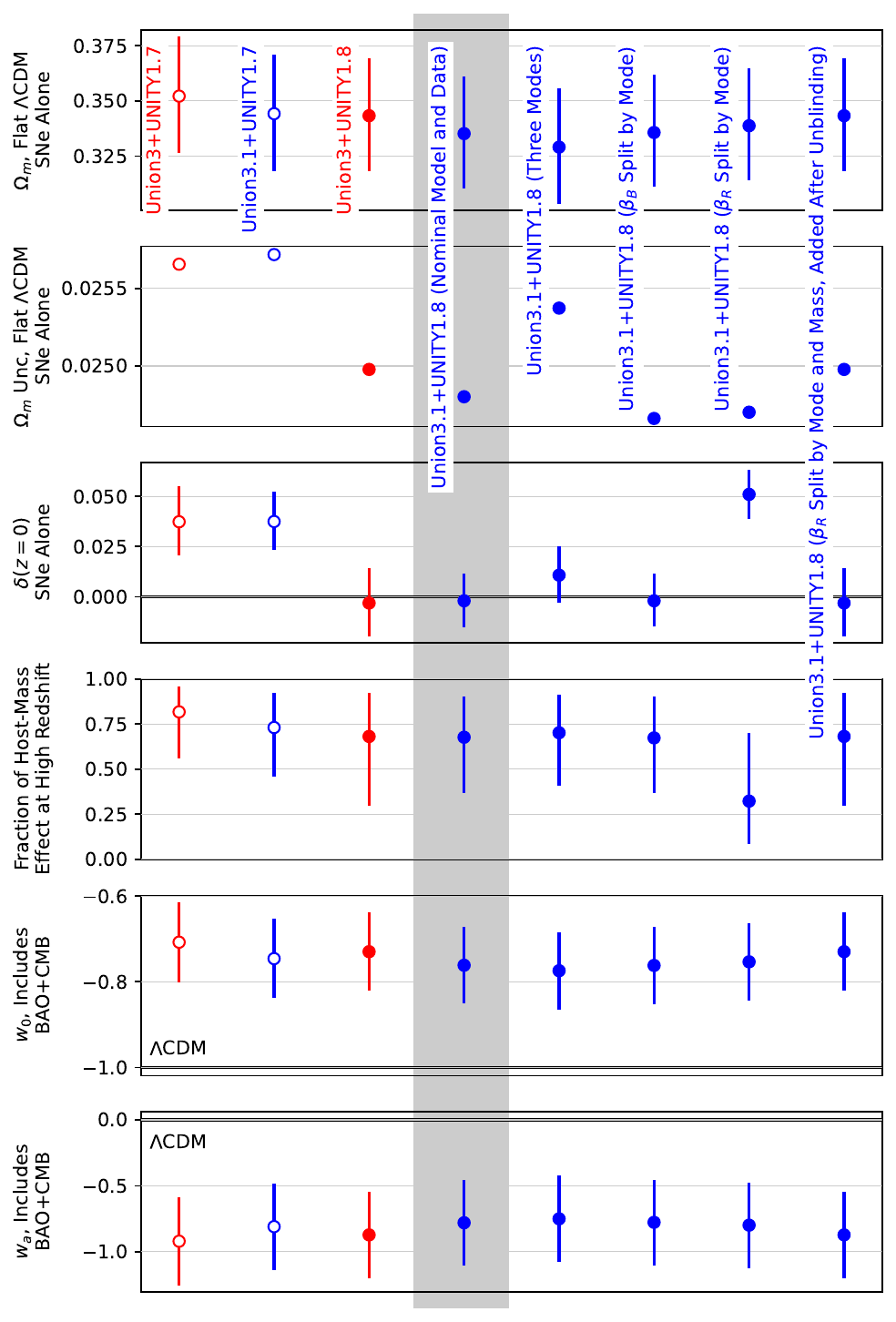}
    \caption{Tracking the inference of standardization and cosmological parameters across the versions of Union and UNITY discussed in this work. The central gray rectangle denotes our suggested nominal UNITY1.8 model. Red (blue) points are based on the Union3 (Union3.1) compilation. Outlined (filled) points correspond to UNITY1.7 (UNITY1.8) runs. 
    Points to the right of the rectangle (except the rightmost one) track the UNITY1.8 variant models that were confirmed to return consistent cosmologies before we unblinded. The leftmost two points show the impact of switching from Union3 to Union3.1 and were taken from \citet{Hoyt2026Mass}. It is important to note that all UNITY1.8 variants that enable different $\beta_R$ values with host-galaxy mass (all filled points except the second rightmost one) obtain a mass step ($\delta$) value consistent with zero. It is also very interesting to note that UNITY1.7 (outlined points) gets cosmological results within uncertainties of UNITY1.8, even though its population assumptions do not match the data as well as UNITY1.8's (only one population mode rather than two). It is also interesting to note that all UNITY1.8 variants return (slightly) smaller $\Omega_m$ uncertainties than UNITY1.7 when run with the same input data. (In other words, all the filled points have smaller $\Omega_m$ uncertainties than the corresponding outlined points.) \label{fig:variants}}
\end{figure*}

Before deciding whether to unblind UNITY1.8, we first verified that each of the variants had minimal impact on the inferred cosmology, namely $\Omega_m$ for flat \LCDM, as well as $w_0$-$w_a$ when including BAO and CMB data (external cosmological constraints are discussed in Section~\ref{sec:cosmoexternal}). The results are shown as the boxed nominal model and the three markers to the right in all rows of \autoref{fig:variants}. The cosmology inferred by UNITY1.8 and the adopted variants matches the nominal model to better than $0.25 \sigma$, despite one of the model variants---the one that splits $\beta_R$ on \Pslow rather than \Phigheff---leading to significant shifts in the host-mass-dependent standardization parameters. If the deviation had exceeded $\sim 0.3 \sigma$, we would have delayed unblinding and continued to evaluate the merits of different models at the SN standardization level before revisiting the unblinding stage. That was not the case and so we locked in the nominal model and unblinded UNITY1.8.

After unblinding, we decided to run another UNITY1.8 variant: having a separate $\beta_R$ for \fasthyphenmode SNe, but still splitting $\beta_R$ for \slowhyphenmode SNe by host-galaxy mass. This model might be motivated if progenitor age were a meaningful predictor of $\beta_R$ among high-stellar-mass host galaxies, i.e., if young SNe in high-stellar-mass hosts had $\beta_R$ values more like young SNe in low-stellar-mass hosts. Interestingly, we find that the \slowhyphenmode SNe hosted by high-stellar-mass galaxies have a $\beta_R$ consistent with the \fasthyphenmode SNe (most of which are in high-stellar-mass galaxies). Thus, $\beta_R$ really does seem to be a stronger function of host-galaxy stellar mass than the slow/fast mode. The results of this model are shown in the right column of Figure~\ref{fig:variants}. In short, the host-mass-step ($\delta(z=0)$) is still consistent with zero (discussed in Section~\ref{sec:hoststandard}), and the cosmological constraints actually show slightly more tension with flat \LCDM than the nominal model. 

We discuss the nominal-model unblinded cosmology results for the remainder of this article.

\clearpage

\section{Supernova Modeling Results and Implications} \label{sec:SNresults}

\begin{figure*}
    \centering
    \includegraphics[width=\textwidth]{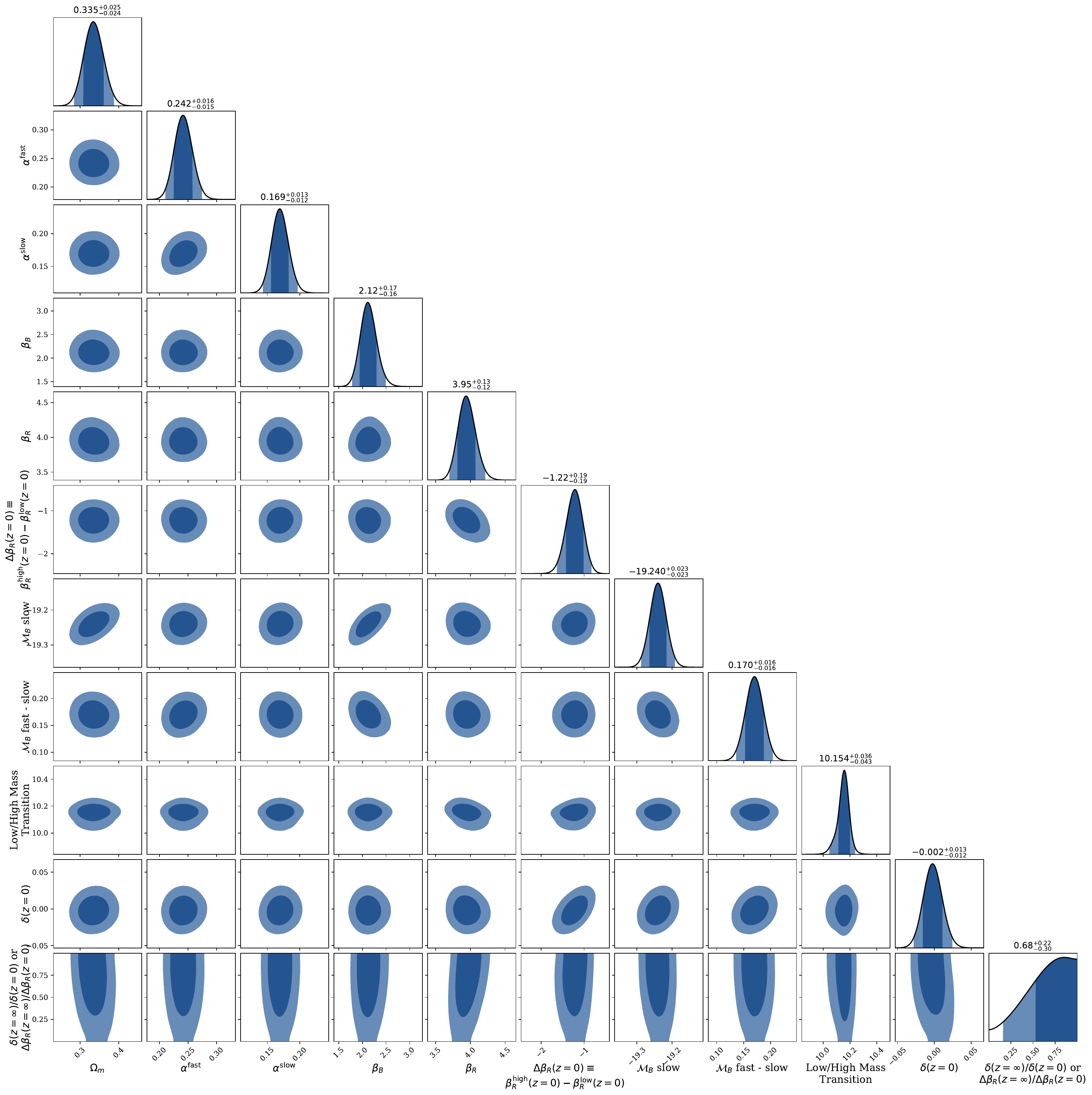}
    \caption{Corner plot of standardization parameters from Equation~\ref{eq:tripp}. We observe significantly different $x_1$-standardization relations ($\alpha$) between the two modes, three distinctly different color-standardization relations ($\beta$), and a host-mass/luminosity step consistent with zero ($\delta(z=0)=0$, thus zero for all redshifts by Equation~\ref{eq:Phigheff}). See Table~\ref{tab:fitparams} for a more complete list of parameters. \label{fig:cornerplot}}
\end{figure*}

\begin{figure}
    \centering
    \includegraphics[width=0.5\textwidth]{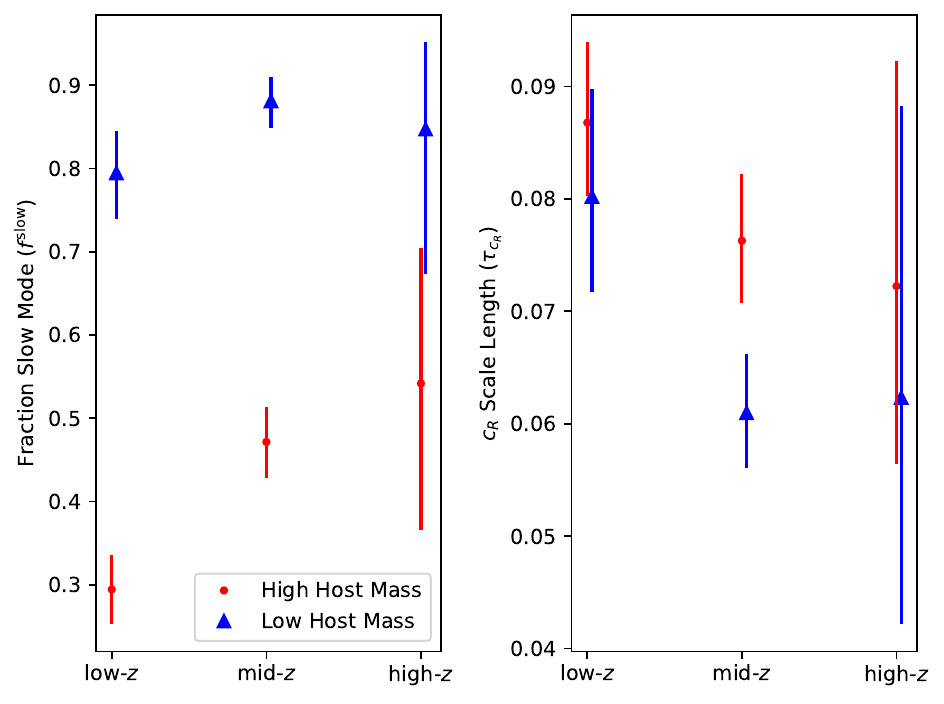}
    \caption{Population parameters that are split by host-galaxy-stellar-mass and redshift bins. We plot the posterior median as well as showing the 15.9th--84.1st percentile range. We see evidence that older populations have a lower fraction in the slow mode ({\bf left panel}) and perhaps a higher \tauc {\bf right panel}.}
    \label{fig:redshiftmass}
\end{figure}

Here, we summarize the SN-focused results from running UNITY1.8 on Union3.1.  Figure~\ref{fig:cornerplot} shows a corner plot of many of the parameters related to standardization, Table~\ref{tab:fitparams} shows 68\% credible intervals for a more complete set, and Figure~\ref{fig:redshiftmass} plots credible intervals for parameters binned in host-galaxy-stellar-mass and redshift bins. Many of UNITY1.8's supernova results agree with previous results in broad strokes, but with some differences and some significant improvements. Our posteriors show strong evidence for two population modes (Section~\ref{sec:twopopevidence}). The $x_1$ standardization varies between the two modes, discussed in Section~\ref{sec:xonestandard}. We continue to see evidence of differences in color standardization between red and blue SNe and between SNe in low-mass and high-mass host galaxies, discussed in Section~\ref{sec:colorstandard}. Importantly, UNITY1.8 eliminates the host-mass/luminosity correlation, discussed in Section~\ref{sec:hoststandard}. The two modes have very different unexplained dispersions, with the \slowhyphenmode SNe having less, discussed in Section~\ref{sec:unexplained}. Finally, it is important to note that UNITY1.8 is an empirical model based on the empirical SALT3 \citep{Kenworthy2021, Taylor2023} light-curve fit results, and some of our conclusions could easily reflect specific properties of SALT3 and not SNe~Ia in general. Validation of our conclusions will require exploration with other light-curve fitters. Ideally, one would bring the light-curve fitting and training inside UNITY so there is no information compression due to light-curve fitting.\footnote{Even more ideally, one would compare observations directly against physical models of SNe~Ia (e.g., \citealt{Hoeflich1996}). However, this will require quantifying and controlling the uncertainties of those models.}

\newcommand{\alphafastvalue}{\ensuremath{0.242^{+0.016}_{-0.015}}\xspace}
\newcommand{\alphaslowvalue}{\ensuremath{0.169^{+0.013}_{-0.012}}\xspace}
\newcommand{\deltaalphavalue}{\ensuremath{0.072\pm 0.017}\xspace}
\newcommand{\betaBvalue}{\ensuremath{2.12^{+0.17}_{-0.16}}\xspace}
\newcommand{\betaRvalue}{\ensuremath{3.95^{+0.13}_{-0.12}}\xspace}
\newcommand{\deltabetaRvalue}{\ensuremath{-1.22\pm 0.19}\xspace}
\newcommand{\betaRhighvalue}{\ensuremath{3.34^{+0.13}_{-0.12}}\xspace}
\newcommand{\betaRlowvalue}{\ensuremath{4.56^{+0.19}_{-0.17}}\xspace}
\newcommand{\MBslowvalue}{\ensuremath{-19.240\pm 0.023}\xspace}
\newcommand{\MBfastminusslowvalue}{\ensuremath{0.170\pm 0.016}\xspace}
\newcommand{\stepmassvalue}{\ensuremath{10.154^{+0.036}_{-0.043}}\xspace}
\newcommand{\deltazerovalue}{\ensuremath{-0.002^{+0.013}_{-0.012}}\xspace}
\newcommand{\deltahvalue}{\ensuremath{0.68^{+0.22}_{-0.30}}\xspace}
\newcommand{\Rxonefastvalue}{\ensuremath{0.827^{+0.039}_{-0.038}}\xspace}
\newcommand{\Rxoneslowvalue}{\ensuremath{0.489\pm 0.026}\xspace}
\newcommand{\xonestarfastvalue}{\ensuremath{-0.916\pm 0.068}\xspace}
\newcommand{\xonestarslowvalue}{\ensuremath{0.547\pm 0.032}\xspace}
\newcommand{\Rcfastvalue}{\ensuremath{0.0571^{+0.0052}_{-0.0054}}\xspace}
\newcommand{\Rcslowvalue}{\ensuremath{0.0391\pm 0.0032}\xspace}
\newcommand{\cstarfastvalue}{\ensuremath{-0.0465\pm 0.0081}\xspace}
\newcommand{\cstarslowvalue}{\ensuremath{-0.0818\pm 0.0067}\xspace}
\newcommand{\outlfracvalue}{\ensuremath{0.0127^{+0.0033}_{-0.0028}}\xspace}
\newcommand{\sigmaintfastvalue}{\ensuremath{0.084^{+0.012}_{-0.013}}\xspace}
\newcommand{\meansigmaintvalue}{\ensuremath{0.1188^{+0.0098}_{-0.0097}}\xspace}
\newcommand{\mBxonecintvarianceonevalue}{\ensuremath{0.38\pm 0.13}\xspace}
\newcommand{\mBxonecintvariancetwovalue}{\ensuremath{0.126^{+0.041}_{-0.035}}\xspace}
\newcommand{\mBxonecintvariancethreevalue}{\ensuremath{0.49\pm 0.12}\xspace}
\newcommand{\tauclowzhighmassvalue}{\ensuremath{0.0868^{+0.0070}_{-0.0063}}\xspace}
\newcommand{\fracxoneslowlowzhighmassvalue}{\ensuremath{0.294^{+0.039}_{-0.040}}\xspace}
\newcommand{\taucmidzhighmassvalue}{\ensuremath{0.0763^{+0.0058}_{-0.0054}}\xspace}
\newcommand{\fracxoneslowmidzhighmassvalue}{\ensuremath{0.472\pm 0.041}\xspace}
\newcommand{\tauchighzhighmassvalue}{\ensuremath{0.072^{+0.020}_{-0.016}}\xspace}
\newcommand{\fracxoneslowhighzhighmassvalue}{\ensuremath{0.54^{+0.16}_{-0.17}}\xspace}
\newcommand{\tauclowzlowmassvalue}{\ensuremath{0.0802^{+0.0094}_{-0.0084}}\xspace}
\newcommand{\fracxoneslowlowzlowmassvalue}{\ensuremath{0.795^{+0.048}_{-0.053}}\xspace}
\newcommand{\taucmidzlowmassvalue}{\ensuremath{0.0610^{+0.0051}_{-0.0047}}\xspace}
\newcommand{\fracxoneslowmidzlowmassvalue}{\ensuremath{0.881^{+0.027}_{-0.030}}\xspace}
\newcommand{\tauchighzlowmassvalue}{\ensuremath{0.062^{+0.026}_{-0.020}}\xspace}
\newcommand{\fracxoneslowhighzlowmassvalue}{\ensuremath{0.85^{+0.10}_{-0.17}}\xspace}

\begin{deluxetable*}{lr}
\caption{68.3\% credible intervals for UNITY1.8 parameters. The posterior median is given and the asymmetric error bars are computed using the 15.9th and 84.1st percentiles. \label{tab:fitparams}}
\tablehead{\colhead{Parameter}  & \colhead{Posterior}}
\startdata
$\scriptM$ \slowhyphenmode & $\MBslow = \MBslowvalue$ (mag) \\
$\scriptM$ \fasthyphenmode $-$ $\scriptM$ \slowhyphenmode & $\MBfastminusslow = \MBfastminusslowvalue$ (mag) \\
\fasthyphenmode $\alpha$ & $\alphafast = \alphafastvalue$ (mag per $x_1$) \\
\slowhyphenmode $\alpha$ & $\alphaslow = \alphaslowvalue$ (mag per $x_1$) \\
$\beta$ for blue, Gaussian-distributed color & $\beta_B = \betaBvalue$ (mag per mag) \\
$\beta$ for red, exponentially distributed color in high-mass hosts & $\betaRH = \betaRhighvalue$ (mag per mag) \\
\phantom{$\beta$ for red, exponentially distributed color} in low-mass hosts & $\betaRL = \betaRlowvalue$ (mag per mag) \\
$\beta_R$ difference with host mass & $\Delta \beta_R \equiv \betaRH - \betaRL =  \deltabetaRvalue $ (mag per mag) \\
log$_{10}$ host-galaxy mass at step &  $= \stepmassvalue$ (log$_{10} \Msol$) \\
host-mass-luminosity step at low redshift & $\delta(z=0) = \deltazerovalue$ (mag) \\
fraction of host-mass correlations remaining at high redshift & $\delta(z=\infty)/\delta(z = 0)=  \deltahvalue$ (mag per mag) \\
& $= \Delta \beta_R (z=\infty) / \Delta \beta_R (z=0)$ \\
\hline
outlier fraction & $f^{\mathrm{outl}} = \outlfracvalue$\\
\slowhyphenmode unexplained dispersion, (Added: averaged over all) SN~samples & $<\sigma^{\mathrm{unexpl.,\, slow}}> = \meansigmaintvalue$ \\
extra \fasthyphenmode unexplained dispersion, added in quadrature & $\sigma^{\mathrm{unexpl.,\, fast}} = \sigmaintfastvalue$ \\
fraction unexplained variance in $m_B$ & $f^{m_B} = \mBxonecintvarianceonevalue$ \\
fraction unexplained variance in $x_1$ & $f^{x_1} = \mBxonecintvariancetwovalue$ \\
fraction unexplained variance in $c$ & $f^c = \mBxonecintvariancethreevalue$ \\
\hline
Gaussian $x_1$ peak, \fasthyphenmode & $\xonestarfast = \xonestarfastvalue$ ($x_1$) \\
\phantom{Gaussian $x_1$ peak,} \slowhyphenmode & $\xonestarslow = \xonestarslowvalue$ ($x_1$) \\
Gaussian $x_1$ width, \fasthyphenmode  & $\Rxonefast = \Rxonefastvalue$ ($x_1$) \\
\phantom{Gaussian $x_1$ width,} \slowhyphenmode & $\Rxoneslow = \Rxoneslowvalue $ ($x_1$) \\
Gaussian $c_B$ peak, \fasthyphenmode & $\cstarfast = \cstarfastvalue$ (mag) \\
\phantom{Gaussian $c_B$ peak,} \slowhyphenmode& $\cstarslow = \cstarslowvalue$ (mag) \\
Gaussian $c_B$ width, \fasthyphenmode & $\Rcfast = \Rcfastvalue$ (mag) \\
\phantom{Gaussian $c_B$ width,} \slowhyphenmode& $\Rcslow = \Rcslowvalue$ (mag) \\
exponential $c_R$ scale, low-$z$, high-mass & $\tauc = \tauclowzhighmassvalue$ (mag) \\
\phantom{exponential $c_R$ scale,} mid-$z$, high-mass & $\tauc = \taucmidzhighmassvalue$ (mag) \\
\phantom{exponential $c_R$ scale,} high-$z$, high-mass & $\tauc = \tauchighzhighmassvalue$ (mag) \\
\phantom{exponential $c_R$ scale,} low-$z$, low-mass & $\tauc = \tauclowzlowmassvalue$ (mag) \\
\phantom{exponential $c_R$ scale,} mid-$z$, low-mass & $\tauc = \taucmidzlowmassvalue$ (mag) \\
\phantom{exponential $c_R$ scale,} high-$z$, low-mass & $\tauc = \tauchighzlowmassvalue$ (mag) \\
fraction in \slowmode, low-$z$, high-mass & $\fslow = \fracxoneslowlowzhighmassvalue$ \\
\phantom{fraction in \slowmode,} mid-$z$, high-mass & $\fslow = \fracxoneslowmidzhighmassvalue$ \\
\phantom{fraction in \slowmode,} high-$z$, high-mass & $\fslow = \fracxoneslowhighzhighmassvalue$ \\
\phantom{fraction in \slowmode,} low-$z$, low-mass & $\fslow = \fracxoneslowlowzlowmassvalue$ \\
\phantom{fraction in \slowmode,} mid-$z$, low-mass & $\fslow = \fracxoneslowmidzlowmassvalue$ \\
\phantom{fraction in \slowmode,} high-$z$, low-mass & $\fslow = \fracxoneslowhighzlowmassvalue$ \\
\enddata
\end{deluxetable*}

\subsection{Statistical Evidence for Two Populations} \label{sec:twopopevidence}

\begin{figure*}[htbp]
    \includegraphics[width=0.5\textwidth]{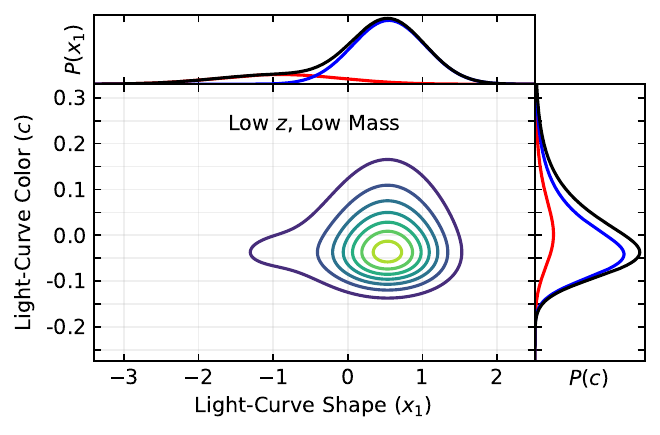}
    \includegraphics[width=0.5\textwidth]{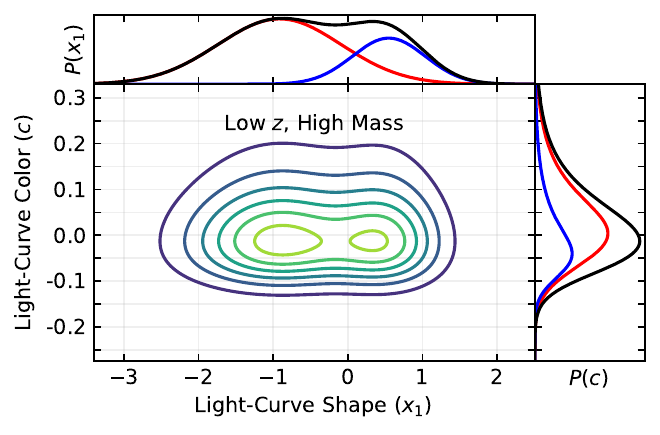}

    \includegraphics[width=0.5\textwidth]{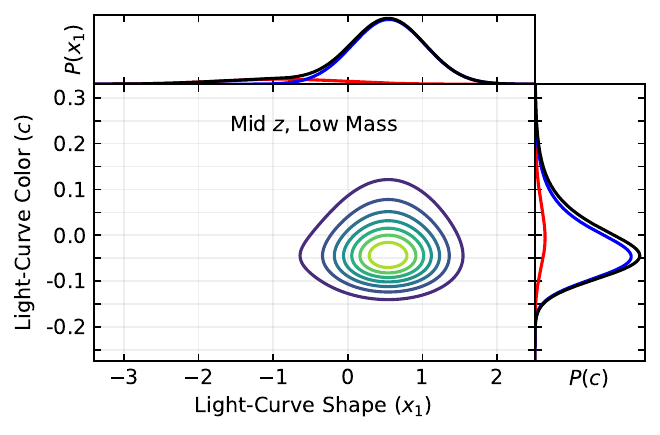}
    \includegraphics[width=0.5\textwidth]{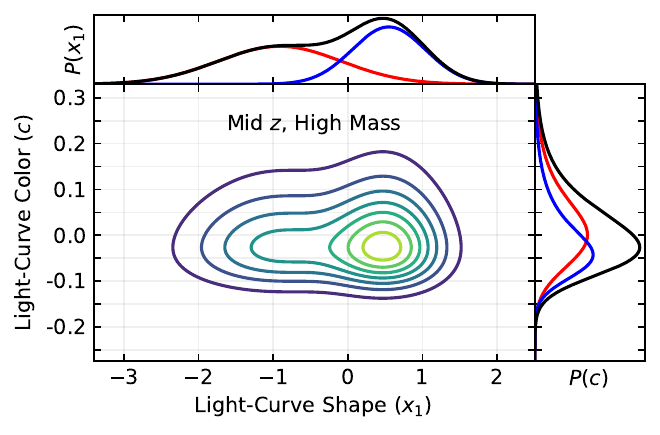}

    \includegraphics[width=0.5\textwidth]{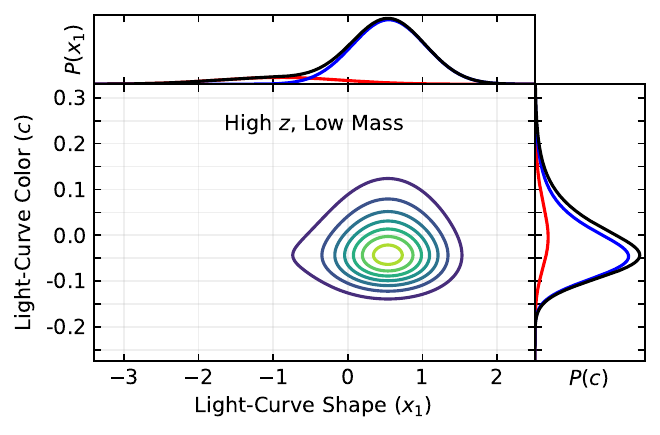}
    \includegraphics[width=0.5\textwidth]{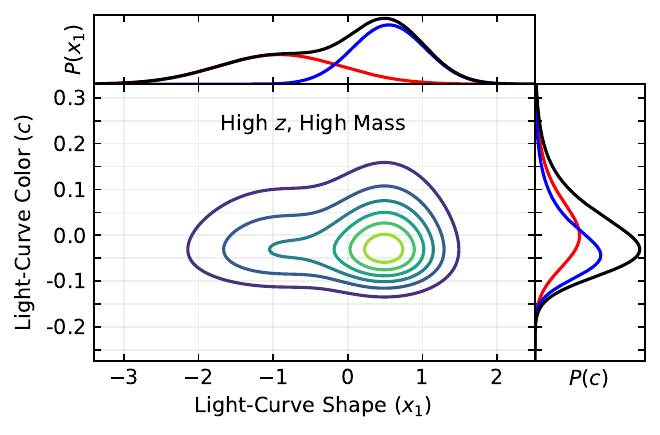}
\caption{Two-mode population models from Union3.1+UNITY1.8 , computed from the posterior median. The red curves show the \slowhyphenmode population, the blue curves show the \fasthyphenmode population, and the black curves show the sum. We also show 2D contours. The $x_1$ and $c_B$ distributions are associated with each mode, while the $c_R$ distributions change with redshift and mass bin. The relative scaling of each mode also changes with the redshift and mass bin, with more \fasthyphenmode SNe at lower redshift and higher mass. \label{fig:pop}}
\end{figure*}

A salient finding from UNITY1.8 is significant evidence for two populations, which is straightforwardly defined as a deviation of the \fslow parameter from \autoref{eq:tripp} either 0 or 1. In the three low-mass bins we find the fraction of SNe contained in the slow population, $\fslow$, to be $\fracxoneslowlowzlowmassvalue$, $\fracxoneslowmidzlowmassvalue$, and $\fracxoneslowhighzlowmassvalue$, in the low, mid, and high, redshift bins, respectively. In the high mass bin, we find $\fslow$ is $\fracxoneslowlowzhighmassvalue$, $\fracxoneslowmidzhighmassvalue$, and $\fracxoneslowhighzhighmassvalue$. The fraction of slow-mode SNe in low-mass hosts remains consistently $\sim0.85$ across our probed redshifts, while the same in high-mass hosts increases with increasing redshift. \textbf{MOVED FROM 3.2: This trend is visualized in \autoref{fig:pop}, which illustrates our two-population $x_1$ and $c$ model, with the lower-$x_1$ mode showing up at higher host stellar mass and lower redshift (older populations, on average).}

\subsection{$x_1$ Population and Standardization} \label{sec:xonestandard}

We find that two modes is a good description of the $x_1$ population as a function of redshift and host-galaxy stellar mass. In our nominal model, we model these two modes with Gaussians and find the values in Table~\ref{tab:fitparams}. We find that the fast mode has peak in $x_1$ lower by 1.5 (\xonestarfastvalue vs. \xonestarslowvalue) and that the fast mode has a wider population distribution width (\Rxonefastvalue vs. \Rxoneslowvalue).

We find evidence for two distinct values of $\alpha$ for the two modes: $\alphafast = \alphafastvalue$, $\alphaslow = \alphaslowvalue$. There is good consistency in the literature for $\alphafast$ (or $\alpha$ for low-$x_1$ SNe), but there is more disagreement for \alphaslow. Our value for \alphafast agrees within uncertainties with the $\alpha$ value measured in bins below, above and in the range $10^{10} < M_{\sun}^{\mathrm{host}} < 10^{11}$ by \citet[][Fig\,7]{Garnavich2023}. \citet{Ginolin2025a} report for a volume-limited ZTF DR2 sample a global $\alpha = 0.161 \pm 0.010$ and for a broken-alpha standardization law with break point $x_1=-0.48 \pm 0.08$, $\alphaslow = 0.083 \pm 0.009$ and $\alphafast = 0.271 \pm 0.011$. In an analysis of sibling SNe in ZTF DR2, \citet{Dhawan_2024arXiv240601434D} found a larger global $\alpha = 0.228 \pm 0.029$ using ZTF DR2 data than did the full ZTF analysis, which they show is due to sibling occurrences preferring lower-$x_1$ (faster declining) SNe. They also enforce an $x_1$ split equal to that found by \citeauthor{Ginolin2025a} and find $\alphaslow = 0.133 \pm 0.072$ and $\alphafast = 0.274 \pm 0.045$. \citet{Newman_2025arXiv250820023N} report an $\alpha=0.209 \pm 0.015$ for a narrow selection of SNe with $-2.0 < x_1 < 0.0$ and $-0.2 < c < 0.1$.

\subsection{Color Population and Standardization} \label{sec:colorstandard}
 Union3.1+UNITY1.8 finds distinct values of $\beta$ between blue SNe, red SNe in low-stellar-mass galaxies, and red SNe in high-stellar-mass galaxies. The bluest SNe have the shallowest magnitude-color relation ($\beta_B = \betaBvalue$). For redder SNe, we find a significant $\beta_R$ split between low- and high-stellar-mass host galaxies: $\beta_R$ in low-mass hosts is \betaRlowvalue, while for high-mass at low redshift, it is \betaRhighvalue (mean $\beta_R=\betaRvalue$, $\Delta \beta_R = \deltabetaRvalue$). Union3+UNITY1.5 found consistent values although with much larger uncertainties due mostly to the bug in the outlier color width parameter limit discussed in their Appendix~B: $\beta_B=2.6\pm 0.4$, $\beta_R=3.62^{+0.21}_{-0.18}$, and $\Delta \beta_R = -1.1^{+0.3}_{-0.4}$. For comparison to extinction-law $R_V$ values, we remind the reader that $R_V \approx R_B - 1$ so one expects $\beta_R \approx 4$ if it is driven by Milky-Way-like dust extinction ($R_V \approx 3.1$, \citealt{Savage1979}).

The values of $\beta_R$ are similar to those from the \citet{Wojtak2023} and \citet{Wojtak2025} models, with \citet{Wojtak2025} finding their $\widehat{R_B}$ values of $4.074^{+0.510}_{-0.494}$ and $3.030\pm0.117$ for their slow and fast decliner subpopulations, compared to our \betaRlowvalue and \betaRhighvalue for $\beta_R$ in low- and high-mass host galaxies. The agreement tightens when we adjust the UNITY model to split $\beta_R$ on $x_1$ mode, similar to \citet{Wojtak2025}, rather than on mass as in our nominal model. This results in values of $\beta_R$ equal to $4.13^{+0.14}_{-0.13}$ and $3.27^{+0.17}_{-0.16}$ for the slow-mode and fast-mode populations, respectively.

In Figure~\ref{fig:c_x1_split}, we plot the latent \cBtrue and \cRtrue values inferred by UNITY for each SN as a function of the SALT3 color parameter \cobs from the Union3.1 light-curve fits. The $y$-coordinate of the top distribution of points is $c_R$, while that of the bottom distribution is $c_B$. Each marker represents a single SN and each SN appears twice on this plot (one in each color distribution). In this plane, we see that fast-declining SNe~Ia are associated with redder (bluer) values of $c_B$ ($c_R$) than slow-declining ones. This finding is consistent with faster-declining SNe occurring more frequently in passive galaxy environments and that they appear to have redder intrinsic colors \citep{Garnavich_2004, Ginolin2025b}. 

\newcommand{\colortruesentence}{We see that \fasthyphenmode/lower $x_1$ SNe have on average redder Gaussian-distributed $c_B$ colors for all values of observed color (\cobs). For red values of the observed color, \fasthyphenmode SNe also tend to have $c_R$ colors more similar to $c_B$ than do \slowhyphenmode SNe.\xspace}

\begin{figure}[htbp]
    \centering
    \includegraphics[width=0.5\textwidth]{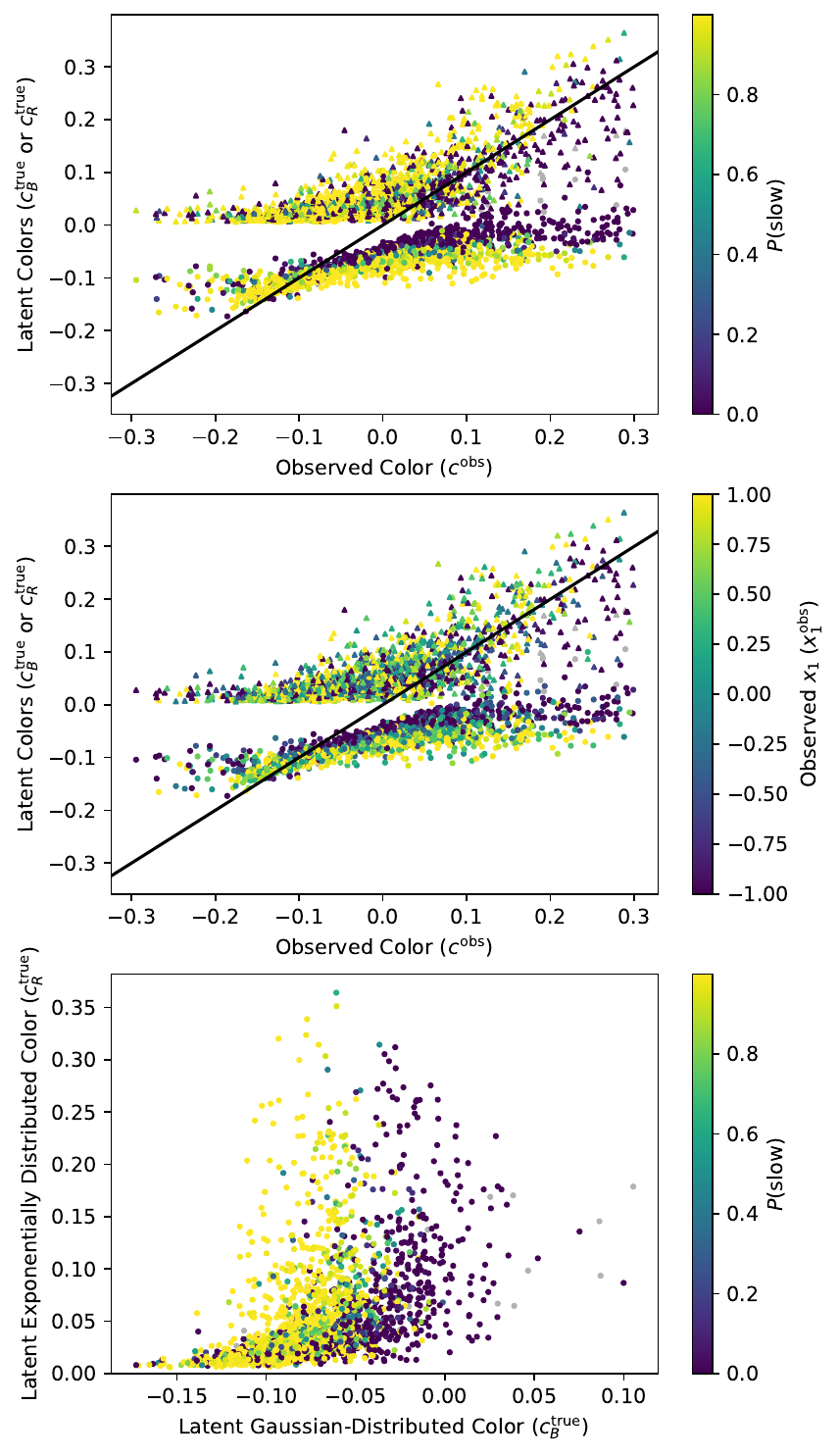}
    \caption{{\bf Top panel:} scatter plot of latent colors ($c_B$ and $c_R$) as a function of observed color, color coded by \Pslow. The {\bf middle panel} shows the same plot but color coded by the observed \lighthypencurvehypencshape parameter ($x_1$). \colortruesentence The {\bf bottom panel} shows \cRtrue plotted as a function of \cBtrue, color coded by \Pslow and clearly shows the difference in $c^*$ for the two modes.}
    \label{fig:c_x1_split}
\end{figure}

\subsection{Host Correlations} \label{sec:hoststandard}

One of our most important results is that we find $\delta(z=0) = \deltaOUNITYOneEight$~mag for Union3.1+UNITY1.8, i.e., there is no longer an explicit luminosity step as a function of host-galaxy stellar mass for unreddened SNe. A host-stellar-mass step consistent with zero is a unique result\footnote{\citet{BroutScolnic2021} find no evidence for a stellar-mass step after applying a model with host-stellar-mass-dependent $R_V$, however the Dark Energy Survey analysis still does find such a step \citep{Vincenzi2024} using the BS21 model updated by \citet{Popovic2023}. However, these are not simultaneous fits and so it is difficult to evaluate the difference between finding a set of parameters that force the stellar-mass step to zero and finding zero as the best fit. In Union3+UNITY1.5, which treats all parameters simultaneously, the inclusion of a split $\beta_R$ reduces the host-stellar-mass step to $0.031 \pm 0.019$~mag, but did not eliminate it. \citet{Hoyt2026Mass} finds \deltaOUNITYOneSeven for Union3.1+UNITY1.7, reproduced in Figure~\ref{fig:variants} for the same input data we use for UNITY1.8.} that may indicate that much of what has been considered the host-mass relation is due to previous models conflating two distinct subtypes of SNe~Ia which have different relative frequencies as a function of host stellar mass and redshift (likely due to different delay-time distributions, e.g., \citealt{Wiseman2021}). Noting that this is true for ``unreddened SNe'' is important as $\beta_R$ still depends on host mass in our nominal model, so host mass still has some effect for reddened SNe. Figure~\ref{fig:variants} shows that making $\beta_R$ a function of the two modes (rather than host stellar mass) gives an apparent luminosity step, so the host galaxy must affect either the color standardization or the magnitude standardization; splitting the SNe by mode is not enough. 

We measure the host-galaxy stellar-mass-step location to be $\stepmassvalue$ (log$_{10} \Msol$), a small but statistically significant difference from $10^{10} \Msol$ (this step location controls the $\beta_R$ stellar-mass-step location as well as the luminosity step location, so it is well constrained). Fortunately for past results, there is very little covariance between this value and $\Omega_m$.

We do not see any evidence of evolution in the host-stellar-mass luminosity and $\beta_R$ relations with redshift, although our constraints are weak and not very Gaussian; at high redshift, we find \deltahvalue compared to the low-redshift value of 1 but this is also fairly consistent with 0 (which would imply the host-galaxy correlations go away at high redshift).

\subsection{Unexplained Dispersions} \label{sec:unexplained}

\newcommand{\smallunexplainedsentence}{This figure shows generally small ($\lesssim 0.05$~magnitudes) unexplained dispersion for the slow-mode SNe measured from rolling surveys and thus have many SN-free ``reference'' images (these surveys also typically use scene-modeling photometry for more accurate uncertainties, \citealt{Holtzman2008, Astier2013}). We denote rolling surveys with an asterisk.\xspace}

\begin{figure*}
    \centering
    \includegraphics[width=0.95\textwidth]{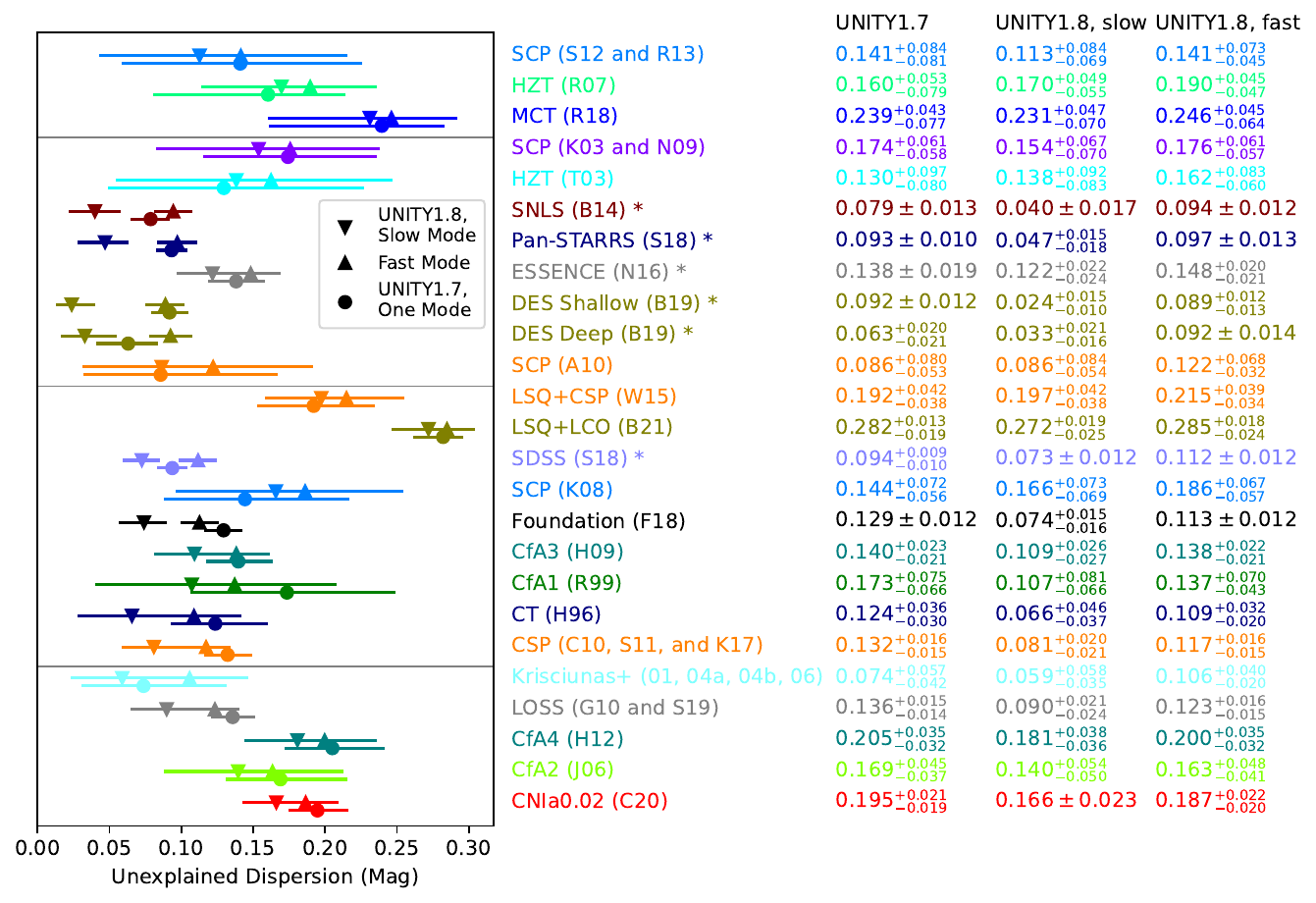}
    \caption{Unexplained dispersion by SN sample for both UNITY1.7 (one population mode) and our recommended UNITY1.8 (with two modes). We find the fast (low-$x_1$) mode has additional unexplained dispersion, although whether this is due to features of the SALT3 model or whether it is a real property of SNe~Ia is not clear. \smallunexplainedsentence}
    \label{fig:unexpl}
\end{figure*}

Figure~\ref{fig:unexpl} shows the 68.3\% credible intervals for the unexplained-dispersion parameters for each SN sample for both UNITY1.7 and UNITY1.8. UNITY1.7 has one unexplained-dispersion parameter for each sample while UNITY1.8 has one parameter per sample for the slow mode, then another parameter $\sigma^{\mathrm{unexpl.,\, fast}} = \sigmaintfastvalue$ for all the samples that is added in quadrature for the \fasthyphenmode SNe. (Our motivation for keeping this value the same for all samples is that we expect most of the variation sample-to-sample to be due to inaccuracies in assigned photometry uncertainties rather than differences in SNe with sample selection.) \smallunexplainedsentence We note that these are inferred values after modeling out the measurement uncertainties and SALT3 model uncertainties. We thus cannot directly tell if the results of our unexplained-dispersion model are intrinsic properties of SNe~Ia or just properties of SALT3. UNITY does not make Hubble-residual predictions, so we do not quote a Hubble-diagram dispersion. (We have also not considered whether it would be possible in a dataset with realistic uncertainties to select a pure sample of unreddened \slowhyphenmode SNe that may be seen to have lower dispersion.)

\begin{figure*}
    \centering
    \includegraphics[width=\textwidth]{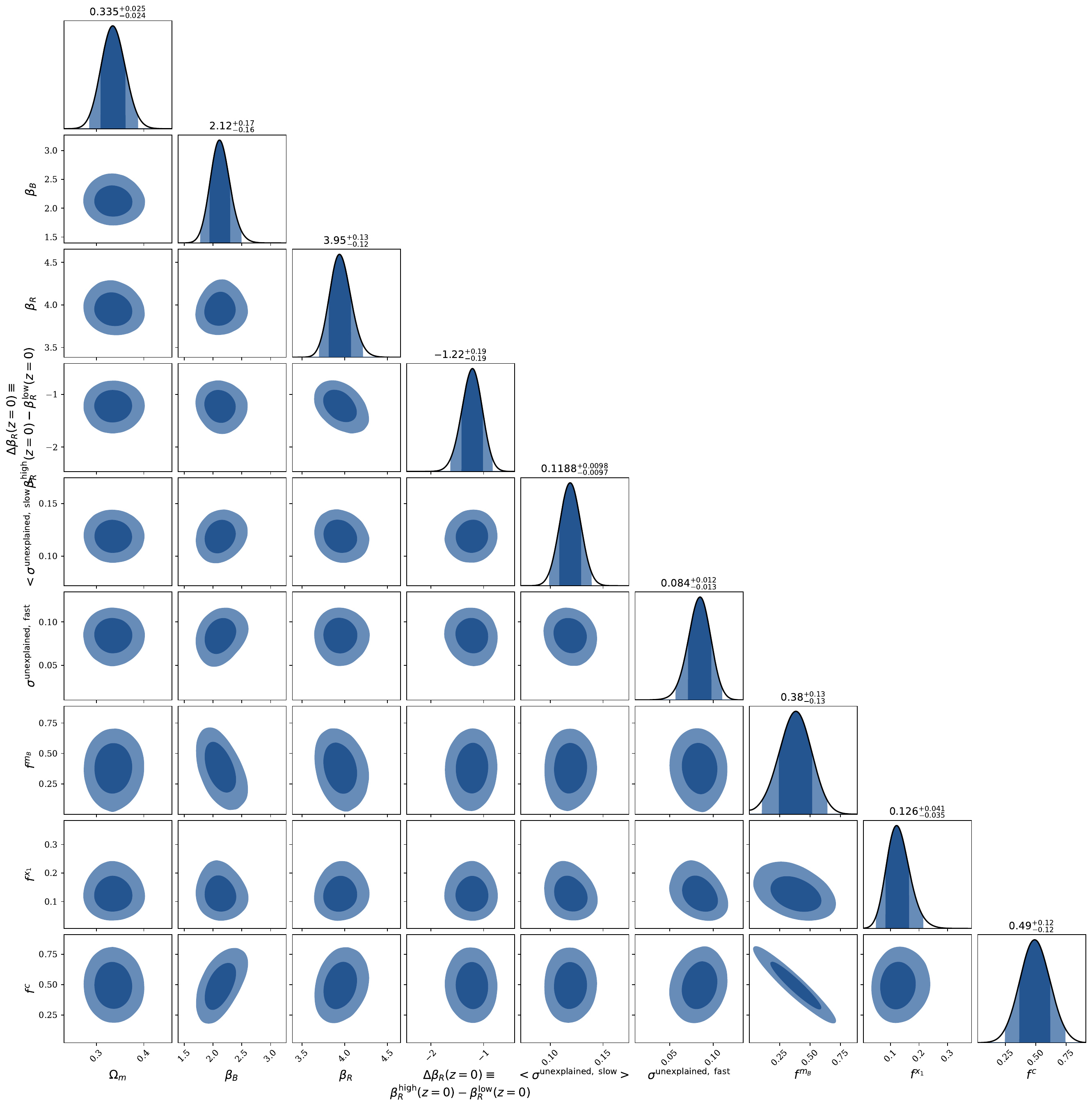}
    \caption{Corner plot of standardization parameters and unexplained dispersion. There is a large degeneracy between the fraction of the unexplained dispersion in magnitude ($f^{m_B}$) and the fraction in color ($f^c$), with the fraction in $x_1$ constrained to be much smaller. As one would expect, putting more unexplained dispersion into color (larger $f^c$) increases $\beta$ (especially $\beta_B$, which is measured with a smaller color baseline than $\beta_R$) because there is more inferred regression dilution. \label{fig:unexplcontour}}
\end{figure*}

Figure~\ref{fig:unexplcontour} shows the allocation of unexplained variance between $m_B$, $x_1$, and $c$. Compared to Union3+UNITY1.5, we find a shift of unexplained dispersion from $c$ to $m_B$, resulting in an almost equal fraction of variance between the two (fraction for $m_B$ is \mBxonecintvarianceonevalue, fraction for $c$ is \mBxonecintvariancethreevalue, with the remaining \mBxonecintvariancetwovalue in $x_1$). Using \citet{kessler2013} Figure~11 to roughly compare against two frequently used scatter models, we find this result is in between the \citet{Guy2010} and \citet{Chotard2011} models, although more similar to \citet{Guy2010}. One of the \citet{Chotard2011} spectral components is aligned with $x_1$, which UNITY1.8 does a better job modeling, so it is not clear whether this is an unexpected result.

\section{Cosmology Results} \label{sec:cosmology}

\subsection{SN Cosmology Results} \label{sec:SNCosmology}

As in Union3+UNITY1.5 \citep{Rubin_2025}, we compute an approximate distance-modulus covariance matrix with UNITY so that our SN results are easily transferable to other analyses without rerunning UNITY. We put a quadratic spline through a series of spline nodes in redshift (shown in Figure~\ref{fig:muvsz}) and model the difference in distance modulus between the data and a flat \LCDM model with $\Om = 0.3$. The values of the spline nodes and their covariance matrix is an excellent approximation to the full UNITY posterior, which can be seen by comparing the frequentist fits in this Section with the corresponding posteriors in Figure~\ref{fig:variants} (and as discussed in \citealt{Rubin_2025}).

\newcommand{\tensionsentence}{The weak preference of the SN data for a higher flat-\LCDM \Om value is mitigated when fitting a $w_0$-$w_a$ model (see also \citealt{OColgain2025}), discussed more in Section~\ref{sec:cosmoexternal}.\xspace}

Figure~\ref{fig:muvsz} plots the distance-modulus residuals against three cosmological models: flat \LCDM with $\Om=0.3$ (the BAO+CMB best fit, also similar to SN+BAO+CMB), flat \LCDM with $\Omega_m = 0.334$ (the SN-only best-fit), and a $w_0$-$w_a$ model which is the best fit including BAO and CMB constraints. \tensionsentence

\begin{figure*}
    \centering
    \includegraphics[width=0.7\textwidth]{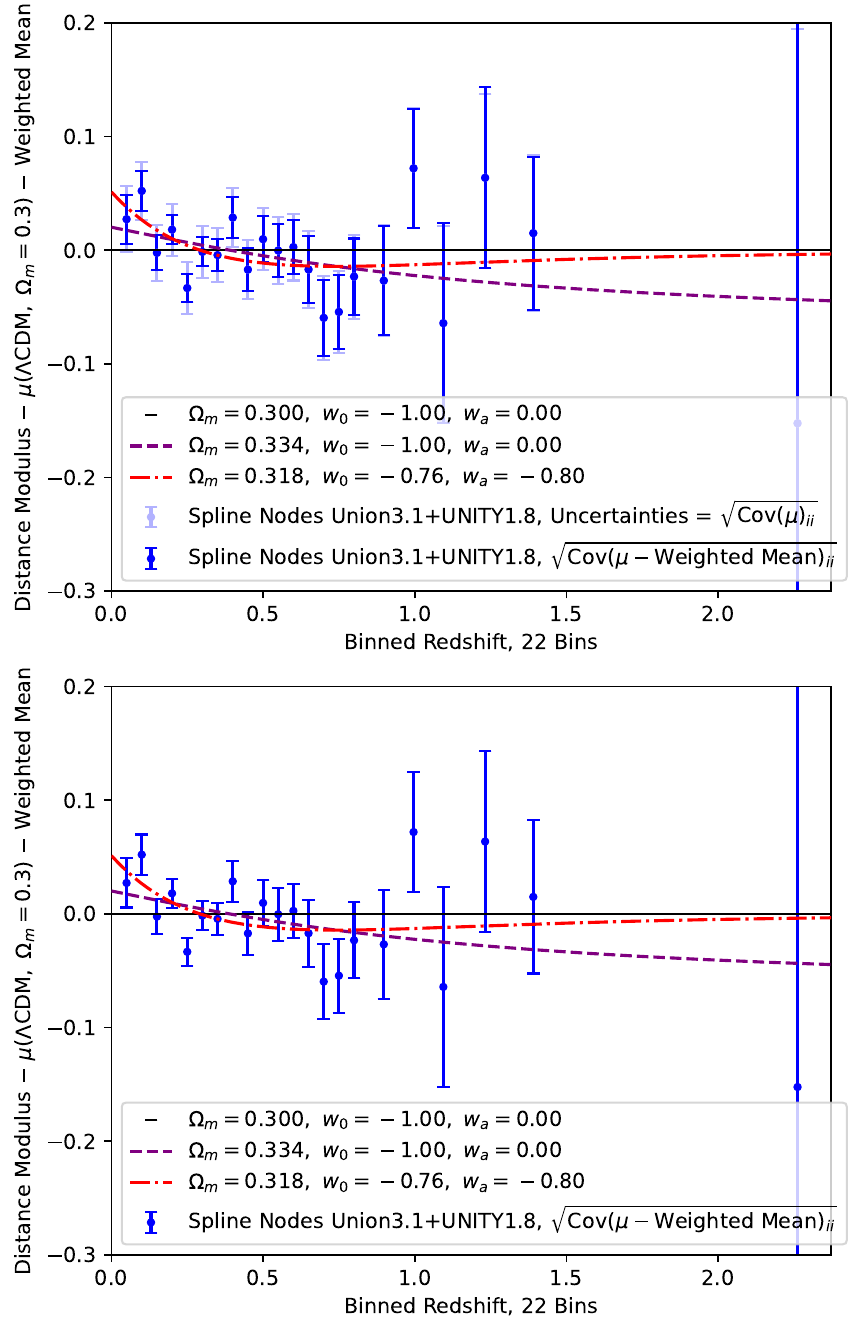}
    \caption{Plots of the UNITY distance-modulus residuals against three cosmological models: flat \LCDM with $\Om=0.3$ (black line), flat \LCDM with $\Omega_m = 0.334$ (the SN-only best-fit, purple dashed line), and a $w_0$-$w_a$ model which is the best fit including BAO and CMB constraints (red dot-dashed line). \tensionsentence}
    \label{fig:muvsz}
\end{figure*}

\newcommand{\contourresultssentence}{In general, the three contours show compatible constraints. The contours are completely enclosed for \LCDM, so we also compute the Figure of Merit (the inverse area of the $2\sigma$ contour); our results show Union3.1+UNITY1.8 has a 6\% higher FoM than Union3.1/UNITY1.7 \citep{Hoyt2026Mass} and 15\% higher FoM than Union3/UNITY1.5 \citep{Rubin_2025}. \xspace}

\begin{figure*}
    \centering
    \includegraphics[width=0.4\textwidth]{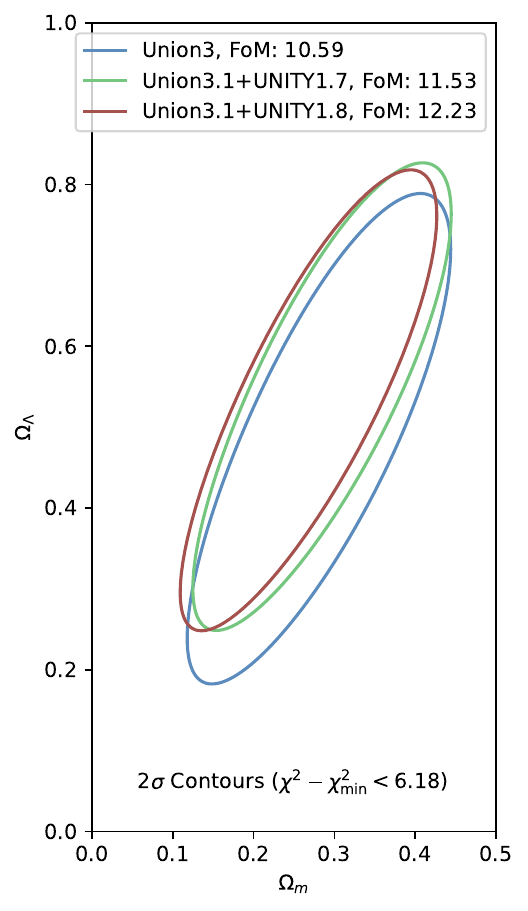}
        \includegraphics[width=0.49\textwidth]{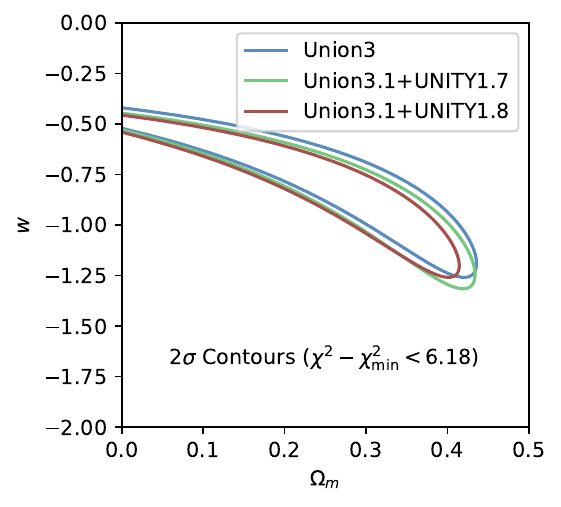}
        \caption{Comparison of SN-only cosmological constraints for Union3+UNITY1.5, Union3.1+UNITY1.7, and Union3.1+UNITY1.8. The {\bf left panel} shows \LCDM while the {\bf right panel} shows flat $\Omega_m$-$w$. In all cases, the 2D 2$\sigma$ contour is plotted (the contour enclosing $\Delta \chi^2 < 6.18$ or 95.4\% confidence). \contourresultssentence \label{fig:SNonlycontours}}
\end{figure*}

We compute frequentist contours ($\Delta \chi^2$ compared to the best fit of 2.296, 6.180, and 11.829 for 68.3\%, 95.4\%, and 99.7\% confidence) by fixing the two parameters shown in the plane and fitting for the others. We use an adaptive-refinement contour code that chooses points to evaluate.\footnote{\url{https://github.com/rubind/adaptive_contour}} Figure~\ref{fig:SNonlycontours} compares such contours for Union3+UNITY1.5, Union3.1+UNITY1.7, and Union3.1+UNITY1.8 for flat \LCDM, and flat \Omw. \contourresultssentence

\subsection{Cosmology Results with External Constraints} \label{sec:cosmoexternal}

\noindent

\begin{figure*}[h!tbp]
    \centering
    \includegraphics[width=0.7 \textwidth]{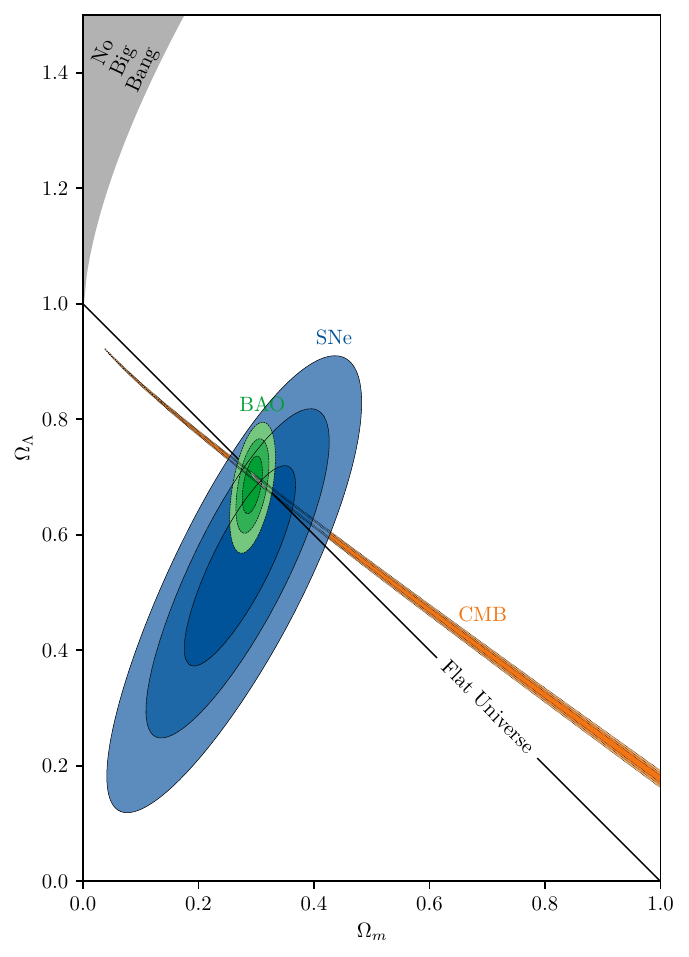}
    \caption{Constraints in the \Om-\OL plane. We show the constraints for SNe (in blue), BAO (in green), CMB (in orange), and combined (in gray). The probe constraints have different orientations; combining especially BAO and CMB makes for a much stronger constraint than either alone.}
    \label{fig:OmOL}
\end{figure*}

\begin{figure*}[h!tbp]
    \centering
    \includegraphics[width=0.7 \textwidth]{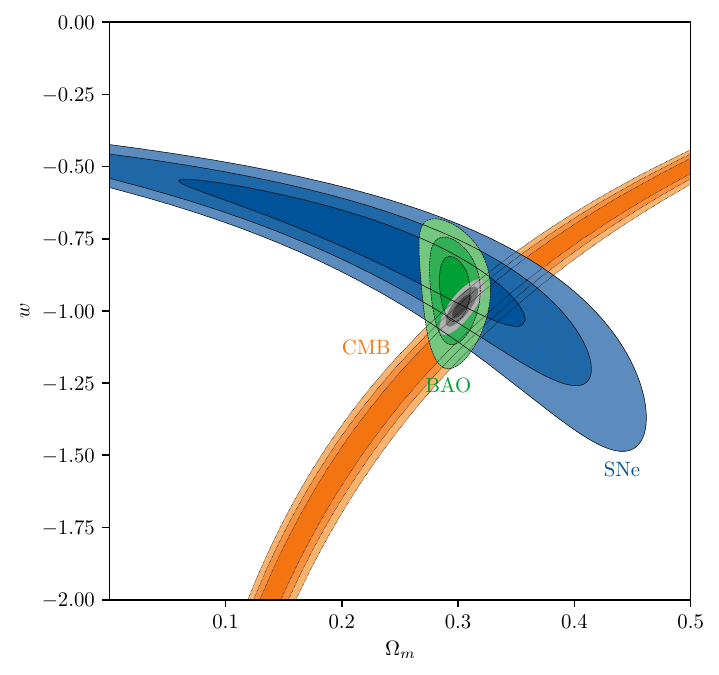}
    \caption{Constraints in the \Om-$w$ plane for flat-universe, constant equation-of-state parameter $w$ models. Again we show the constraints for SNe (in blue), BAO (in green), CMB (in orange), and combined (in gray). The probe constraints have different orientations and thus combine together for a much stronger net measurement.}
    \label{fig:Omw}
\end{figure*}

\begin{figure}[h!tbp]
    \centering
    \includegraphics[width=0.49 \textwidth]{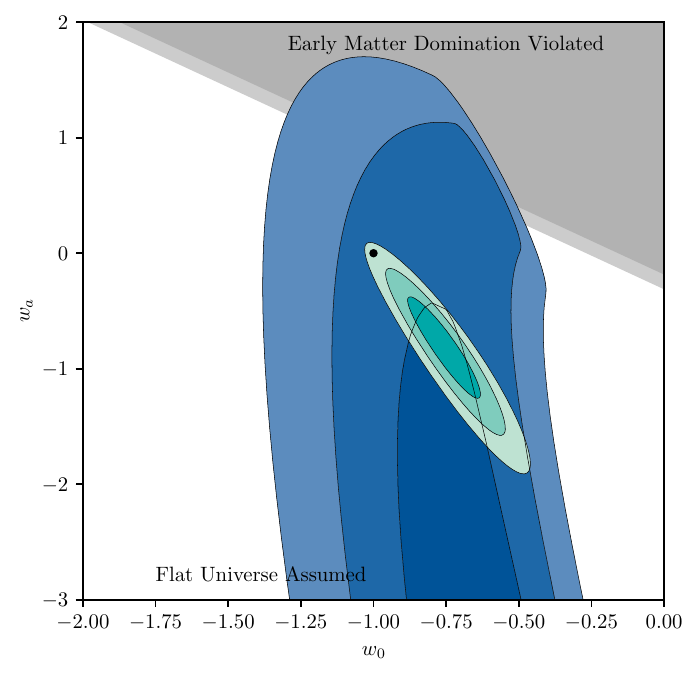}
    \caption{Constraints in the $w_0$-$w_a$ plane for flat-universe. Again we show the constraints for SNe (in blue), but now show SNe+BAO+CMB combined in teal. The SN-only constraints are consistent with the combined constraints. We indicate \LCDM with a black dot.}
    \label{fig:wwaSNe}
\end{figure}

\begin{figure*}[h!tbp]
    \centering
    \includegraphics[width=0.49 \textwidth]{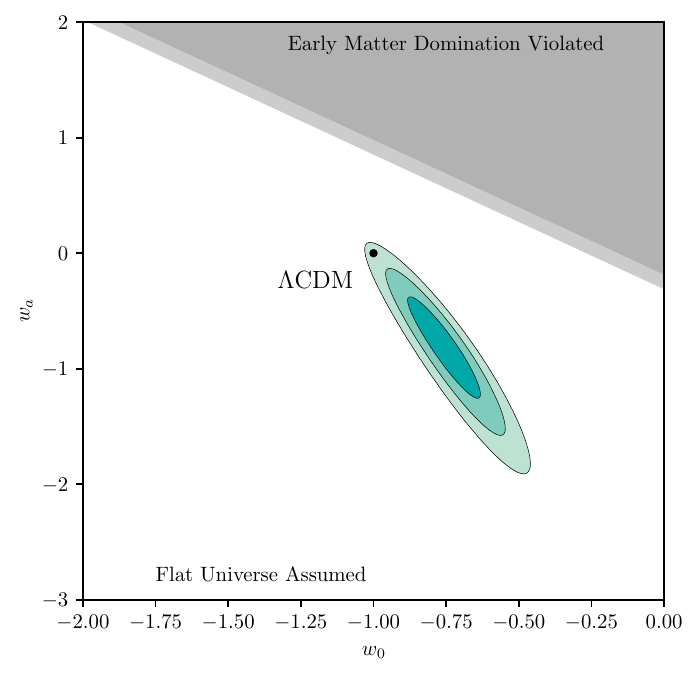}
    \includegraphics[width=0.49 \textwidth]{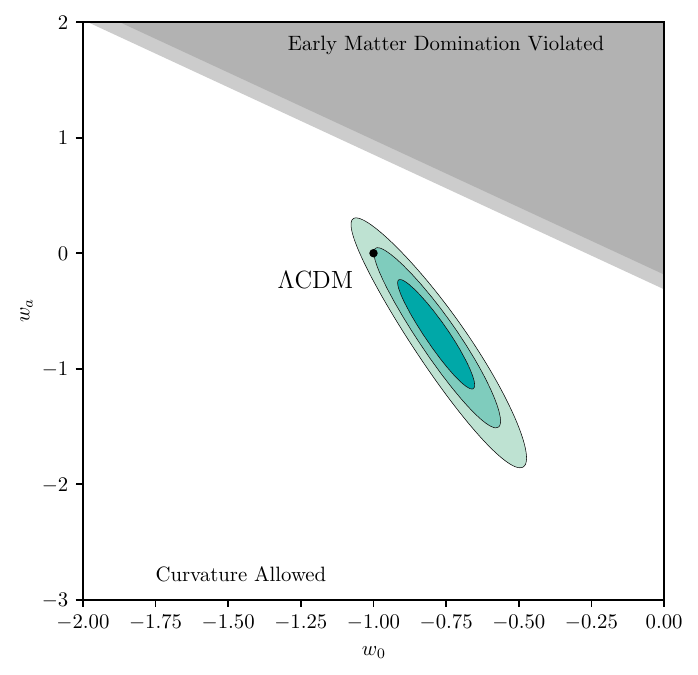}
    \caption{Constraints in the $w_0$-$w_a$ plane combining SNe, BAO, and CMB. The {\bf left panel} shows the 1, 2, and 3$\sigma$ contours assuming a flat universe, while the {\bf right panel} also fits for curvature. We also mark off the parameter space where early matter domination would begin to be violated (the shaded regions show 1\% and 10\% of the matter density at $z=1100$ assuming $\Omega_m = 0.3$). The contours show weak tension with \LCDM (indicated with a black dot).}
    \label{fig:w0wa}
\end{figure*}

\begin{figure*}[h!tbp]
    \centering
    \includegraphics[width=0.49 \textwidth]{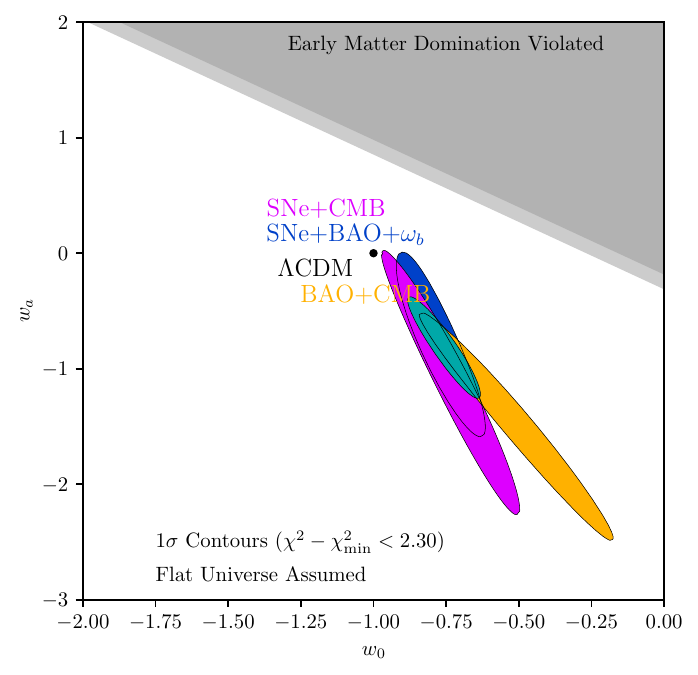}
    \includegraphics[width=0.49 \textwidth]{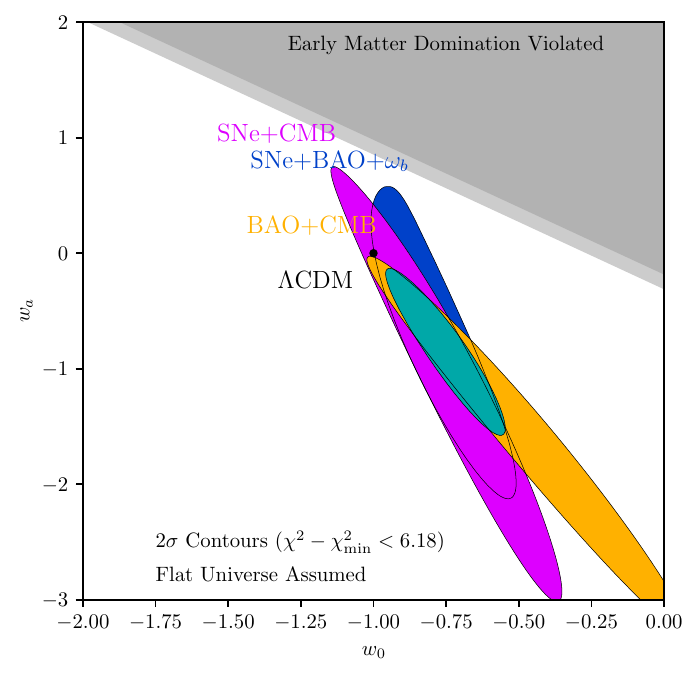}
    \caption{Constraints in the $w_0$-$w_a$ plane. The {\bf left panel} shows the $1 \sigma$ contours and the right panel shows $2 \sigma$ contours, both assuming a flat universe. The contours show constraints from BAO+CMB, SNe+CMB, and SNe+BAO (as well as the same BAO+CMB+SNe contour from Figure~\ref{fig:w0wa}). Once again, we indicate \LCDM with a black dot. The contour for each pair of probes is angled differently, showing how the constraints get stronger when all three probes are combined together. The results from each pair of probes are compatible with the results that use all three.  
    \label{fig:w0waindiv}}
\end{figure*}

Following \citet{Rubin_2025}, we include BAO, CMB, and/or $H_0$ constraints and consider flat \LCDM, \LCDM with curvature, flat $w$CDM, flat $w_0$-$w_a$, and $w_0$-$w_a$ with curvature. Table~\ref{tab:cosmoconstraints} presents our best-fit values, $1\sigma$ uncertainties, and $\chi^2$ values. Figure~\ref{fig:OmOL} shows our constraints in the \Om-\OL plane. Figure~\ref{fig:Omw} shows our constraints in the \Om-$w$ plane. As a sanity check of the SN results, Figure~\ref{fig:wwaSNe} shows the combined constraints and the SN-only constraints for flat-universe $w_0$-$w_a$. Figure~\ref{fig:w0wa} shows our results in the $w_0$-$w_a$ plane (with and without allowing spatial curvature). Figure~\ref{fig:w0waindiv} investigates which individual data combinations drive our $w_0$-$w_a$ constraints..

\begin{deluxetable*}{lccccccc}
 \tablehead{
 \colhead{Probes} & \colhead{$\chi^2$ (DoF)}  & \colhead{$h$} & \colhead{$\Omega_m$} & \colhead{$\Omega_k$} & \colhead{$w$ or $w_0$} & \colhead{$w_a$} & \colhead{DETF FoM} }
 \startdata
\hline
 \multicolumn{8}{c}{Flat $\Lambda$CDM}\\ 
 \hline
SNe & 28.8 (20)  & \nodata  &  $0.334^{+0.025}_{-0.024}$  & \nodata  & \nodata  & \nodata  & \nodata  \\ 
SNe+CMB & 29.3 (21)  &  $0.673^{+0.006}_{-0.006}$  &  $0.317^{+0.008}_{-0.008}$  & \nodata  & \nodata  & \nodata  & \nodata  \\ 
BAO+CMB & 14.7 (14)  &  $0.685^{+0.003}_{-0.003}$  &  $0.300^{+0.004}_{-0.004}$  & \nodata  & \nodata  & \nodata  & \nodata  \\ 
SNe+BAO+$\omega_b$ & 41.2 (33)  &  $0.685^{+0.006}_{-0.006}$  &  $0.301^{+0.008}_{-0.008}$  & \nodata  & \nodata  & \nodata  & \nodata  \\ 
SNe+BAO+CMB & 45.4 (35)  &  $0.685^{+0.003}_{-0.003}$  &  $0.301^{+0.004}_{-0.004}$  & \nodata  & \nodata  & \nodata  & \nodata  \\ 
SNe+BAO+CMB+$H_0^{\mathrm{TRGB}}$ & 46.6 (36)  &  $0.685^{+0.003}_{-0.003}$  &  $0.300^{+0.004}_{-0.004}$  & \nodata  & \nodata  & \nodata  & \nodata  \\ 
SNe+BAO+CMB+$H_0^{\mathrm{Ceph.}}$ & 72.5 (36)  &  $0.689^{+0.003}_{-0.003}$  &  $0.295^{+0.003}_{-0.003}$  & \nodata  & \nodata  & \nodata  & \nodata  \\ 
\hline
 \multicolumn{8}{c}{Open $\Lambda$CDM}\\ 
 \hline
SNe & 27.7 (19)  & \nodata  &  $0.274^{+0.063}_{-0.065}$  &  $\phantom{-}0.172^{+0.175}_{-0.166}$  & \nodata  & \nodata  & \nodata  \\ 
SNe+CMB & 28.8 (20)  &  $0.654^{+0.027}_{-0.025}$  &  $0.334^{+0.027}_{-0.026}$  &  $-0.004^{+0.006}_{-0.007}$  & \nodata  & \nodata  & \nodata  \\ 
BAO+CMB & 10.8 (13)  &  $0.687^{+0.003}_{-0.003}$  &  $0.302^{+0.004}_{-0.004}$  &  $\phantom{-}0.002^{+0.001}_{-0.001}$  & \nodata  & \nodata  & \nodata  \\ 
SNe+BAO+$\omega_b$ & 40.1 (32)  &  $0.669^{+0.016}_{-0.016}$  &  $0.293^{+0.011}_{-0.011}$  &  $\phantom{-}0.039^{+0.038}_{-0.038}$  & \nodata  & \nodata  & \nodata  \\ 
SNe+BAO+CMB & 41.1 (34)  &  $0.687^{+0.003}_{-0.003}$  &  $0.303^{+0.004}_{-0.004}$  &  $\phantom{-}0.003^{+0.001}_{-0.001}$  & \nodata  & \nodata  & \nodata  \\ 
SNe+BAO+CMB+$H_0^{\mathrm{TRGB}}$ & 42.0 (35)  &  $0.688^{+0.003}_{-0.003}$  &  $0.302^{+0.004}_{-0.004}$  &  $\phantom{-}0.003^{+0.001}_{-0.001}$  & \nodata  & \nodata  & \nodata  \\ 
SNe+BAO+CMB+$H_0^{\mathrm{Ceph.}}$ & 64.9 (35)  &  $0.692^{+0.003}_{-0.003}$  &  $0.298^{+0.004}_{-0.004}$  &  $\phantom{-}0.003^{+0.001}_{-0.001}$  & \nodata  & \nodata  & \nodata  \\ 
\hline
 \multicolumn{8}{c}{Flat $w$CDM}\\ 
 \hline
SNe & 27.2 (19)  & \nodata  &  $0.240^{+0.082}_{-0.109}$  & \nodata  &  $-0.776^{+0.158}_{-0.177}$  & \nodata  & \nodata  \\ 
SNe+CMB & 28.3 (20)  &  $0.663^{+0.012}_{-0.012}$  &  $0.325^{+0.012}_{-0.012}$  & \nodata  &  $-0.960^{+0.041}_{-0.041}$  & \nodata  & \nodata  \\ 
BAO+CMB & 14.3 (13)  &  $0.690^{+0.009}_{-0.009}$  &  $0.296^{+0.007}_{-0.007}$  & \nodata  &  $-1.022^{+0.037}_{-0.039}$  & \nodata  & \nodata  \\ 
SNe+BAO+$\omega_b$ & 37.0 (32)  &  $0.656^{+0.015}_{-0.016}$  &  $0.297^{+0.009}_{-0.009}$  & \nodata  &  $-0.898^{+0.049}_{-0.049}$  & \nodata  & \nodata  \\ 
SNe+BAO+CMB & 45.1 (34)  &  $0.681^{+0.007}_{-0.007}$  &  $0.303^{+0.006}_{-0.006}$  & \nodata  &  $-0.983^{+0.028}_{-0.028}$  & \nodata  & \nodata  \\ 
SNe+BAO+CMB+$H_0^{\mathrm{TRGB}}$ & 46.5 (35)  &  $0.684^{+0.006}_{-0.006}$  &  $0.301^{+0.005}_{-0.005}$  & \nodata  &  $-0.994^{+0.026}_{-0.027}$  & \nodata  & \nodata  \\ 
SNe+BAO+CMB+$H_0^{\mathrm{Ceph.}}$ & 66.2 (35)  &  $0.701^{+0.006}_{-0.006}$  &  $0.287^{+0.004}_{-0.004}$  & \nodata  &  $-1.059^{+0.024}_{-0.024}$  & \nodata  & \nodata  \\ 
\hline
 \multicolumn{8}{c}{Flat $w_0$-$w_a$}\\ 
 \hline
SNe & 24.4 (18)  & \nodata  &  $0.443^{+0.061}_{-0.088}$  & \nodata  &  $-0.489^{+0.419}_{-0.304}$  &  $-5.45^{+3.55}_{-4.75}$  & \nodata  \\ 
SNe+CMB & 26.1 (19)  &  $0.677^{+0.013}_{-0.014}$  &  $0.312^{+0.014}_{-0.012}$  & \nodata  &  $-0.728^{+0.154}_{-0.159}$  &  $-1.11^{+0.75}_{-0.76}$  & 1.61  \\ 
BAO+CMB & \phantom{0}7.9 (12)  &  $0.646^{+0.019}_{-0.019}$  &  $0.341^{+0.022}_{-0.021}$  & \nodata  &  $-0.532^{+0.230}_{-0.210}$  &  $-1.40^{+0.61}_{-0.69}$  & 1.81  \\ 
SNe+BAO+$\omega_b$ & 34.8 (31)  &  $0.671^{+0.014}_{-0.016}$  &  $0.322^{+0.015}_{-0.017}$  & \nodata  &  $-0.774^{+0.104}_{-0.099}$  &  $-0.79^{+0.52}_{-0.53}$  & 1.98  \\ 
SNe+BAO+CMB & 35.9 (33)  &  $0.668^{+0.008}_{-0.008}$  &  $0.318^{+0.008}_{-0.008}$  & \nodata  &  $-0.760^{+0.084}_{-0.082}$  &  $-0.79^{+0.28}_{-0.30}$  & 5.65  \\ 
SNe+BAO+CMB+$H_0^{\mathrm{TRGB}}$ & 39.1 (34)  &  $0.674^{+0.007}_{-0.007}$  &  $0.313^{+0.007}_{-0.007}$  & \nodata  &  $-0.805^{+0.079}_{-0.077}$  &  $-0.69^{+0.27}_{-0.29}$  & 6.20  \\ 
SNe+BAO+CMB+$H_0^{\mathrm{Ceph.}}$ & 64.7 (34)  &  $0.698^{+0.006}_{-0.006}$  &  $0.291^{+0.006}_{-0.005}$  & \nodata  &  $-0.984^{+0.068}_{-0.066}$  &  $-0.30^{+0.24}_{-0.26}$  & 8.02  \\ 
\hline
 \multicolumn{8}{c}{Open $w_0$-$w_a$}\\ 
 \hline
SNe+CMB & 24.5 (18)  &  $0.565^{+0.072}_{-0.040}$  &  $0.448^{+0.070}_{-0.096}$  &  $-0.045^{+0.031}_{-0.020}$  &  $-0.518^{+0.351}_{-0.269}$  &  $-4.75^{+3.02}_{-3.88}$  & \nodata  \\ 
BAO+CMB & \phantom{0}7.4 (11)  &  $0.652^{+0.022}_{-0.021}$  &  $0.337^{+0.023}_{-0.022}$  &  $\phantom{-}0.001^{+0.002}_{-0.002}$  &  $-0.599^{+0.249}_{-0.231}$  &  $-1.19^{+0.68}_{-0.76}$  & \nodata  \\ 
SNe+BAO+$\omega_b$ & 34.8 (30)  &  $0.670^{+0.018}_{-0.017}$  &  $0.321^{+0.017}_{-0.018}$  &  $\phantom{-}0.004^{+0.049}_{-0.048}$  &  $-0.774^{+0.105}_{-0.101}$  &  $-0.80^{+0.54}_{-0.57}$  & 1.57  \\ 
SNe+BAO+CMB & 34.9 (32)  &  $0.671^{+0.008}_{-0.008}$  &  $0.318^{+0.008}_{-0.008}$  &  $\phantom{-}0.001^{+0.001}_{-0.001}$  &  $-0.787^{+0.089}_{-0.087}$  &  $-0.68^{+0.31}_{-0.32}$  & 5.23  \\ 
SNe+BAO+CMB+$H_0^{\mathrm{TRGB}}$ & 37.7 (33)  &  $0.676^{+0.008}_{-0.008}$  &  $0.313^{+0.007}_{-0.007}$  &  $\phantom{-}0.002^{+0.001}_{-0.001}$  &  $-0.834^{+0.083}_{-0.082}$  &  $-0.57^{+0.29}_{-0.31}$  & 5.74  \\ 
SNe+BAO+CMB+$H_0^{\mathrm{Ceph.}}$ & 60.7 (33)  &  $0.701^{+0.006}_{-0.006}$  &  $0.292^{+0.006}_{-0.005}$  &  $\phantom{-}0.003^{+0.001}_{-0.001}$  &  $-1.023^{+0.072}_{-0.070}$  &  $-0.11^{+0.27}_{-0.28}$  & 7.34  \\ 
 \enddata
 \caption{Constraints on cosmological parameters. The SN $\chi^2$ values are based on spline-interpolated distances (with 22 nodes, so SN DoF = $22- N_{\mathrm{fit}}$), so they are much smaller than the number of SNe (\nTotUNITY). As discussed in the text, the only major tension we see between data combinations is with the Cepheid-based $H_0$ measurement (i.e., the ``Hubble tension''); generally the probes are consistent with each other. This table also shows \evidencestrength tension with flat \LCDM, with flat $w_0$-$w_a$ reducing the SN+BAO+CMB $\chi^2$ by \DeltaChiSqSNeBAOCMB.
 \label{tab:cosmoconstraints}}
\end{deluxetable*}

\newcommand{\Deltachisquared}{\ensuremath{\Delta \chi^2}\xspace}

\clearpage

In short, our most notable finding is that we continue to see \evidencestrength evidence for time-varying dark energy as parameterized with the $w_0$-$w_a$ model. In fact, we find central values similar to Union3 (which used SDSS+BOSS+eBOSS BAO + 6dF), with the mild SN update towards \LCDM ($w_0= -1$, $w_a=0$) canceled by the DESI move away from \LCDM \citep{DESICollaboration2025DR2}. However, our uncertainties are smaller, mostly due to the DESI improvement over earlier BAO measurements. Evaluating this tension with the 2D $\chi^2$ values (for $w_0$/$w_a$), we find an increase from $\SigmaTwoDBAOCMB \sigma$ ($\Deltachisquared = \DeltaChiSqBAOCMB$) to $\SigmaTwoDSNeBAOCMB \sigma$ ($\Deltachisquared = \DeltaChiSqSNeBAOCMB$) when including SN constraints. When also including TRGB-calibrated $H_0$, the significance falls slightly to $\SigmaTwoDSNeBAOCMBHOT \sigma$ ($\Deltachisquared = \DeltaChiSqSNeBAOCMBHOT$). As seen in other analyses, the Cepheid calibrated $H_0$ measurements have severe tension with the early universe measurements for all cosmological models considered.

\subsubsection{Impact of CMB Choice on Results}

\newcommand{\UnionCMB}{\texttt{Commander}+\texttt{SimAll}+\texttt{Plik}\xspace}
\newcommand{\DESICMB}{\texttt{Commander}+\texttt{SimAll}+\texttt{CamSpec}, Planck and ACT DR6 CMB lensing\xspace}
\newcommand{\PoPCMB}{\texttt{Commander}+\texttt{LoLLiPoP}+\texttt{HiLLiPoP}, Planck PR4 lensing}
\newcommand{\compressionparams}{\ensuremath{\{R,\ \theta_*,\ \Omega_bh^2\}}\xspace}

Our nominal result finds somewhat lower evidence for time-evolving dark energy than, e.g., the DESI results ($3.8\sigma$ for Union3+BAO+CMB, \citealt{DESICollaboration2025DR2}). Some of this difference is due to the assumed CMB constraints. As described in \citet{Rubin_2025}, we make our own CMB compression\footnote{In other words, replacing the full CMB likelihood with a constraint on the shift parameter, angular scale, and baryon density \compressionparams.} based on a chain from Planck PR3 (\texttt{base\_w\_plikHM\_TTTEEE\_lowl\_lowE}, \citealt{PlanckCollaboration2020}), which uses the \texttt{Plik} temperature (TT),
polarization (EE) and cross (TE) power spectra (for $\ell \geq 30$) combined with \texttt{SimAll}+\texttt{Commander} (for
$\ell < 30$). In contrast, DESI retains \texttt{SimAll}+\texttt{Commander} (for
$\ell < 30$) but combines with \texttt{CamSpec} for $\ell \geq 30$ \citep{rosenberg22}, and  also includes Planck and ACT DR6 CMB lensing \citep{Madhavacheril2024ACTDR6}. DESI also uses the full CMB likelihood.

\begin{figure}
    \centering
    \includegraphics[width=0.4\textwidth]{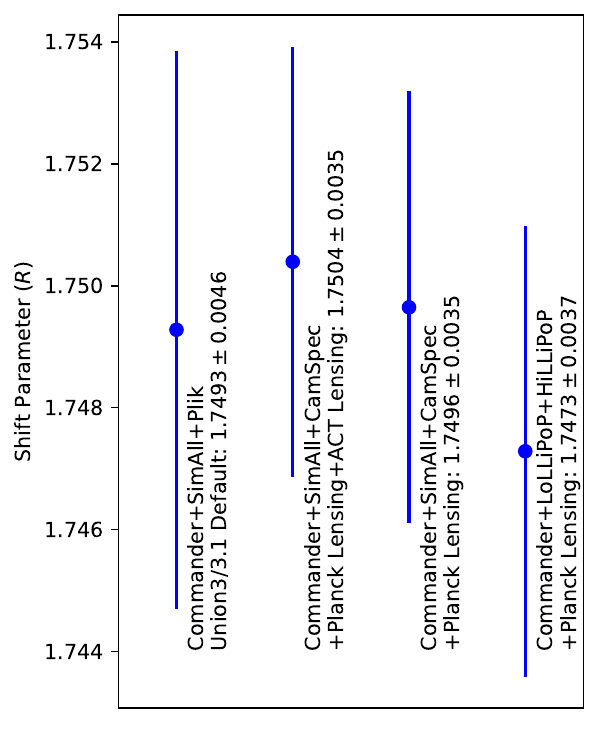}
    \caption{Comparison of our shift-parameter measurements derived from MCMC chains. The choice of CMB likelihood and data has an impact on the inferred shift parameter with a peak-to-peak range close to the quoted uncertainties.}
    \label{fig:shiftparameter}
\end{figure}

\autoref{tab:SNCMB} presents the range of significance against a cosmological constant under the $w_0$-$w_a$ parametrization. The table is broken into three blocks for three major Union+UNITY releases, Union3+UNITY1.5 \citep{Rubin_2025}, Union3.1+UNITY1.7 \citep{Hoyt2026Mass}, and Union3.1+UNITY1.8 (this work). Comparing the top row of the topmost block, based on the older Union3+UNITY1.5 SN measurements, with the other two rows of the same block emphasizes the impact of the new DESI-DR2 BAO measurements on increasing tension with $\Lambda$CDM. The bottom two rows of all three blocks in the table use the exact same assumptions and methodologies, allowing us trace directly the impact of the different SN measurements on $w_0$-$w_a$ constraints. Going from Union3+UNITY1.5 to Union3.1 and either UNITY1.7 or UNITY1.8 decreases tension against a cosmological constant by $0.4\sigma$ and $0.5\sigma$, respectively. In the last block, we explore a broader range of CMB assumptions, comparing \texttt{Commander}+\texttt{SimAll}+\texttt{CamSpec} with \texttt{Commander}+\texttt{LoLLiPop} \citep{Tristram2021} +\texttt{HiLLiPop} \citep{Tristram2024}, both with Planck PR4 lensing; this difference is $0.3\sigma$ on the level of tension.\footnote{When making compressed likelihoods, we generated \texttt{Cobaya} \citep{Torrado2021} input files using the online input generator: \url{https://cobaya-gui.streamlit.app} for \texttt{CAMB} \citep{Lewis2011}, except for the DESI combination (\DESICMB), where we used DESI's input file. We ran each chain until the \citet{Gelman1992} $\hat{R}$ statistic was below 1.005. We discarded the first 30\% of the chain as burn in and computed \compressionparams for each MCMC sample, taking each sample's weight into account.} Figure~\ref{fig:shiftparameter} shows that there are significant differences (comparable to the quoted uncertainties) in the inferred shift parameter values for these combinations.

\renewcommand{\arraystretch}{1.2}
\newcommand{\chisig}[2]{$\Delta \chi^2 = #1$, $#2\,\sigma$}

\begin{deluxetable*}{p{2cm} p{4.2cm} p{2 cm} p{2 cm} l p{2.5 cm}}[htbp]
\caption{Table showing how different SN and CMB analyses impact the observed deviations from a cosmological constant in the $w_0$-$w_a$ plane. \label{tab:SNCMB}}
    \tablehead{
  \colhead{} &
  \multicolumn{3}{c}{CMB} &
  \colhead{} &
  \colhead{} \\
  \cline{2-4} 
  \colhead{BAO} &
  \colhead{Planck Likelihoods} &
  \colhead{Lensing} &
  \colhead{Treatment} &
  \colhead{Significance} &
  \colhead{Reference}
}
\startdata
\noalign{\vskip 3pt}
\multicolumn{6}{c}{\textbf{Union3+UNITY1.5}} \\
\noalign{\vskip 3pt}
\hline
 6dF+SDSS+ BOSS+eBOSS & \makecell[lt]{\texttt{Commander}+\texttt{SimAll} (low-$\ell$) \\ \texttt{Plik} (high-$\ell$) } & No lensing & Compressed \compressionparams & \chisig{\phantom{1}7.3}{2.2} & \makecell[lt]{\cite{Rubin_2025}} \\
 DESI-DR2 +6dF & \makecell[lt]{\texttt{Commander}+\texttt{SimAll} (low-$\ell$) \\ \texttt{CamSpecNPIPE} (high-$\ell$)} & Planck+ ACT-DR6 & Compressed \compressionparams & \chisig{15.6}{3.5} & This Work \\
 DESI-DR2 & \makecell[lt]{\texttt{Commander}+\texttt{SimAll} (low-$\ell$) \\ \texttt{CamSpecNPIPE} (high-$\ell$)} & Planck+ ACT-DR6 & Full & \chisig{17.4}{3.8} & \cite{DESICollaboration2025DR2}\\
\hline
\noalign{\vskip 3pt}
\multicolumn{6}{c}{\textbf{Union3.1+UNITY1.7}} \\
\noalign{\vskip 3pt}
\hline
 DESI-DR2 +6dF & \texttt{Commander}+\texttt{SimAll} (low-$\ell$) \texttt{CamSpecNPIPE} (high-$\ell$) & Planck+ ACT-DR6 & Compressed \compressionparams & \chisig{12.4}{3.1} & This Work \\
 DESI-DR2\tablenotemark{a} & \texttt{Commander}+\texttt{SimAll} (low-$\ell$) \texttt{CamSpecNPIPE} (high-$\ell$) & Planck+ ACT-DR6 & Full & \chisig{14.4}{3.4} & \citet{Hoyt2026Mass} \\
\hline
\noalign{\vskip 3pt}
\multicolumn{6}{c}{\textbf{Union3.1+UNITY1.8}} \\
\noalign{\vskip 3pt}
\hline
 DESI-DR2 +6dF & \texttt{Commander}+\texttt{SimAll} (low-$\ell$) \texttt{Plik} (high-$\ell$)& No lensing & Compressed \compressionparams & $\Delta \chi^2 = \phantom{1}\DeltaChiSqSNeBAOCMB$, $\SigmaTwoDSNeBAOCMB \sigma$ & This Work \\
 DESI-DR2 +6dF & \texttt{Commander}+\texttt{LoLLiPoP} (low-$\ell$) \texttt{HiLLiPoP} (high-$\ell$) & Planck PR4  & Compressed \compressionparams & \chisig{\phantom{1}9.4}{2.6} & This Work \\
 sDESI-DR2 +6dF & \texttt{Commander}+\texttt{SimAll} (low-$\ell$) \texttt{CamSpecNPIPE} (high-$\ell$) & Planck PR4  & Compressed \compressionparams & \chisig{11.3}{2.9} & This Work \\
 DESI-DR2 +6dF & \texttt{Commander}+\texttt{SimAll} (low-$\ell$) \texttt{CamSpecNPIPE} (high-$\ell$) & Planck+ ACT-DR6 & Compressed \compressionparams & \chisig{12.0}{3.0} & This Work \\
 DESI-DR2\tablenotemark{a} & \texttt{Commander}+\texttt{SimAll} (low-$\ell$) \texttt{CamSpecNPIPE} (high-$\ell$) & Planck+ ACT-DR6 & Full & \chisig{14.0}{3.3} & \citet{Hoyt2026Mass} \\
\enddata
\tablecomments{Each section of the table combines a different version of the Union+UNITY SN cosmology analysis with BAO measurements and different, comparably valid choices for the adopted Planck CMB likelihood such as the treatment of noise in the maps. Note that 6dF is a very low redshift BAO point and has a small effect on inferred cosmological parameters because SNe dominate the constraining power at low redshifts.}
\tablenotetext{a}{High redshift datasets, assumptions, and methods adopted exactly as per the nominal BAO+CMB+SN analysis of \citet{DESICollaboration2025DR2}. Comparing with their published result (topmost block of table) directly traces the impact that the new Union+UNITY SN analyses have on their reported $w_0$-$w_a$ tension with $\Lambda$CDM.}
\end{deluxetable*}

\renewcommand{\arraystretch}{1}

\section{Conclusions} \label{sec:conclusion}

This work introduces UNITY1.8, a Bayesian hierarchical model that incorporates two modes of normal SNe~Ia. Each SN is only probabilistically assigned to each mode, and the relative fractions are fit for and allowed to vary in redshift and host-galaxy stellar-mass bins. The two modes have separate absolute magnitudes, $x_1$ and ``intrinsic'' (assumed Gaussian distributed) color distributions, and $x_1$ standardization coefficients.

Using tests with the predictive posterior distributions, we show evidence that UNITY1.8 matches the data better than older, single-mode UNITY models. In simulated-data testing, we do find small ($\sim 0.1$ in $x_1$) Eddington-like biases on $x_1$ population parameters. However the $x_1$ population parameters were $\sim 1\%$ of the total cosmological-uncertainty variance in Union3 and we see no evidence of bias on cosmological parameters, so we consider this acceptable.

Encouragingly, we find that UNITY1.8 makes the host-stellar-mass/luminosity step consistent with zero for unreddened SNe. The only remaining host-stellar-mass dependence is in the color-standardization coefficient of red SNe ($\beta_R$).

We find very low unexplained dispersions for the \slowhyphenmode SNe from mid-redshift SN datasets that have rolling light curves and scene-modeling photometry. Whether this is an artifact of SALT3 (e.g., an overestimated uncertainty model), or whether it is an intrinsic property of SNe~Ia will need further investigation.

We find UNITY1.8 decreases the cosmological-parameter uncertainties compared to earlier models. We still see some tension with flat \LCDM, finding \evidencestrength evidence for time-varying dark energy models. The strength of this tension is dependent on the choice of CMB likelihood and how it is treated, varying from $2.6\sigma$ to $3.3\sigma$.

Finally, we find it strongly encouraging that a substantially different standardization two-population model yields consistent cosmological constraints with prior work that assumed a single population.

\begin{acknowledgments}

We thank Paul Shah and Andrei Cuceu for discussions of CMB constraints. We also thank the referee for thoughtful feedback. This research uses services or data provided by the Astro Data Lab, which is part of the Community Science and Data Center (CSDC) Program of NSF NOIRLab. NOIRLab is operated by the Association of Universities for Research in Astronomy (AURA), Inc. under a cooperative agreement with the U.S. National Science Foundation. The technical support and advanced computing resources from University of Hawai`i Information Technology Services - Research Cyberinfrastructure, funded in part by the National Science Foundation CC* awards \#2201428 and \#2232862 are gratefully acknowledged. Support for this work was provided by NASA through grant number HST-AR-16631.001-A from the Space Telescope Science Institute, which is operated by AURA, Inc., under NASA contract NAS 5-26555. This work was supported in part by the Director, Office of Science, Office of High Energy Physics of the U.S. Department of Energy under Contract No. DE-AC02-05CH11231.

\end{acknowledgments}

\software{
Astropy \citep{Astropy},
CAMB \citep{Lewis2011},
Cobaya \citep{Torrado2021},
Extinction \citep{barbary_kyle_2016_804967},
Matplotlib \citep{matplotlib}, 
Numpy \citep{numpy},
PairV \citep{Davis2011},
PyStan \citep{allen_riddell_2018_1456206},
SciPy \citep{scipy},
SNCosmo \citep{sncosmo},
Stan \citep{Carpenter2017}
}

\clearpage

\appendix

\section{Simulated-Data Testing}
\label{sec:simdataappendix}

Simulated-data testing is an essential part of a complex analysis to ensure the correctness of the results. As noted above, we performed our simulated-data testing before unblinding the cosmological parameters from the real data. Our simulated-data testing is based on that done for Union3+UNITY1.5 \citep{Rubin_2025} with a few differences. In short, this testing simulated different rolling-survey datasets: one at very low redshift similar to the Lick Observatory Supernova Search SNe \citep{Ganeshalingam2010}, one at low redshift similar to Foundation \citep{Foley2018}, one at mid redshift similar to the Dark Energy Survey deep tier \citep{DES2016} with 600 SNe, and one at high redshift similar to the Multi-Cycle Treasury programs \citep{Riess2018} with an average of 78 SNe after selection cuts. The high-redshift dataset was purely magnitude limited (as assumed by UNITY), but the others had a simulated spectroscopic observer who picked out the brightest SNe available to observe on a given date. This makes for a more difficult (but realistic) test for UNITY. Each realization draws the zeropoint values from a normal distribution of width 0.01~mag, draws $\delta(z=\infty)/\delta(z=0)$ uniformly between 0 and 1, and draws the distribution of unexplained dispersion in ($f^{m_B}$, $f^{x_1}$, $f_c$) from a 3D simplex. The updates are the following:

\begin{itemize} 
\item To improve the realism now that UNITY will be used for Hubble-constant measurements, we now simulate two magnitude-limited nearby datasets, one in $griz$ filters with a depth of 20~AB at $5\sigma$ in all filters (the same as in Union3), but also another one with a depth of 19~AB and light curves in $BVRI$ filters. As can be seen in Tables~\ref{tab:simdata} and \ref{tab:simdataw}, after simulated spectroscopic selection, this dataset is about 1.5 magnitudes shallower. 

\item For both nearby datasets, we simulate 300 Hubble-flow SNe and 20 calibrator SNe. These calibrators have a simple distance-ladder covariance matrix of (0.05 magnitudes)$^2$ along the diagonal and (0.02 magnitudes)$^2$ zeropoint uncertainty shared among all entries; these values are roughly similar to \citealt{Riess_2022}. We also assume a mean host stellar mass of $10.4\pm1$ (not $10\pm1$) for the calibrator SNe to make sure that we can standardize across different host-galaxy selections as will be required for the Hubble constant.

\item We simulate from a two-$x_1$-mode model with parameters given in Tables~\ref{tab:simdata} and \ref{tab:simdataw}, which are similar to the values found for the real data (Section~\ref{sec:SNresults}). This enables a test of UNITY1.8, but also shows us what happens when UNITY1.7 runs on data simulated assuming a two-mode model.

\item A minor change from the Union3+UNITY1.5 testing is the inclusion of photometric uncertainties in the magnitudes used for spectroscopic selection. This was left out of the old testing to be a harder test for UNITY1.5 to pass. The bias due to selection effects scales as $(\mathrm{corrected\ Hubble\ dispersion})^2/(\mathrm{uncorrected\ dispersion})$, and if the photometric uncertainties are included in the selection, they are added in quadrature to the denominator. So sharp cuts in magnitude mean worse selection effect biases than are generally present in the real data.

\end{itemize}

Similarly to the testing in Union3, we perform a realistic test and do not test UNITY with data simulated from UNITY. For example, the simulated data generation simulates an observer who selects bright SNe to spectroscopically confirm, but UNITY models selection as an error function in magnitude. The simulated data assumes an exponential distribution of red color \cRtrue, but UNITY approximates this with a mixture of four Gaussians (\citealt{Rubin_2025}, Appendix~C). 

\newcommand{\simdatatablestenence}{For each parameter value from UNITY, we quote up to four numbers. The first number is the median of the 100 posteriors.}

Table~\ref{tab:simdata} shows the results of our simulated-data testing assuming flat \LCDM and Table~\ref{tab:simdataw} shows the results assuming flat $w_0$-$w_a$ (with \Om fixed). Each row is a parameter in the analysis while the columns compare the performance of UNITY1.8 (our new recommended analysis) and UNITY1.7. For each of the 100 realizations, for each UNITY parameter, we compute the posterior median (as an estimate of the best-fit value) and one half the difference between the 84.1st percentile and the 15.9th percentile as an estimate of the uncertainty. Then, for each UNITY parameter, we quote (up to) four numbers in the table. The first is the mean of the best-fit values. The second is the mean of the uncertainties. In the parentheses, we quote the mean of the pulls, the mean best fit minus the mean true value divided by the mean uncertainty. For the parameters where we do see a bias, this gives a sense of how significant the bias is as a fraction of the uncertainty. The second number in the parentheses is the uncertainty on the mean pull. In other words, it is the RMS of the best-fit values divided by $\sqrt{N^{\mathrm{realizations}}} = 10$ divided by the mean uncertainty. For Gaussian-distributed parameters with correctly estimated uncertainties and 100 realizations, this should be $0.1 \sigma$. Note that the true value does not have to exist in order to compute the uncertainty on the mean pull, i.e., a single $\alpha$ value does not exist in the simulations, but one can still ask if UNITY1.7 recovers a consistent $\alpha$ value (within uncertainties) across the 100 realizations.

\newcommand{\simtablecaption}{We show results for both UNITY1.8 and UNITY1.7, run on the same set of input simulations. As described in the text, we show (up to) four numbers for each parameter, for each UNITY model. The first is the mean of the best-fit values, the second is the mean uncertainty over the 100 realizations. In parentheses, we show the mean pull and the uncertainty on the mean pull.\xspace}

\begin{deluxetable*}{llrr}
\label{tab:simdata}
\tabletypesize{\scriptsize}
\tablehead{\colhead{Parameter} & \colhead{Input} & \colhead{UNITY1.8, Two-$x_1$ Modes} & \colhead{UNITY1.7}}
\startdata
\multicolumn{4}{c}{Cosmology Parameters}\\ 
\hline 
$H_0$ & $71.000$ & $70.864 \pm 1.017\ (-0.13 \pm 0.10) \sigma$ & $71.006 \pm 1.037\ (+0.01 \pm 0.09) \sigma$\\
$\Omega_m$ & $0.300$ & $0.304 \pm 0.017\ (+0.23 \pm 0.08) \sigma$ & $0.313 \pm 0.020\ (+0.69 \pm 0.08) \sigma$\\
\hline 
\multicolumn{4}{c}{Standardization Parameters}\\ 
\hline 
$\alpha$ & \nodata & \nodata & $0.158 \pm 0.011\ (\cdots \pm 0.14) \sigma$\\
$\alpha$ slow & $0.170$ & $0.141 \pm 0.023\ (-1.30 \pm 0.13) \sigma$ & \nodata\\
$\alpha$ fast $-$ slow & $0.070$ & $0.068 \pm 0.023\ (-0.09 \pm 0.11) \sigma$ & \nodata\\
$\mathcal{M}_B$ fast $-$ slow & $0.180$ & $0.159 \pm 0.024\ (-0.88 \pm 0.09) \sigma$ & \nodata\\
$\beta_B$ & $2.100$ & $2.074 \pm 0.238\ (-0.11 \pm 0.10) \sigma$ & $2.062 \pm 0.247\ (-0.15 \pm 0.11) \sigma$\\
$\beta_{RL}$ & $4.400$ & $4.427 \pm 0.190\ (+0.14 \pm 0.09) \sigma$ & $4.200 \pm 0.188\ (-1.06 \pm 0.09) \sigma$\\
$\beta_{RH}$ & $3.200$ & $3.247 \pm 0.199\ (+0.24 \pm 0.09) \sigma$ & $3.334 \pm 0.255\ (+0.53 \pm 0.10) \sigma$\\
$\delta(z=0)$ & $0.000$ & $-0.006 \pm 0.015\ (-0.42 \pm 0.09) \sigma$ & $0.024 \pm 0.015\ (+1.54 \pm 0.08) \sigma$\\
$\delta(z=\infty)/\delta(z=0)$ & $\mathcal{U}(0,\ 1)$ & $0.428 \pm 0.262\ (-0.26 \pm 0.07) \sigma$ & $0.426 \pm 0.284\ (-0.25 \pm 0.05) \sigma$\\
step mass & $10.000$ & $10.000 \pm 0.064\ (-0.00 \pm 0.12) \sigma$ & $9.997 \pm 0.069\ (-0.05 \pm 0.11) \sigma$\\
\hline 
\multicolumn{4}{c}{Population Parameters}\\ 
\hline 
$x_1^*$ Fast & $-0.900$ & $-0.765 \pm 0.114\ (+1.18 \pm 0.10) \sigma$ & \nodata\\
$R^{x_1}$ Fast & $0.800$ & $0.909 \pm 0.072\ (+1.51 \pm 0.10) \sigma$ & \nodata\\
$x_1^*$ Slow & $0.500$ & $0.496 \pm 0.038\ (-0.10 \pm 0.10) \sigma$ & \nodata\\
$R^{x_1}$ Slow & $0.500$ & $0.544 \pm 0.046\ (+0.98 \pm 0.14) \sigma$ & \nodata\\
$c^*$ Fast & $-0.050$ & $-0.049 \pm 0.008\ (+0.06 \pm 0.09) \sigma$ & \nodata\\
$R^c$ Fast & $0.060$ & $0.061 \pm 0.006\ (+0.12 \pm 0.10) \sigma$ & \nodata\\
$c^*$ Slow & $-0.080$ & $-0.078 \pm 0.005\ (+0.42 \pm 0.10) \sigma$ & \nodata\\
$R^c$ Slow & $0.040$ & $0.041 \pm 0.003\ (+0.24 \pm 0.11) \sigma$ & \nodata\\
\fslow low-$z$, high mass & $0.500$ & $0.510 \pm 0.065\ (+0.16 \pm 0.12) \sigma$ & \nodata\\
\fslow mid-$z$, high mass & $0.500$ & $0.485 \pm 0.063\ (-0.24 \pm 0.10) \sigma$ & \nodata\\
\fslow high-$z$, high mass & $0.500$ & $0.438 \pm 0.144\ (-0.43 \pm 0.10) \sigma$ & \nodata\\
\fslow low-$z$, low mass & $0.800$ & $0.801 \pm 0.048\ (+0.03 \pm 0.10) \sigma$ & \nodata\\
\fslow mid-$z$, low mass & $0.800$ & $0.809 \pm 0.046\ (+0.21 \pm 0.09) \sigma$ & \nodata\\
\fslow high-$z$, low mass & $0.800$ & $0.725 \pm 0.123\ (-0.61 \pm 0.10) \sigma$ & \nodata\\
$m_{50}$ Very Low-$z$ & \nodata & $16.187 \pm 0.090\ (\cdots \pm 0.12) \sigma$ & $16.192 \pm 0.090\ (\cdots \pm 0.12) \sigma$\\
$\sigma_m$ Very Low-$z$ & \nodata & $0.246 \pm 0.040\ (\cdots \pm 0.14) \sigma$ & $0.248 \pm 0.041\ (\cdots \pm 0.14) \sigma$\\
$\sigma^{\mathrm{unexpl}}$ Very Low-$z$ & $0.080$ & $0.068 \pm 0.017\ (-0.69 \pm 0.10) \sigma$ & $0.105 \pm 0.011\ (+2.22 \pm 0.11) \sigma$\\
$m_{50}$ Low-$z$ & \nodata & $17.650 \pm 0.103\ (\cdots \pm 0.09) \sigma$ & $17.649 \pm 0.104\ (\cdots \pm 0.10) \sigma$\\
$\sigma_m$ Low-$z$ & \nodata & $0.273 \pm 0.046\ (\cdots \pm 0.14) \sigma$ & $0.275 \pm 0.046\ (\cdots \pm 0.14) \sigma$\\
$\sigma^{\mathrm{unexpl}}$ Low-$z$ & $0.080$ & $0.069 \pm 0.013\ (-0.82 \pm 0.11) \sigma$ & $0.101 \pm 0.009\ (+2.21 \pm 0.12) \sigma$\\
$m_{50}$ Mid-$z$ & \nodata & $23.134 \pm 0.058\ (\cdots \pm 0.11) \sigma$ & $23.136 \pm 0.059\ (\cdots \pm 0.11) \sigma$\\
$\sigma_m$ Mid-$z$ & \nodata & $0.215 \pm 0.026\ (\cdots \pm 0.13) \sigma$ & $0.215 \pm 0.027\ (\cdots \pm 0.13) \sigma$\\
$\sigma^{\mathrm{unexpl}}$ Mid-$z$ & $0.080$ & $0.071 \pm 0.011\ (-0.82 \pm 0.12) \sigma$ & $0.101 \pm 0.008\ (+2.66 \pm 0.11) \sigma$\\
$m_{50}$ High-$z$ & \nodata & $25.984 \pm 0.228\ (\cdots \pm 0.06) \sigma$ & $25.984 \pm 0.233\ (\cdots \pm 0.06) \sigma$\\
$\sigma_m$ High-$z$ & \nodata & $0.303 \pm 0.143\ (\cdots \pm 0.04) \sigma$ & $0.306 \pm 0.144\ (\cdots \pm 0.04) \sigma$\\
$\sigma^{\mathrm{unexpl}}$ High-$z$ & $0.080$ & $0.066 \pm 0.031\ (-0.44 \pm 0.08) \sigma$ & $0.097 \pm 0.028\ (+0.63 \pm 0.10) \sigma$\\
$\sigma^{\mathrm{unexpl}}$ fast & $0.080$ & $0.098 \pm 0.018\ (+0.98 \pm 0.10) \sigma$ & \nodata\\
$f^{m_B}$ & Simplex & $0.472 \pm 0.236\ (+0.64 \pm 0.05) \sigma$ & $0.468 \pm 0.162\ (+0.90 \pm 0.10) \sigma$\\
$f^{x_1}$ & Simplex & $0.230 \pm 0.155\ (-0.86 \pm 0.07) \sigma$ & $0.291 \pm 0.090\ (-0.79 \pm 0.16) \sigma$\\
$f^{c}$ & Simplex & $0.254 \pm 0.188\ (-0.33 \pm 0.06) \sigma$ & $0.225 \pm 0.157\ (-0.58 \pm 0.07) \sigma$\\
$f^{\mathrm{outl}}$ & 0.008--0.029 & $0.012 \pm 0.003\ (-2.09 \pm 0.09) \sigma$ & $0.015 \pm 0.004\ (-0.93 \pm 0.09) \sigma$\\
\enddata
\caption{Simulated-data results assuming flat \LCDM. \simtablecaption}
\end{deluxetable*}

\begin{deluxetable*}{llrr}
\label{tab:simdataw}
\tabletypesize{\scriptsize}
\tablehead{\colhead{Parameter} & \colhead{Input} & \colhead{UNITY1.8, Two-$x_1$ Modes} & \colhead{UNITY1.7}}
\startdata
\multicolumn{4}{c}{Cosmology Parameters}\\ 
\hline 
$H_0$ & $71.000$ & $70.870 \pm 1.030\ (-0.13 \pm 0.09) \sigma$ & $71.085 \pm 1.050\ (+0.08 \pm 0.09) \sigma$\\
$w_0$ & $-1.000$ & $-0.993 \pm 0.119\ (+0.06 \pm 0.09) \sigma$ & $-1.034 \pm 0.123\ (-0.28 \pm 0.09) \sigma$\\
$w_0 + 0.15\;w_a$ & $-1.000$ & $-0.999 \pm 0.048\ (+0.02 \pm 0.08) \sigma$ & $-0.982 \pm 0.053\ (+0.35 \pm 0.09) \sigma$\\
$w_a$ & $0.000$ & $-0.029 \pm 0.735\ (-0.04 \pm 0.09) \sigma$ & $0.367 \pm 0.738\ (+0.50 \pm 0.08) \sigma$\\
\hline 
\multicolumn{4}{c}{Standardization Parameters}\\ 
\hline 
$\alpha$ & \nodata & \nodata & $0.158 \pm 0.011\ (\cdots \pm 0.14) \sigma$\\
$\alpha$ slow & $0.170$ & $0.140 \pm 0.022\ (-1.34 \pm 0.13) \sigma$ & \nodata\\
$\alpha$ fast $-$ slow & $0.070$ & $0.067 \pm 0.022\ (-0.13 \pm 0.11) \sigma$ & \nodata\\
$\mathcal{M}_B$ fast $-$ slow & $0.180$ & $0.159 \pm 0.024\ (-0.88 \pm 0.09) \sigma$ & \nodata\\
$\beta_B$ & $2.100$ & $2.073 \pm 0.239\ (-0.11 \pm 0.10) \sigma$ & $2.065 \pm 0.249\ (-0.14 \pm 0.11) \sigma$\\
$\beta_{RL}$ & $4.400$ & $4.433 \pm 0.190\ (+0.17 \pm 0.09) \sigma$ & $4.204 \pm 0.188\ (-1.04 \pm 0.09) \sigma$\\
$\beta_{RH}$ & $3.200$ & $3.253 \pm 0.199\ (+0.27 \pm 0.09) \sigma$ & $3.336 \pm 0.247\ (+0.55 \pm 0.10) \sigma$\\
$\delta(z=0)$ & $0.000$ & $-0.007 \pm 0.015\ (-0.44 \pm 0.09) \sigma$ & $0.023 \pm 0.015\ (+1.53 \pm 0.08) \sigma$\\
$\delta(z=\infty)/\delta(z=0)$ & $\mathcal{U}(0,\ 1)$ & $0.433 \pm 0.264\ (-0.24 \pm 0.07) \sigma$ & $0.438 \pm 0.287\ (-0.20 \pm 0.05) \sigma$\\
step mass & $10.000$ & $9.998 \pm 0.065\ (-0.03 \pm 0.12) \sigma$ & $9.997 \pm 0.069\ (-0.04 \pm 0.12) \sigma$\\
\hline 
\multicolumn{4}{c}{Population Parameters}\\ 
\hline 
$x_1^*$ Fast & $-0.900$ & $-0.766 \pm 0.114\ (+1.18 \pm 0.10) \sigma$ & \nodata\\
$R^{x_1}$ Fast & $0.800$ & $0.913 \pm 0.072\ (+1.56 \pm 0.10) \sigma$ & \nodata\\
$x_1^*$ Slow & $0.500$ & $0.496 \pm 0.038\ (-0.10 \pm 0.10) \sigma$ & \nodata\\
$R^{x_1}$ Slow & $0.500$ & $0.546 \pm 0.045\ (+1.02 \pm 0.14) \sigma$ & \nodata\\
$c^*$ Fast & $-0.050$ & $-0.049 \pm 0.008\ (+0.09 \pm 0.09) \sigma$ & \nodata\\
$R^c$ Fast & $0.060$ & $0.061 \pm 0.006\ (+0.14 \pm 0.10) \sigma$ & \nodata\\
$c^*$ Slow & $-0.080$ & $-0.078 \pm 0.005\ (+0.43 \pm 0.10) \sigma$ & \nodata\\
$R^c$ Slow & $0.040$ & $0.041 \pm 0.003\ (+0.27 \pm 0.11) \sigma$ & \nodata\\
\fslow low-$z$, high mass & $0.500$ & $0.510 \pm 0.065\ (+0.15 \pm 0.12) \sigma$ & \nodata\\
\fslow mid-$z$, high mass & $0.500$ & $0.486 \pm 0.063\ (-0.23 \pm 0.10) \sigma$ & \nodata\\
\fslow high-$z$, high mass & $0.500$ & $0.436 \pm 0.145\ (-0.44 \pm 0.10) \sigma$ & \nodata\\
\fslow low-$z$, low mass & $0.800$ & $0.801 \pm 0.048\ (+0.03 \pm 0.11) \sigma$ & \nodata\\
\fslow low-$z$, low mass & $0.800$ & $0.810 \pm 0.046\ (+0.22 \pm 0.09) \sigma$ & \nodata\\
\fslow low-$z$, low mass & $0.800$ & $0.723 \pm 0.123\ (-0.62 \pm 0.10) \sigma$ & \nodata\\
$m_{50}$ Very Low-$z$ & \nodata & $16.189 \pm 0.091\ (\cdots \pm 0.12) \sigma$ & $16.197 \pm 0.089\ (\cdots \pm 0.12) \sigma$\\
$\sigma_m$ Very Low-$z$ & \nodata & $0.246 \pm 0.041\ (\cdots \pm 0.14) \sigma$ & $0.247 \pm 0.041\ (\cdots \pm 0.14) \sigma$\\
$\sigma^{\mathrm{unexpl}}$ Very Low-$z$ & $0.080$ & $0.068 \pm 0.017\ (-0.70 \pm 0.10) \sigma$ & $0.105 \pm 0.011\ (+2.19 \pm 0.11) \sigma$\\
$m_{50}$ Low-$z$ & \nodata & $17.652 \pm 0.103\ (\cdots \pm 0.09) \sigma$ & $17.652 \pm 0.103\ (\cdots \pm 0.10) \sigma$\\
$\sigma_m$ Low-$z$ & \nodata & $0.273 \pm 0.046\ (\cdots \pm 0.14) \sigma$ & $0.275 \pm 0.046\ (\cdots \pm 0.14) \sigma$\\
$\sigma^{\mathrm{unexpl}}$ Low-$z$ & $0.080$ & $0.069 \pm 0.013\ (-0.86 \pm 0.11) \sigma$ & $0.101 \pm 0.009\ (+2.18 \pm 0.12) \sigma$\\
$m_{50}$ Mid-$z$ & \nodata & $23.132 \pm 0.059\ (\cdots \pm 0.11) \sigma$ & $23.135 \pm 0.060\ (\cdots \pm 0.11) \sigma$\\
$\sigma_m$ Mid-$z$ & \nodata & $0.215 \pm 0.026\ (\cdots \pm 0.13) \sigma$ & $0.216 \pm 0.027\ (\cdots \pm 0.13) \sigma$\\
$\sigma^{\mathrm{unexpl}}$ Mid-$z$ & $0.080$ & $0.071 \pm 0.011\ (-0.81 \pm 0.12) \sigma$ & $0.102 \pm 0.008\ (+2.68 \pm 0.11) \sigma$\\
$m_{50}$ High-$z$ & \nodata & $25.984 \pm 0.234\ (\cdots \pm 0.06) \sigma$ & $25.997 \pm 0.235\ (\cdots \pm 0.06) \sigma$\\
$\sigma_m$ High-$z$ & \nodata & $0.305 \pm 0.144\ (\cdots \pm 0.04) \sigma$ & $0.304 \pm 0.146\ (\cdots \pm 0.04) \sigma$\\
$\sigma^{\mathrm{unexpl}}$ High-$z$ & $0.080$ & $0.066 \pm 0.031\ (-0.44 \pm 0.08) \sigma$ & $0.097 \pm 0.028\ (+0.62 \pm 0.10) \sigma$\\
$\sigma^{\mathrm{unexpl}}$ fast & $0.080$ & $0.098 \pm 0.018\ (+0.97 \pm 0.10) \sigma$ & \nodata\\
$f^{m_B}$ & Simplex & $0.486 \pm 0.237\ (+0.69 \pm 0.06) \sigma$ & $0.462 \pm 0.160\ (+0.88 \pm 0.10) \sigma$\\
$f^{x_1}$ & Simplex & $0.223 \pm 0.155\ (-0.90 \pm 0.07) \sigma$ & $0.295 \pm 0.091\ (-0.75 \pm 0.16) \sigma$\\
$f^{c}$ & Simplex & $0.247 \pm 0.188\ (-0.36 \pm 0.06) \sigma$ & $0.226 \pm 0.158\ (-0.56 \pm 0.07) \sigma$\\
$f^{\mathrm{outl}}$ & 0.008--0.029 & $0.012 \pm 0.003\ (-2.07 \pm 0.09) \sigma$ & $0.015 \pm 0.004\ (-0.93 \pm 0.10) \sigma$\\
\enddata
\caption{Simulated-data results assuming flat $w_0$-$w_a$. \simtablecaption}
\end{deluxetable*}

\section{BAO Results} \label{sec:BAOresults}

\newcommand{\zprime}{\ensuremath{z^{\prime}}\xspace}
\newcommand{\zeff}{\ensuremath{z_{\mathrm{eff}}}\xspace}
\newcommand{\rdours}{\ensuremath{r_d^{\mathrm{ours}}}\xspace}
\newcommand{\rdfidours}{\ensuremath{r_{d, \mathrm{fid}}^{\mathrm{ours}}}\xspace}
\newcommand{\rdfid}{\ensuremath{r_{d, \mathrm{fid}}}\xspace}

Table~\ref{tab:BAOresults} presents the BAO distances used in this work. We standardize the sound horizon ($r_d$) values to the assumptions used in the cosmology analysis following \citet{Rubin_2025}. In short, BAO distances are measured with respect to the sound horizon:
\begin{equation}
    r_d = \int_{z_d}^{\infty} \frac{c_s (\zprime) d \zprime}{H(\zprime)} \;.
\end{equation}

Unfortunately, different approximations can scale $r_d$ by a cosmology-independent constant and different results are quoted relative to different fiducial cosmologies. Denoting \rdours to be $r_d$ using our approximations (described in more detail in \citealt{Rubin_2025}), we rescale all BAO measurements to be on a consistent scale. For example, for the DESI BGS measurement, we write
\begin{equation}
    D_V(\zeff = 0.295) \left[ \frac{\rdfidours}{\rdours} \right] = 1168.189\pm11.032~\mathrm{Mpc} \;. \nonumber
\end{equation}
where \rdfidours has to be computed for the same cosmological parameters as each BAO analysis used to compute their \rdfid.

\begin{deluxetable*}{llccccccccccccccccc}
\label{tab:BAOresults}
\tabletypesize{\scriptsize}
\rotate
\tablehead{
  \colhead{Measurement} & \colhead{Type} & \colhead{$z$} & \colhead{Our $r_d$} & \colhead{Distance} & \multicolumn{14}{c}{Covariance Matrix}
}
\startdata
6dF & $D_V$ & 0.106 & 153.745 & 457.574 & 417.279 & 0 & 0 & 0 & 0 & 0 & 0 & 0 & 0 & 0 & 0 & 0 & 0 & 0 \\
BGS & $D_V$ & 0.295 & 150.617 & 1168.189 & 0 & 121.700 & 0 & 0 & 0 & 0 & 0 & 0 & 0 & 0 & 0 & 0 & 0 & 0 \\
LRG1 & $D_M$ & 0.51 & 150.617 & 1998.659 & 0 & 0 & 603.392 & 20.432 & 0 & 0 & 0 & 0 & 0 & 0 & 0 & 0 & 0 & 0 \\
LRG1 & $H$ & 0.51 & 150.617 & 93.224 & 0 & 0 & 20.432 & 3.284 & 0 & 0 & 0 & 0 & 0 & 0 & 0 & 0 & 0 & 0 \\
LRG2 & $D_M$ & 0.706 & 150.617 & 2552.159 & 0 & 0 & 0 & 0 & 677.818 & 18.691 & 0 & 0 & 0 & 0 & 0 & 0 & 0 & 0 \\
LRG2 & $H$ & 0.706 & 150.617 & 104.763 & 0 & 0 & 0 & 0 & 18.691 & 3.158 & 0 & 0 & 0 & 0 & 0 & 0 & 0 & 0 \\
LRG3+ELG1 & $D_M$ & 0.934 & 150.617 & 3173.614 & 0 & 0 & 0 & 0 & 0 & 0 & 499.866 & 11.756 & 0 & 0 & 0 & 0 & 0 & 0 \\
LRG3+ELG1 & $H$ & 0.934 & 150.617 & 115.535 & 0 & 0 & 0 & 0 & 0 & 0 & 11.756 & 1.598 & 0 & 0 & 0 & 0 & 0 & 0 \\
ELG2 & $D_M$ & 1.321 & 150.617 & 4059.831 & 0 & 0 & 0 & 0 & 0 & 0 & 0 & 0 & 2187.865 & 45.501 & 0 & 0 & 0 & 0 \\
ELG2 & $H$ & 1.321 & 150.617 & 143.775 & 0 & 0 & 0 & 0 & 0 & 0 & 0 & 0 & 45.501 & 5.024 & 0 & 0 & 0 & 0 \\
QSO & $D_M$ & 1.484 & 150.617 & 4488.010 & 0 & 0 & 0 & 0 & 0 & 0 & 0 & 0 & 0 & 0 & 12496.646 & 357.834 & 0 & 0 \\
QSO & $H$ & 1.484 & 150.617 & 159.020 & 0 & 0 & 0 & 0 & 0 & 0 & 0 & 0 & 0 & 0 & 357.834 & 40.985 & 0 & 0 \\
Lya & $D_M$ & 2.33 & 150.617 & 5734.745 & 0 & 0 & 0 & 0 & 0 & 0 & 0 & 0 & 0 & 0 & 0 & 0 & 6100.358 & 93.002 \\
Lya & $H$ & 2.33 & 150.617 & 236.116 & 0 & 0 & 0 & 0 & 0 & 0 & 0 & 0 & 0 & 0 & 0 & 0 & 93.002 & 7.633 \\
\enddata
\caption{BAO distances for this analysis. The first line is from 6dF \citep{Beutler2011}; the others are from DESI DR2 \citep{DESICollaboration2025DR2}.}
\end{deluxetable*}

\section{Parameters and Priors} \label{sec:priors}

Table~\ref{tab:Priors} presents a list of the parameters and priors in UNITY1.8. As with previous versions, we generally prefer using weakly informative priors and parameter limits much larger than the posteriors.

\begin{deluxetable*}{p{7 cm}p{10 cm}}
 \tablehead{
 \colhead{Prior} & \colhead{Parameter Description}  }
 \startdata
 \hline
 \multicolumn{2}{c}{Standardization Coefficients}\\
 \hline
$\MBslow \sim \mathcal{N}(-19,\ 0.5^2)$ & Slow-population absolute magnitude (one per sample if making per-sample residual plots) \\
$\MBfastminusslow \sim \mathcal{U}(-1,\ 1)$ & Fast minus slow absolute-magnitude offset \\
$H_0 \sim \mathcal{U}(50,\ 100)$ & Hubble constant \\
$\alphafast = \tan(\alpha^{\mathrm{angle,\; fast}}), \quad \alpha^{\mathrm{angle,\; fast}} \sim \mathcal{U}(0,\ 0.35)$ & $x_1$ standardization coefficient for fast mode \\
$\alphaslow = \tan(\alpha^{\mathrm{angle,\; slow}}), \quad \alpha^{\mathrm{angle,\; slow}} \sim \mathcal{U}(0,\ 0.35)$ & $x_1$ standardization coefficient for slow mode \\
$\beta_B = \tan(\beta_B^{\mathrm{angle}}), \quad \beta_B^{\mathrm{angle}} \sim \mathcal{U}(-1.4,\ 1.4)$ & $c_B$ standardization coefficient \\
$\beta_{R}^{\mathrm{low}} = \tan(\beta^{\mathrm{angle,\; low}}_R), \quad \beta^{\mathrm{angle,\; low}}_R \sim \mathcal{U}(0,\ 1.4)$ & $c_R$ standardization coefficient in low-stellar-mass hosts \\
$\beta_{R}^{\mathrm{high}} = \tan(\beta^{\mathrm{angle,\; high}}_R), \quad \beta^{\mathrm{angle,\; high}}_R \sim \mathcal{U}(0,\ 1.4)$ & $c_R$ standardization coefficient in high-stellar-mass hosts \\
$\log_{10} M_{\mathrm{step}} \sim \mathcal{U}(9.0,\ 10.5)$ & Host-mass step location, affects both $\delta$ and $\Delta \beta_R$ \\
$\delta(z=0) \sim \mathcal{N}(0,\ 0.2^2)$ & Host-mass/luminosity standardization coefficient at $z=0$ \\
$\delta(z=\infty)/\delta(z=0) \sim \mathcal{U}(0,\ 1)$ & Fractional high-$z$ asymptote of host-mass step, also affects $\Delta \beta_R$ \\
 \hline
 \multicolumn{2}{c}{Cosmology Parameters}\\
 \hline
$\Omega_m \sim \mathcal{U}(0,\ 1)$ & Unless $\Omega_m$ fixed, or $j_0$-$q_0$ or distance-modulus-spline models active\\
$w$ or $w_0 \sim \mathcal{U}(-2,\ 0)$ & Dark-energy equation-of-state parameter ($w$CDM or $w_0$--$w_a$) \\
$w_a \sim \mathcal{U}(-5,\ 5)$ & Time variation of dark energy ($w_0$-$w_a$ only) \\
$q_0 \sim \mathcal{U}(-2,\ 2)$ & Deceleration parameter ($q_0$-$j_0$ model only) \\
$j_0 \sim \mathcal{U}(-5,\ 5)$ & Jerk parameter ($q_0$-$j_0$ model only) \\
$(\Omega_m,\ w_0,\ w_a) \sim \mathcal{N}_{3D}(\mu_{\mathrm{BAO+CMB}},\ \Sigma_{\mathrm{BAO+CMB}})$ & External BAO+CMB prior if included ($w_0$--$w_a$ mode only) \\
$\mu \sim \mathcal{N}(0,\ 1^2) + \mu(\mathrm{Flat}\Lambda\mathrm{CDM},\ \Omega_m = 0.3)$ & Distance-modulus spline nodes (when active) \\
 \hline
 \multicolumn{2}{c}{Unexplained Dispersion}\\
 \hline
$\sigma^{\text{unexplained}}_j  \sim \mathcal{U}(0.01,\ 0.3)$ & Sample-dependent unexplained dispersion \\
$\sigma^{\text{unexplained,\; fast}}  \sim \mathcal{U}(0.01,\ 0.15)$ & Extra unexplained dispersion for fast population \\
$\sigma^{\text{unexplained,\; calib}}  \sim \mathcal{U}(0.01,\ 0.3)$ & Extra unexplained dispersion for calibrators (when fitting $H_0$) \\
$f^{m_B}, f^{x_1}, f^c \sim$ 3D Unit Simplex & Fraction of unexplained dispersion in $\{m_B, x_1, c\}$ \\
 \hline
 \multicolumn{2}{c}{Population Parameters}\\
 \hline
$x_{1,i}^{\mathrm{true}} \sim (1-f^{\mathrm{slow}}_{km})\,\mathcal{N}(\xonestarfast,\ (\Rxonefast)^2) + f^{\mathrm{slow}}_{km}\,\mathcal{N}(\xonestarslow,\ (\Rxoneslow)^2)$ & Two-Gaussian latent $x_1$ distribution \\
$\xonestarfast \sim \mathcal{N}(-1,\ 2^2)\,\mathcal{U}(-5,\ -0.5)$ & Mean of fast latent $x_1$ component \\
$\Rxonefast \sim \mathcal{N}(1,\ 2^2)\,\mathcal{U}(0.1,\ 2)$ & Width of fast latent $x_1$ component \\
$\xonestarslow \sim \mathcal{N}(1,\ 2^2)\,\mathcal{U}(-0.5,\ 5)$ & Mean of slow latent $x_1$ component \\
$\Rxoneslow \sim \mathcal{N}(1,\ 2^2)\,\mathcal{U}(0.1,\ 2)$ & Width of slow latent $x_1$ component \\
$f^{\mathrm{slow}}_{km} \sim \mathcal{U}(0.01,\ 0.99)$ & Redshift- and mass-dependent slow-population fraction \\
$c_{B,i}^{\mathrm{true}} \sim (1-f^{\mathrm{slow}}_{km})\,\mathcal{N}(\cstarfast,\ (\Rcfast)^2) + f^{\mathrm{slow}}_{km}\,\mathcal{N}(\cstarslow,\ (\Rcslow)^2)$ & Latent blue-color distribution \\
$\cstarfast \sim \mathcal{N}(-0.1,\ 0.2^2)\,\mathcal{U}(-0.5,\ 0.5)$ & Mean of latent $c_B$ distribution for fast population \\
$\cstarslow \sim \mathcal{N}(-0.1,\ 0.2^2)\,\mathcal{U}(-0.5,\ 0.5)$ & Mean of latent $c_B$ distribution for slow population \\
$\Rcfast \sim \mathcal{N}(0.1,\ 0.2^2)\,\mathcal{U}(0.01,\ 0.2)$ & Width of latent $c_B$ distribution for fast population \\
$\Rcslow \sim \mathcal{N}(0.1,\ 0.2^2)\,\mathcal{U}(0.01,\ 0.2)$ & Width of latent $c_B$ distribution for slow population \\
$c_{R,i}^{\mathrm{true}} = \mathrm{ExponentialApprox}(1/\tau_{c_R,\;km})$ & Latent red-color \\
$\tau_{c_R,\;km} \sim \mathcal{N}(0.1,\ 0.2^2)\,\mathcal{U}(0.01,\ 0.2)$ & Scale of latent red-color distribution, redshift- and mass-dependent \\
 \hline
 \multicolumn{2}{c}{Outlier Population Parameters}\\
 \hline
$f^{\mathrm{outl}} \sim \mathrm{LogNormal}(\log(0.02),\ 0.5^2)\ \mathcal{U}(0.001,\ 0.1)$ & Fraction of outliers \\
$\sigma_{m_B}^{\mathrm{outl}} \sim \mathcal{N}(0.5,\ 0.5^2)\,\mathcal{U}(0.2,\ 2)$ & Outlier width in $m_B$ \\
$\sigma_{x_1}^{\mathrm{outl}} \sim \mathcal{N}(3,\ 3^2)\,\mathcal{U}(1,\ 10)$ & Outlier width in $x_1$ \\
$\sigma_{c_B}^{\mathrm{outl}} \sim \mathcal{N}(0.5,\ 0.5^2)\,\mathcal{U}(0.1,\ 1)$ & Outlier width in $c_B$ \\
$\sigma_{c_R,\mathrm{unit}}^{\mathrm{outl}} \sim \mathcal{N}(10,\ 3^2)\,\mathcal{U}(1,\ 10)$ & Outlier width in $c_R/\tau_{c_R,\;k}$ \\
$\sigma_{c_R}^{\mathrm{outl}} \equiv \sigma_{c_R,\mathrm{unit}}^{\mathrm{outl}} \; \tau_{c_R,\;k}$ & Outlier width in $c_R$ \\
 \hline
 \multicolumn{2}{c}{Systematic Uncertainties}\\
 \hline
$\Delta \mathrm{sys}_l \sim \mathcal{N}(0,\ 1^2)$ & Normalized calibration/systematics perturbations \\
$m^{\mathrm{cut}}_j \sim \mathcal{N}(m^{\mathrm{estimated}}_j,\ 0.5^2)\,\mathcal{U}(14,\ 30)$ & Median survey depth \\
$\sigma^{\mathrm{cut}}_j \sim \mathcal{N}(\sigma^{\mathrm{estimated}}_j,\ 0.25^2)\,\mathcal{U}(0.1,\ 3)$ & Survey-depth width \\
\enddata
 \caption{Priors used in this analysis. A subscript of $i$ indicates per-SN values, a subscript of $j$ indicates per-dataset values, a subscript of $k$ indicates redshift range (low-$z$, mid-$z$, or high-$z$), a subscript of $l$ indicates other (global) multiple values, and a subscript of $m$ indicates host-galaxy stellar mass (low or high). \label{tab:Priors}}
\end{deluxetable*}

{}

\end{document}